\documentclass[12pt,letterpaper]{article}

\usepackage[includeheadfoot,
            marginratio={1:1,2:3}, 
            width=412pt,
            height=688pt,]{geometry}
            \usepackage{tikz}
\usepackage{tikz-feynman}
\usepackage[utf8]{inputenc}
\usepackage{amsmath}
\usepackage{amsfonts}
\usepackage{amsthm}
\usepackage{amssymb}
\usepackage{graphicx}
\usepackage{booktabs}
\usepackage{empheq}
\usepackage{paralist}
\usepackage{cite}
\usepackage[hidelinks]{hyperref}
\usepackage{xcolor}
\usepackage{tensor}
\usepackage[export]{adjustbox}
\usepackage{braket}
\usepackage{tikz}
\usepackage{tikz-cd}
\usepackage{yfonts}
\usepackage{enumitem}
\usepackage{physics}
\usepackage{mdframed}
\usepackage{subcaption}
\usepackage{array}
\usetikzlibrary{decorations.pathreplacing,calc}

\numberwithin{equation}{section}
\numberwithin{table}{section}

\newcommand{\nc}{\newcommand}
\nc{\lb}{\llbracket}
\nc{\rb}{\rrbracket}
\nc{\gl}{\llbracket}
\nc{\gr}{\rrbracket}
\nc{\bbR}{\mathbb{R}}
\nc{\bbC}{\mathbb{C}}
\nc{\bbZ}{\mathbb{Z}}
\nc{\cO}{\mathcal{O}}
\nc{\cS}{\mathcal{S}}
\nc{\cM}{\mathcal{M}}
\nc{\cT}{\mathcal{T}}
\nc{\cX}{\mathcal{X}}
\nc{\cQ}{\mathcal{Q}}
\nc{\cD}{\mathcal{D}}
\nc{\cC}{\mathcal{C}}
\nc{\cG}{\mathcal{G}}
\nc{\cF}{\mathcal{F}}
\nc{\cI}{\mathcal{I}}
\nc{\pd}{\partial}
\nc{\eps}{\epsilon}
\nc{\la}{\lambda}
\newcommand\beq{\begin{equation}}
\newcommand\eeq{\end{equation}}
\nc{\del}{\partial}
\nc{\tri}{\hspace{-31pt}\vartriangle\hspace{-31pt}}
\nc{\blacktri}{\blacktriangle}
\nc{\eq}[1]{\begin{equation}
                     \begin{split} #1 \end{split}
                     \end{equation}}
\nc{\ul}{\underline}
\nc{\ov}{\overline}
\nc{\fa}{\hat}
\nc{\fb}{\MakeUppercase}
\nc{\fc}{\tilde }
\nc{\Lie}{{\cal L}} 
\nc{\lambdabar}{{\mkern0.75mu\mathchar '26\mkern -9.75mu\lambda}}

\newcommand{\R}{\mathbb{R}}

\renewcommand{\S}{\mathbb{S}}

\renewcommand{\L}{\mathcal{L}}

\newcommand{\A}{\mathcal{A}}

\newcommand{\N}{\mathbb{N}}
\newcommand{\Z}{\mathbb{Z}}

\newcommand{\D}{\mathcal{D}}
\newcommand{\E}{\mathcal{E}}

\newcommand{\F}{\mathcal{F}}

\newcommand{\av}{\alpha_v}
\newcommand{\nuv}{\nu_v}
\newcommand{\Nv}{N_{\rm v}}

\newmuskip\pFqmuskip

\newcommand*\pFq[7][8]{
  \begingroup 
  \pFqmuskip=#1mu\relax
  \mathchardef\normalcomma=\mathcode`,
  \mathcode`\,=\string"8000
  \begingroup\lccode`\~=`\,
  \lowercase{\endgroup\let~}\pFqcomma
  {}_{#2}{#3}_{#4}{\left[\left.\genfrac..{0pt}{}{#5}{#6}\right|#7\right]}
  \endgroup
}
\newcommand{\pFqcomma}{{\normalcomma}\mskip\pFqmuskip}

\allowdisplaybreaks[2]

\tikzset{connect/.style={rounded corners=#1,
        to path= ($(\tikztostart)!-#1!(\tikztotarget)!-#1!-90:(\tikztotarget)$) -- ($(\tikztotarget)!-#1!(\tikztostart)!-#1!90:(\tikztostart)$) --
        ($(\tikztotarget)!-#1!(\tikztostart)!#1!90:(\tikztostart)$) -- ($(\tikztostart)!-#1!(\tikztotarget)!-#1!90:(\tikztotarget)$) -- cycle (\tikztotarget)
}}
\tikzset{connect/.default=4mm}

\tikzset{connectup/.style={rounded corners=#1,
        to path= ($(\tikztostart)!-#1!(\tikztotarget)!-#1!-90:(\tikztotarget)$) -- ($(\tikztotarget)!-#1!(\tikztostart)!-#1!90:(\tikztostart)$) --
        ($(\tikztotarget)!-#1!(\tikztostart)!#1!90:(\tikztostart)$) -- ($(\tikztostart)!-#1!(\tikztotarget)!-#1!90:(\tikztotarget)$)  (\tikztotarget)
}}
\tikzset{connectup/.default=4mm}

\tikzset{connectsmall/.style={rounded corners=#1,
        to path= ($(\tikztostart)!-#1!(\tikztotarget)!-#1!-90:(\tikztotarget)$) -- ($(\tikztotarget)!-#1!(\tikztostart)!-#1!90:(\tikztostart)$) --
        ($(\tikztotarget)!-#1!(\tikztostart)!#1!90:(\tikztostart)$) -- ($(\tikztostart)!-#1!(\tikztotarget)!-#1!90:(\tikztotarget)$) -- cycle (\tikztotarget)
}}
\tikzset{connectsmall/.default=1.5mm}

\begin{document}

\vspace*{.6cm}
\begin{center}
{\Large
A Reduction Algorithm for Cosmological Correlators:\\[.3cm]
Cuts, Contractions, and Complexity
}

\vspace{.6cm}
\end{center}

\vspace{0.35cm}
\begin{center}
 Thomas W.~Grimm, Arno~Hoefnagels, and Mick van Vliet
\end{center}

\vspace{1cm}
\begin{center} 
\vspace{0.5cm} 
\emph{
Institute for Theoretical Physics, Utrecht University,\\ 
Princetonplein 5, 3584 CC Utrecht, 
The Netherlands } \\
   
\vspace{0.2cm}

\vspace{0.3cm}
\end{center}

\vspace{01cm}


\begin{abstract}
\noindent
Cosmological correlators are fundamental observables in an expanding universe and are highly non-trivial functions even at tree-level.
In this work, we uncover novel structures in the space of such tree-level correlators that enable us to develop a new recursive algorithm for their explicit computation. We begin by formulating cosmological correlators as solutions to GKZ systems and develop a general strategy to construct additional differential operators, called reduction operators, when a GKZ system is reducible. Applying this framework, we determine all relevant reduction operators, and show that they can be used to build up the space of functions needed to represent the correlators. Beyond relating different integrals, these operators also yield a large number of algebraic relations, including cut and contraction relations between diagrams. This implies a significant reduction in the number of functions needed to represent each tree-level cosmological correlator.  We present first steps to quantify the complexity of our reduction algorithm by using the Pfaffian framework. 
While we focus on tree-level cosmological correlators, our approach provides a blueprint for other perturbative settings.

\end{abstract}

\clearpage

\tableofcontents


\newpage

\parskip=.2cm

\section{Introduction}

The evaluation of observables in quantum field theory is generally a challenging task, and considerable research is dedicated to developing new techniques to address it. This challenge becomes even more pronounced in non-flat spacetimes, such as in cosmological settings that describe an expanding  universe. In these scenarios, the natural observables are cosmological correlators, which are evaluated on a fixed time-slice \cite{weinberg_quantum_2005,benincasa_amplitudes_2022,lee_records_2024}. 
Computing these correlators in perturbation theory requires supplementing the standard Feynman rules with additional time integrals at each vertex, since virtual particles can be created at any point in the prior evolution of the universe. Consequently, even tree-level computations of cosmological correlators present significant challenges. Understanding the general mathematical structure of these correlators is an essential aspect of describing the history of our universe and forms an active topic of research \cite{baumann_bootstrapping_2020,sleight_bootstrapping_2020,arkani-hamed_cosmological_2020,baumann_cosmological_2020,baumann_cosmological_2021,pajer_boostless_2020,goodhew_cosmological_2021,kuhne_faces_2022,pimentel_boostless_2022,hogervorst_nonperturbative_2023,wang_bootstrapping_2023,arkani-hamed_differential_2023,jazayeri_parity_2023,jazayeri_shapes_2023,benincasa_geometry_2024,chowdhury_subtle_2024,choudhury_primordial_2024,fan_cosmological_2024,aoki_cosmological_2024,kuhne_faces_2024,albayrak_perturbative_2024,liu_dispersive_2024,arkani-hamed_differential_2024,baumann_new_2024,lee_records_2024,werth_cosmoflow_2024,anninos_sitter_2024,melville_sitter_2024,goodhew_cosmological_2024,werth_spectral_2024,fevola_algebraic_2025,de_physical_2024,qin_cosmological_2024,cespedes_massive_2025}. A key guiding principle in this endeavor is to seek the simplest possible algorithm of representing and calculating these correlators. In this work, we take a significant step forward by introducing a new approach that systematically explores the space of all tree-level correlators with a focus on minimizing the computational complexity.

A powerful method for describing cosmological correlators is to capture them in a system of differential equations, which has recently led to many new insights  \cite{pinol_cosmological_2023,artico_integrationbyparts_2024,de_cosmology_2024,he_differential_2024,baumann_kinematic_2024,chen_multivariate_2024,liu_massive_2024,gasparotto_differential_2025}. These differential equation representations often reveal much of the deep underlying geometric structure among cosmological correlators. A notable example is the kinematic flow representation of \cite{arkani-hamed_differential_2023}, which provides a systematic 
way to construct a first-order system of differential equations that is satisfied by the cosmological correlators. As pointed out in \cite{Grimm:2024mbw} this system is special in that it represents the correlator as a Pfaffian function, constructed from finitely many building block functions. This Pfaffian representation comes with a well-established notion of complexity \cite{Khovanskii,GV}, which then allows one to upper-bound the number of poles in such a correlator \cite{Grimm:2024mbw}. It turns out that these bounds overshoot the physical expectations, which indicates that the kinematic flow representation must miss many simplifying relations. This motivates the search for another representation that more directly takes into account the special properties of such tree-level correlators. 

In this work we start our search for a simpler representation of tree-level cosmological correlators by viewing them as solutions to a system of differential equations known as a GKZ system, named after Gelfand, Kapranov, and Zelevinsky \cite{gelfand_generalized_1990,gelfand_hypergeometric_1991,gelfand_discriminants_1994}.  These systems are well-understood mathematically, and have been widely applied also in the Feynman amplitude literature (see, for example, \cite{delacruz_feynman_2019,feng_feynman_2022,klausen_hypergeometric_2020,klemm_lloop_2020,klausen_kinematic_2022,chestnov_macaulay_2022,ananthanarayan_feyngkz_2023,tellander_cohenmacaulay_2023,chestnov_restrictions_2023,caloro_ahypergeometric_2023,henn_dmodule_2024,levkovich-maslyuk_yangian_2024}). At first, the use of GKZ systems seems counter-productive, since one has to introduce a host of new variables in addition to the physical kinematic variables. However, as was first observed in \cite{Grimm:2024tbg}, these GKZ systems are not generic but rather feature a large amount of reducibility. The reducibility of GKZ systems has been linked in the mathematical literature \cite{saito_irreducible_2011,beukers_irreducibility_2011,schulze_resonance_2012} to the existence of a so-called resonant faces. In this work we will review the general criteria for reducibility and then show how the resulting data, most prominently each resonant face $F$, is used to construct a new differential operator $Q^{F}$.\footnote{It was shown in \cite{Grimm:2024tbg} that their existence can be inferred by using techiques from D-modules and the study of the Euler-Koszul complex.} The operators $Q^F$  were termed \textit{reduction operators} in \cite{Grimm:2024tbg}, and it was shown that they posses several remarkable properties when acting on the solution space of the GKZ system.\footnote{ In particular, they annihilate a subset of solutions to the GKZ system and allow for mapping solutions at differing parameters to eachother.} In contrast to \cite{Grimm:2024tbg}, our construction of $Q^F$ will be fully algorithmic, which opens the possibility to apply this construction in many other settings. In this work, however, we will then turn to applying the construction of reduction operators to the GKZ system for tree-level cosmological correlators.  

For the GKZ system associated to cosmological correlators we find two classes of reduction operators. The first consists of operators that are first-order in the derivatives, while the second will contain higher-derivative terms. We are able to show that appropriate combinations of these operators only depend on the physical variables and have numerous remarkable properties. Particularly, the physical first-order operators suffice to build first-order differential systems whose solutions form a basis for the cosmological correlators with a fixed number of external momenta. The resulting first-order differential equation can be written as a matrix equation of the form $\dd \mathcal{I} = A \mathcal{I}$, with an upper-triangular matrix $A$. Thus, this construction can be seen as a replacement of the kinematical flow algorithm of \cite{arkani-hamed_differential_2023}. Furthermore, this implies that also in this representation of the correlators they are Pfaffian functions with a quantitative notion of complexity. This implies that the analysis of \cite{Grimm:2024mbw}, which was carried out for the kinematical flow algorithm, can now be applied to this new Pfaffian system.  

A crucial part of our analysis is to additionally use the higher-order reduction operators. We find that the they imply that the basis functions obtained in the first-order reduction are related by extra algebraic relations. We will term the algorithm arising from using all reduction operators to determine a minimal basis the \textit{recursive reduction algorithm}.
Taking symmetries into account as well, we are able to establish a bound on the minimal number of functions one needs to represent a correlator and show a severe reduction in complexity. For example, for the double-exchange diagram, it turns out that merely four basis functions suffice to parametrize the correlator. While we will fall short of fully evaluating the computational complexity of the recursive reduction algorithm, it is apparent from the counting alone that it is significantly simpler than any known approach to the problem. 

Examining the action of the reduction operators on the level of the Feynman diagrams, we find that they implement relations and simplifications in an intuitive diagrammatical way. To see this, we introduce the notion of a tube and a tubing of a diagram as in \cite{arkani-hamed_differential_2023}. A tube is simply a collection of adjacent vertices of the diagram, while a tubing consists of a collection of tubes. The latter can be used to index the basis integrals under consideration. We show that the reduction operators relate different integrals of this type, essentially by removing a tube from a tubing. This action can then be combined with the fact that in the absence of certain tubes, the associated integral might be related to a contracted or cut Feynman diagram. In the former case, the reduction operator ensures that an edge joining two vertices can be removed, while in the latter case the Feynman diagram splits and the integral factorizes. We comment on the relation of these operations to the locality and singularity structure of the underlying physical theory.

The outline of this work is as follows. In section \ref{sec:cosmo} we review some basics of cosmological correlators, explaining the model which we work with and setting the stage for the rest of the paper. 
In section \ref{sec:GKZ} we introduce GKZ systems and descibe the conditions of when they are reducible. This general discussion includes the sketch of an algorithm of how the reduction operators can be found. We then apply these techniques and determine the reduction operators for the GKZ system associated to tree-level cosmological correlators.
In section \ref{sec:physics} we then explain how the reduction operators connect different integral contributions to the cosmological correlators. We also show that their action can lead to contractions or cuts of Feynman diagrams. The first application of these special properties is presented in section \ref{sec:differentialchain}, where we describe how to use the first-order reduction operators to obtain a basis of functions closed under partial differentiation. In section \ref{sec:complexity} we then present the full recursive reduction algorithm. We apply also the higher-order reduction operators, determine the induced algebraic relations, and present a counting of minimal basis functions. Finally, in section \ref{sec:conclusion} we summarize our conclusions and provide an outlook. Some technicalities are deferred to appendices \ref{red_op_phys} and \ref{ap:matrixrep}.

\section{Cosmological correlators}\label{sec:cosmo}
The aim of this section is to briefly review the necessary preliminaries on cosmological correlators, as well as to introduce the notation that we will use to describe them. For a more extensive review on cosmological correlators, we refer to \cite{benincasa_amplitudes_2022,lee_records_2024}.

\subsection{General aspects}\label{sec:general}

In our study of cosmological correlators, we will focus on a particular  model introduced in~\cite{arkani-hamed_cosmological_2015}. In particular, we will consider the coefficients of the wave-function of the universe in a perturbative expansion.

\paragraph{The model.} In this work, we focus on a conformally coupled scalar field $\phi$ in an FLRW spacetime with a power-law scale factor $a(\eta) = (\eta/\eta_0)^{-(1+\epsilon)}$. We assume that $\phi$ has general polynomial interactions, so that it is described by the action
\begin{equation} \label{model_action}
    S = \int \dd^4 x \,\sqrt{-g} \left(-\frac{1}{2}(\pd \phi)^2 - \frac{1}{12}R\phi^2 -\sum_{p=3}^D \frac{\lambda_p}{p!}\phi^p\right) \,.
\end{equation}
Here $R$ is the Ricci scalar, and the $\lambda_p$ are the couplings of the various interactions. This model has been studied extensively \cite{arkani-hamed_cosmological_2015,arkani-hamed_emergence_2018,benincasa_cosmological_2019,kuhne_faces_2022,benincasa_geometry_2024,benincasa_asymptotic_2024,de_cosmology_2024,glew_amplitubes_2025}, and it has the advantage of being fairly general while still having a rich mathematical structure in its observables.

\paragraph{Wavefunction coefficients.} To describe the observables of this theory, one frequently employs the notion of a wavefunction of the universe \cite{maldacena_graviton_2011,maldacena_nongaussian_2003,benincasa_flatspace_2018,benincasa_wavefunctionals_2022,creminelli_nonperturbative_2024,stefanyszyn_there_2024,albayrak_perturbative_2024,lee_records_2024}. For a given state $\ket{\Psi}$, this wavefunction is a functional $\Psi[\phi]=\bra{\phi}\ket{\Psi}$ which quantifies the overlap with a spatial field configuration $\ket{\phi}$. This wavefunction admits an expansion of the form  
\begin{equation}
    \Psi[\phi] = \exp\Bigg(\sum_{n\geq 2} \int \dd^3 \vec k_1\cdots \dd^3 \vec k_n \,\psi_n(\vec k_1,\ldots,\vec k_n) \,\phi_{k_1}\cdots\phi_{k_n}  \Bigg) \,.
\end{equation}
Here the $\phi_{k_i}$ are the Fourier modes of the field configuration $\ket{\phi}$, and the wavefunction $\Psi[\phi]$ is hereby encoded in the \textit{wavefunction coefficients} $\psi_n(k_1,\ldots,k_n)$. These wavefunction coefficients, which are directly related to the cosmological correlators of interest, may be found perturbatively by using special Feynman rules that account for the expanding spacetime \cite{arkani-hamed_cosmological_2017}. In particular, by implementing a conformal rescaling, the evaluation of a wavefunction coefficient can be recast in terms of a time integral over a flat space wavefunction coefficient with time-dependent couplings.\footnote{In what follows, we will also use the term `wavefunction coefficient' to indicate the contribution coming from a single Feynman graph.}

\paragraph{Kinematic variables.} In this work, we focus on tree-level Feynman graphs. 
Although these functions can depend on all external momenta $\vec k_1,\ldots,\vec k_n$, their dependence is actually restricted to specific combinations of these momenta. We refer to these combinations as the kinematic variables, and they are defined as follows. Every vertex with label $v$ in the diagram comes with a vertex energy 
\begin{equation}
    X_v = \sum_{i} |\vec k_i|,
\end{equation}
with the sum running over all external propagators attached to the vertex. Meanwhile, every internal propagator is associated to an internal energy variable $Y$ given by the energy flowing over that edge, which can be written in terms of the external momenta $\vec k_i$. For example, consider the tree-level single-exchange diagram
\begin{equation}
\includegraphics[valign=c]{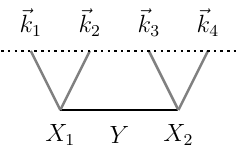}\end{equation}
Here the kinematic variables are given by $X_1=|\vec k_1|+|\vec k_2|$, $X_2=|\vec k_3|+|\vec k_4|$, and $Y = |\vec k_1 + \vec k_2 |  = |\vec k_3 +k_4|$, and the external gray lines are emanating from the spatial slice on which the state $\ket{\Psi}$ is located.

The integrals associated to a particular diagram can be obtained schematically as follows. First, one writes down a factor of $e^{i\eta_v (X_v+x_v)}$ for each vertex $v$. Secondly, one inserts the bulk-to-bulk propagator $G_e(Y_e,\eta_{e_1},\eta_{e_2})$ for each edge $e$, where $\eta_{e_1}$ and $\eta_{e_2}$ are the variables associated to the vertices connected to the edge. The particular form of this propagator will not be important for us, but let us note that it is a solution the Green's function equation
\begin{equation}
    (\partial_{\eta_{e_1}}^2+Y_e^2)G_e=(\partial_{\eta_{e_2}}^2+Y_e^2)G_e=-i\delta(\eta_{e_1}-\eta_{e_2})\,.
\end{equation}
We will use this in section~\ref{sec:physics} to show that certain combinations of reduction operators correspond to contractions of an edge in a diagram. 

Finally, the value of the diagram is then given by the following integral
\begin{equation}\label{eq:wavefunctionpropas}
    \psi_G(X,Y;\epsilon)=\int_{\R_+^{\Nv}}\dd^{\Nv} x  \int_{\R_-^{\Nv}}\dd^{\Nv}\eta \prod_{v=1}^{\Nv} x_v^{\alpha_v-1} e^{i\eta_v(X_v+x_v)}\prod_{e=1}^{N_{\rm e}}G_e(Y_e,\eta_{e_1},\eta_{e_2})\,,
\end{equation}
where $\Nv$ is the number of vertices,  $N_{\rm e}$ the number of edges, 
and $\alpha_v$ depends on $\epsilon$ and the order of the interaction at the vertex $v$. In particular for an interaction of order $k$, $\alpha_v$ is given by 
\beq \label{alpha_nu-form}
   \alpha_v=(4-k)(1+\epsilon)
\eeq
with $\epsilon$ determining the FLRW scale factor. The variables $x_v$ effectively parameterize shifts in the kinematic variables $X_v$, and integrating over these shifts accounts for working in the FLRW spacetime. These integrals admit convenient diagrammatic interpretations as sums over so-called tubings of the diagram, which we will discuss now.

\subsection{Integrals from graph tubings}\label{sec:tubings}

We will now discuss how to convert Feynman graphs contributing to cosmological correlators\footnote{From now on, we refer to wavefunction coefficients as cosmological correlators, since they are directly related \cite{maldacena_nongaussian_2003,goodhew_cosmological_2021,benincasa_amplitudes_2022,stefanyszyn_there_2024}.} into mathematical expressions, which will take the form of twisted integrals of the rational functions of the kinematic variables. Instead of presenting the precise physical Feynman rules, which can be found in \cite{arkani-hamed_cosmological_2017}, we will directly use the diagrammatic language of graph tubings, which were introduced in the setting of cosmological correlators in \cite{arkani-hamed_differential_2024}. These capture much of the structure of the cosmological correlators and will set the stage for the use of the theory of GKZ systems. Note that different definitions of tubings for cosmological correlators exist in the literature (see e.g. \cite{glew_amplitubes_2025}). 
Here, we adopt the definition given in~\cite[Sec 2]{arkani-hamed_differential_2024}.

\paragraph{Graph tubings and index sets.} Given a Feynman graph with the external propagators removed, a \textit{tube} is defined as a subset of adjacent vertices. Diagrammatically, this is denoted by encircling the corresponding vertices. For example, the single-exchange diagram has the following three tubes:

\begin{equation}\label{eq:singexchtubes}
\includegraphics[valign=c]{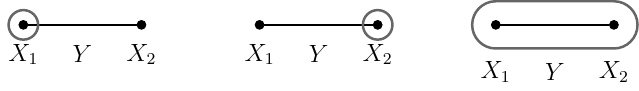}
\end{equation}
These tubes are particularly useful, since they are in one-to-one correspondence with the singularities of the flat space wavefunction coefficient. A \textit{tubing} of a graph is a collection of one or more non-intersecting tubes, and a \textit{complete tubing} is a tubing to which no more tube can be added without violating the non-intersecting condition.

Tubings can be represented in various ways. In this paper, we will regularly switch between the purely graphical representation used in \cite{arkani-hamed_differential_2024} and the representation of tubes in terms of index sets, where each tube corresponds to the index set of the vertices it contains. For example, the tubings above corresponds to the sets $\{1\}$, $\{2\}$ and $\{1,2\}$ respectively. Then, tubings can be represented simply as sets of tubes, or in other words, sets of index sets. In this representation, the tubing containing all of the tubes in~\eqref{eq:singexchtubes} is denoted by $\{\{1\},\{2\},\{1,2\}\}$.

This representation will have a number of advantages for us. For example, one can now easily sum over the vertices $v$ in a tubing. However, the major reason for introducing tubings as index sets is that the subset structure of the tubes now becomes manifest. In a tubing, a tube $T$ may be graphically fully contained in another tube $T'$, and this will correspond directly to containment of the subsets $T\subseteq T'$. This allows us to conveniently consider all tubes $T'$ contained in a tube $T$, or conversely, all tubes $T'$ contained in $T$. Such collections of tubes will play an important role throughout this paper, but in particular when obtaining the reduction operators in section~\ref{GKZ-red-gen}. Furthermore, we can consider the \textit{successor} or \textit{precursor} of a tube $T$, defined  as the minimal tube containing $T$ and the maximal tube contained in $T$, respectively. These will play an important role in section~\ref{sec:differentialchain}.

\paragraph{Integrals from tubings.}

As shown in~\cite{arkani-hamed_differential_2024}, the tubing of a graph can be used to obtain the associated wave-function coefficient. To be specific, we define for every tube $T$ a polynomial $p_T$ by setting
\begin{equation} \label{special_pol_cc}
    p_T(X,Y,x)=\sum_{v\in T} (X_v+x_v) +\sum_e Y_e  \,.
\end{equation}
In this expression the first sum is over all the vertices enclosed by the tube, while the second sum includes all edges that cross the tube. The variable $x_v$ will play the role of an integration variable. To every graph tubing $\cT$, we now associate an integral
\begin{equation}\label{eq:tubingintegral_tX}
    I_\mathcal{T}(X,Y;\alpha)=\int_{\R_+^{\Nv}}\dd^{\Nv} x \, \frac{\prod_{v=1}^{\Nv} x_v^{ \av-1}}{\prod_{T\in \mathcal{T}}p_T(X,Y,x)}\,,
\end{equation}
where the index $v$ runs over the $N_{\rm v}$ vertices in a diagram, and the $\alpha_v$ are variable weights associated to each vertex specified in \eqref{alpha_nu-form}. These integrals will be the key object of interest in describing the structure of cosmological correlators. 
The cosmological correlator associated to a graph $G$ is then recovered as  \cite{arkani-hamed_differential_2024}
\begin{equation}\label{eq:psiG}
    \psi_G(X,Y;\epsilon) =  \sum_{\cT\, \text{complete}} I_\cT(X,Y;\alpha) \,,
\end{equation}
where the sum is over all complete tubings $\cT$ of $G$. 

\paragraph{Convenient variables and permutations.}

Moving forward, we will consider the integrals above in a slightly different perspective. Instead of considering the variables $X_v$ and $Y_e$, we will combine these into variables $z^{(T)}$ for each tube $T$ in the tubing. In particular, we will define
\begin{equation}
    z^{(T)}=\sum_{v\in T} X_v+\sum_e Y_e\,,
\end{equation}
where, as in equation~\eqref{special_pol_cc}, the index $v$ runs over all the vertices enclosed by the tube, while the sum over $e$ is over the edges that cross the tube. These variables are particularly convenient as the polynomials $p_T$ can be written as 
\begin{equation}
    p_T(z,x)=z^{(T)}+\sum_{v\in T} x_v\,.
\end{equation}
Furthermore, the $z^{(T)}$ will map more naturally to GKZ variables defined in the following section.

This change of variables also has another interesting consequence. Rewriting the integral in equation~\eqref{eq:tubingintegral_tX} in terms of these new variables, we see that the only data necessary to define it is combinatorial, namely the data of which tube encircles which vertex. All other diagrammatical data can be re-instated by replacing the $z^{(T)}$ with their definitions in terms of $X_v$ and $Y_e$, as well as choosing the particular values the $\alpha_v$ take. A corollary to this is that many different tubings and diagrams may take the same form after making this replacement. 

For example,  the double-exchange correlator has two complete tubings, given by
\begin{equation}\label{eq:doubleexchangetubings}
\includegraphics[valign=c]{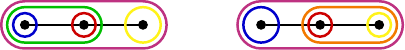}
\end{equation}
In principle, these will lead to different integrals $I_\cT$. However, denoting $T_{\rm b}$, $T_{\rm y}$, $T_{\rm g}$ and $T_{\rm o}$ for the blue, yellow, green and orange tubes respectively, one can make the replacements
\begin{equation}
    z^{(T_{\rm b})}\longleftrightarrow z^{(T_{\rm y})}\,, \quad z^{(T_{\rm g})} \longrightarrow z^{(T_{\rm o})}\,,
\end{equation}
and relate the integral of the left tubing to the right one. We will greatly extend this reasoning in section~\ref{sec:complexity}, where we show that many such relations exist and obtain a minimal set of integrals needed to express the actual correlators. Note that the described abstraction requires keeping track of all relations. However, this is more than compensated for by the significant reduction in the number of integrals to compute, which is an immense advantage in practice.

\section{GKZ perspective and reduction operators}\label{sec:GKZ}

In this section we continue analyzing the perturbative expansion of cosmological correlators and describe how the integral expressions for the tree-level contributions can be used to define certain systems of differential equations, known as GKZ systems. 
In order do to so we first give a brief general introduction to such systems in section \ref{GKZ_basics+cosm}. 
We then discuss how tree-level cosmological correlators are encoded via 
GKZ systems in section \ref{GKZ_for_cosmo}. These GKZ systems have the  feature that they are reducible. We dedicate section \ref{GKZ-red-gen} to describe the reduction of GKZ systems in general, and argue that this yields additional differential equations encoded by reduction operators. This general formalism is applied explicitly to cosmological correlators in section \ref{cosmic_red_op}, where we determine the form of the resulting reduction operators.

\subsection{General GKZ systems} \label{GKZ_basics+cosm}

In the following we first briefly introduce the mathematical framework of GKZ systems. Starting from a suitable integral, we will obtain the differential equations it satisfies, thereby obtaining the GKZ system it defines.

\paragraph{The GKZ integrals and their defining data.} To begin with, we recall that a GKZ system is a type of system of differential equations, associated to every integral of the form
\begin{equation}\label{eq:gengkzint}
   I(z;\alpha,\beta)= \int_\Gamma \dd^{\Nv} x\; \frac{\prod_{v=1}^{\Nv} x_v^{\alpha_v-1}}{\prod_{j=1}^k p_j(z,x)^{\beta_j}}\, .
\end{equation}
Here $\Gamma$ is an integration cycle, $\alpha_v$ and $\beta_j$ are complex numbers, and $z$ and $x$ collectively denote a set of complex variables $z_{j,m}$ and $x_v$. The polynomials $p_j$ in the denominator are assumed to take the form
\begin{equation} \label{p-form}
    p_j(z,x)= \sum_m z_{j,m} \prod_{v=1}^{\Nv} x_v^{(a_{j,m})_v}\, ,
\end{equation}
where the $a_{j,m}$ are vectors describing the powers of $x_v$ in each term of $p_j$. Note that the integral is a function of the coefficients $z_{j,m}$ and the GKZ system is a set of differential equations in these variables that is solved by this integral.

Before we can state the GKZ differential equations, it is convenient to first introduce a matrix $\A_j$ for each polynomial $p_j$, which is formed by taking the vectors $a_{j,m}$ to be the column vectors of the matrix $\A_j$. The defining information of the GKZ system is then best represented by combining the matrices $\A_j$, $j=1,\ldots,k$ and the vectors $\alpha_v,\beta_j$, $v=1\ldots,{\Nv}$ in the following way:
\begin{equation} \label{Anu-def}
    \A\coloneqq \begin{pmatrix} \mathbf{1} & \mathbf{0} &\cdots\\
    \mathbf{0} & \mathbf{1} & \cdots \\
    \vdots&\vdots&\ddots\\
    \A_1 & \A_2 &\cdots 
    \end{pmatrix}\ , \qquad \nu\coloneqq \begin{pmatrix} 
 \beta \\ \alpha\end{pmatrix} .
\end{equation}
where $\mathbf{1}$ denotes a row vector with $1$ at every entry and the $\mathbf{0}$ denotes a row vector of zeroes to fill in the remaining entries. By construction, both the matrix $\A$ and the vector $\nu$ have $\Nv+k$ rows. We denote the number of columns of $\A$ by $N$, and write
\beq \label{column_split}
    \A = (a_1,a_2,...,a_N)\ , 
\eeq
where $a_I$ are the $\Nv+k$-dimensional column vectors. Finally, note that to each column vector $a_I$ in \eqref{column_split} there is an associated variable $z_{j,m}$, which will henceforth
denoted by $z_I$.

\paragraph{GKZ differential equations.}

The GKZ differential equations can be separated into two subsets, the \textit{toric equations} and the \textit{Euler equations}, which take the form 
\beq \label{GKZ-diff_eq}
    \L_{u,v}f(z;\nu)=0\ ,\qquad  \quad(\E_J+\nu_J)f(z;\nu)=0\ ,
\eeq
where $\L_{u,v}$ are the toric operators, $\E_J$ are the Euler operators, and $f(z;\nu)$ is a $z_I$-dependent solution at parameter $\nu$.
To define the toric operators we first have to find vectors $u$ and $v$ in $\N^N$ satisfying $\A u=\A v$. This can be done systematically and yields a basis of dimension $\text{dim}\ker_\Z(\A)$. 
Any such $u$ and $v$ defines 
a toric operator 
\begin{equation}\label{eq:toricopdef}
   \L_{u,v}\coloneqq \prod_{I=1}^N \partial_I^{u_I}-\prod_{I=1}^N\partial_I^{v_I}\ ,
\end{equation} 
where $\partial_I\coloneqq \partial/\partial z_I$.

On the other hand, the $k+n$ Euler operators $\E_J$ are defined 
directly from the matrix $\A$ by setting 
\beq \label{def-Euler-op}
  \E_J \coloneqq \sum_{I=1}^N \A_{JI} \theta_I\, ,
\eeq
where the $\theta_I$ denote the homogeneous $z_I$-derivatives, defined as $\theta_I\coloneqq z_I \partial_I$.

We will often interpret these operators as giving rise to equivalence relations on differential operators. This is possible since, when acting on solutions of the GKZ systems, these operators will vanish. We will split these equivalence relations in two parts. First, we will consider only the toric operators $\L_{u,v}$, and obtain an equivalence relation that implies that, for all $u$ and $v$ as above, 
\begin{equation}
    \L_{u,v}\simeq 0\,,
\end{equation}
where we have introduced the notation $\simeq$ to denote this equivalence relation. Note that here, we have \textit{not} imposed that the Euler equations also hold. If we also consider these equations, we will explicitly write
\begin{equation}
    \L_{u,v} \simeq_{\E+\nu} 0\, ,\quad \E_J+\nu_J\simeq_{\E+\nu} 0\,,
\end{equation}
where now we introduced the notation $\simeq_{\E+\nu}$ for the equivalence relations from both the Euler and the toric equations.

The integral~\eqref{eq:gengkzint} will be a particular solution to the differential equations \eqref{GKZ-diff_eq} that depends on the choice of integration cycle $\Gamma$. To systematically describe all such solutions, one can first determine a complete basis of solutions $f_{d}(z;\nu)$, and expand\footnote{The particular coefficients can either be fixed numerically or by evaluating the integral in specific limits for the $z_I$.}
\begin{equation} \label{phys_amplitude}
    I(z;\alpha,\beta)=\sum_{d=1}^D c_d(\Gamma;\nu)f_{d}(z;\nu)\, ,
\end{equation}
where $D$ is the dimension of the solution space associated to the GKZ system. 
Finding the solutions $f_d$ 
is a non-trivial task and different methods might be more suitable depending on the form of $\A,\nu$. One general approach is to obtain a complete basis of convergent series expansions \cite{saito_Grobner_2000,cattani_three_2006}. 
Alternatively, one can make an Ansatz that automatically solves the Euler equations and then focus on solutions to the toric equations \cite{hosono_gkzCY_1996,hosono_gkzapp_1996}.

\subsection{GKZ systems for cosmological correlators} \label{GKZ_for_cosmo}

Having discussed general GKZ systems, we are now ready to apply these tools to cosmological correlators. In particular, we will provide the toric and Euler operators for a general correlator explicitly. This allows us in section~\ref{cosmic_red_op} to obtain reduction operators for general cosmological correlators.

\paragraph{Cosmological correlators as GKZ integrals.} 
First, we have to cast cosmological correlators in the form of a general GKZ integral. Recall that, to each such a correlator, we could study its complete tubings $\mathcal{T}$, and for each complete tubing obtain an integral
\begin{equation}
    I_\mathcal{T}(X,Y;\alpha)=\int_{\R_+^{\Nv}}\dd^{\Nv} x \, \frac{\prod_{v=1}^{\Nv} x_v^{ \av-1}}{\prod_{T\in \mathcal{T}}p_T(X,Y,x)}\,.
\end{equation}
Notice that this is exactly of the form of a general GKZ integral, except that the polynomials $p_T$ do not have arbitrary coefficients $z_{j,m}$. However, one can simply lift the polynomials to functions of $z$ by defining
\begin{equation}\label{eq:ptz_t}
    p_T(z,x)=z^{(T)}+\sum_{v\in T} z^{(T)}_vx_v\, ,
\end{equation}
where we promoted the coefficients of $p_T(X,Y)$ in equation~\eqref{special_pol_cc} to variables $( z^{(T)},z^{(T)}_v)$. 
Let us stress that the $x$-independent term in \eqref{eq:ptz_t} is parametrized by the variable $z^{(T)}$ without an index. This direction 
is special, since the polynomials in the physical variables are recovered 
when setting 
\beq \label{physical_slice}
z^{(T)}\big|_{\rm phys} = \sum_{v\in T} X_v + \sum_e Y_e\ , \qquad z^{(T)}_{v}\big|_{\rm phys}=1\ , 
\eeq
where the second sum is over all edges that cross the tube $T$ as in section~\ref{sec:tubings}. 
We will also refer to this identification as the restriction to the physical slice. The GKZ integral associated to the complete tubing $\mathcal{T}$ is then given by
\begin{equation}\label{eq:tubingintegral_t}
    I_\mathcal{T}(z;\alpha)=\int_{\R_+^N}\dd^{\Nv} x \, \frac{\prod_{v=1}^{\Nv} x_v^{\alpha_v-1}}{\prod_{T\in \mathcal{T}}p_T(z,x)}\,.
\end{equation}
This integral will then define the GKZ system of differential equations for us.

As an example, let us consider the single-exchange diagram with the tubing\footnote{The GKZ system and reduction operators of this example were studied in much more detail in~\cite{grimm_reductions_2024}.}
\begin{equation}
\includegraphics[valign=c]{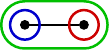}
\end{equation}
where we will denote the blue, red and green tubes as $T_{\rm b}$, $T_{\rm r}$ and $T_{\rm g }$ respectively. We will label the vertices by $v=1,2$ such that $T_{\rm b}=\{1\}$, $T_{\rm r}=\{2\}$ and $T_{\rm g }=\{1,2\}$. From these tubes, applying equation~\eqref{eq:ptz_t} results in the polynomials 
\begin{equation}\label{eq:singexchangepolys}
\begin{array}{rl}
    p_{T_{\rm b}}&=z^{(T_{\rm b})}+z^{(T_{\rm b})}_1 x_1\, ,\\
    p_{T_{\rm r}}&=z^{(T_{\rm r})}+z^{(T_{\rm r})}_2 x_2\, ,\\
    p_{T_{\rm g }}&=z^{(T_{\rm g })}+z^{(T_{\rm g })}_1 x_1+z^{(T_{\rm g })}_2 x_2\, 
\end{array}
\end{equation}
which can be inserted into equation~\eqref{eq:tubingintegral_t} to obtain the GKZ integral for the single-exchange diagram.

\paragraph{GKZ data for cosmological correlators.} 

We are now in the position to obtain the GKZ data for a general cosmological correlator. Recall that it consists of the matrix $\A$ and the parameter $\nu$. To obtain the matrix $\A$, recall that one first constructs a matrix $\A_T$ for each polynomial. For each tube $T$, these matrices are obtained by taking the exponents in $x_v$ for each term of $p_T$ and combining these as the column vectors of the matrix $\A_T$. One then combines these matrices into the matrix $\A$ as in \eqref{Anu-def} by identifying 
\beq \label{eq:Ajorder}
  \A_j = \A_{T_j}\ , \qquad j=1,\ldots,|\mathcal{T}|\, ,
\eeq
where $|\mathcal{T}|$ is the number of tubes  $T_j \in \mathcal{T}$. The parameter $\nu$ can be read off immediately by comparing
\eqref{eq:tubingintegral_t} and \eqref{eq:gengkzint}, resulting in
\begin{equation} \label{nu_cos}
    \nu=(\underbrace{1,\cdots,1}_{\vert \mathcal{T}\vert \text{ times}},\alpha_1,\cdots,\alpha_{\Nv})^{\rm T}\,,
\end{equation}
where we recall that $\Nv$ is the number of integration variables $x_v$, and the $\alpha_v$ are given in \eqref{alpha_nu-form} and depend on $\epsilon$ and the order of the interaction. Note that in this equation, $\rm T$ denotes the transpose and does not refer to a tube. It is interesting to note that much of the following general discussion does not depend on the precise value of $\alpha_v$.

Returning to the example of the single-exchange integral, we can simply read off the exponents of each term in equation~\eqref{eq:singexchangepolys} to obtain the matrices
\begin{equation}
    \begin{array}{rlrlrl}
       \A_{T_{\rm b}}  &=  \begin{pmatrix}
           0 & 1 \\
           0 & 0
       \end{pmatrix},&
       \A_{T_{\rm r}}  &= \begin{pmatrix}
           0 & 0 \\
           0 & 1
       \end{pmatrix},&
        \A_{T_{\rm g}} &=  \begin{pmatrix}
           0 & 1 & 0\\
           0 & 0 & 1
       \end{pmatrix}
    \end{array}
\end{equation}
for each tube. These matrices can be combined into the matrix $\A$, and this matrix together with the parameter $\nu$ then defines the GKZ system of the single-exchange integral. In particular, these are given by
\begin{equation}\label{eq:singexchangeA}
    \A =\left(
\begin{array}{ccccccc}
 1 & 1 & 0 & 0 & 0 & 0 & 0 \\
 0 & 0 & 1 & 1 & 0 & 0 & 0 \\
 0 & 0 & 0 & 0 & 1 & 1 & 1 \\
 0 & 1 & 0 & 0 & 0 & 1 & 0 \\
 0 & 0 & 0 & 1 & 0 & 0 & 1 \\
\end{array}
\right)\, , \qquad  \nu =\begin{pmatrix}1\\ 1\\ 1\\ \alpha_1\\ \alpha_2\end{pmatrix}\ ,
\end{equation}
where the bottom two rows correspond to the matrices $\A_T$ and we recall that $\alpha_1$ and $\alpha_2$ are the twists of the integration variables $x_1$ and $x_2$ respectively. Note that in this construction, we needed to fix an ordering of the tubes. Here, we have chosen 
\begin{equation}
    T_1=T_{\rm b}\, ,\quad T_2=T_{\rm r}\, ,\quad T_3 = T_{\rm g}
\end{equation}
although clearly, the chosen ordering is arbitrary.

\paragraph{Structure in the GKZ data.} 
Let us study the general structure of the matrices $\A$ which we construct for cosmological correlators. Every column vector arises from a particular term in a polynomial $p_T$ for some $T$, and each term corresponds to either a vertex or the constant term. In fact, for every tube $T$ there 
is a set of vectors 
\beq \label{aT-split}
   a^{(T)},\ a^{(T)}_v  \ \ \text{with}\ \ v \in T\ ,
\eeq
and combining these vectors for every tube $T$ we obtain the matrix $\A$. We will collectively denote these column vectors $a^{(T)}_m$, where $m=v$ if the column vector arises from a term in $p_T$ with a vertex, while the index $m$ is removed when it arises from the constant term. Note that these are associated with the coordinates $z^{(T)},z^{(T)}_v$, 
in accordance with \eqref{p-form} and~\eqref{eq:ptz_t}. Labeling 
the tubes as $T_j$ with $j = 1,...,|\mathcal{T}|$ as above, we thus split $\mathcal{A}$ as
\beq \label{cAinvectors}
   \mathcal{A} = \Big( a^{(T_1)} \ \underbrace{a^{(T_1)}_{v_1}}_{v_1 \in T_1} \ |\ a^{(T_2)} \ \underbrace{a^{(T_2)}_{v_2}}_{v_2 \in T_2} \ | \ldots \Big)  \ .
\eeq
To not clutter the notation, we will mostly use the notation \eqref{aT-split}, where it is understood that the index $v$ is associated to the tube $T$.

Comparing \eqref{p-form} and~\eqref{eq:ptz_t} we can now read off the column vectors $a_m^{(T)}$ for any tree-level cosmological correlator. Since the polynomials $p_T$ are all linear $x_v$, these vectors will only consist of ones and zeros. We see that its components split into two parts. First, we have the components of $a_m^{(T)}$ that are introduced when combining the matrices $\mathcal{A}_T$ together, which will consist of the first $\vert \mathcal{T} \vert$ entries. Then, for $1\leq j \leq \vert \mathcal{T}\vert$ we find that the $j$-th entry of $a_m^{(T)}$ is $1$ if $T=T_j$ and $0$ otherwise, where $T_j$ refers to the ordering of tubes we choose when constructing $\A$ in equation~\eqref{eq:Ajorder} or \eqref{cAinvectors}.
The remaining rows admit a similar structure, but now accounts for which vertices appear in the tube $T$. For $a^{(T)}$ this part is zero. However, for the other column vectors $a^{(T)}_v$, $v \in T$, we find that the $\vert \mathcal{T}\vert +v'$-entry of $a^{(T)}_v$ is $1$ if $v=v'$ and zero otherwise. To conclude, we can write the column vectors of 
$\A$ as
\begin{equation} \label{aT-def}
    a^{(T)}=\left(\begin{array}{c} e^{(T)}\\
    \mathbf{0}\end{array}\right) \ , \qquad a^{(T)}_v=\left(\begin{array}{c} e^{(T)} \\ e_v\end{array}\right)\ ,
\end{equation}
where $e^{(T)}$ is a $\vert \mathcal{T}\vert$-dimensional unit vector in the direction associated to $T$, $e_v$ is a $\Nv$-dimensional unit vector in the $v$-th direction, and $\mathbf{0}$ is the $\Nv$-dimensional zero vector. Note that the precise form of these vectors depends on the ordering of $T_j$ and $x_v$ that we have chosen.

To return to our example of the single-exchange integral, we see that \eqref{aT-def}
implies that $\mathcal{A}$ takes the form 
\begin{align}
   \mathcal{A} =& \left(\begin{array}{cc|cc|ccc} a^{(T_{\rm b})} & a^{(T_{\rm b})}_1 & a^{(T_{\rm r})} & a^{(T_{\rm r})}_2 & a^{(T_{\rm g})}& a^{(T_{\rm g})}_1 & a^{(T_{\rm g})}_2 \end{array} \right) \\
   =&  \left(\begin{array}{cc|cc|ccc} e^{(T_{\rm b})} & e^{(T_{\rm b})} & e^{(T_{\rm r})} & e^{(T_{\rm r})} & e^{(T_{\rm g})} & e^{(T_{\rm g})} & e^{(T_{\rm g})} \\ 
   \mathbf{0} &  e_1 &  \mathbf{0} &  e_2 &  \mathbf{0} &  e_1 & e_2
   \end{array} \right) \ . \nonumber
\end{align}
Clearly, upon inserting the unit vectors, we recover the matrix $\mathcal{A}$ given in \eqref{eq:singexchangeA}.

\paragraph{GKZ systems for cosmological correlators.}
We are now in the position to determine the toric and Euler operators associated to $\A,\nu$. Recall that the toric operators arose from vectors $u$ and $v$ in $\N^N$ satisfying $\A u=\A v$. Equivalently, these arise from the relations between the column vectors over the integers. In particular, note that from equation~\eqref{aT-def} it follows that, for any $T$ and any $v$ and $v'$ in $T$, we have
\begin{equation}\label{eq:ATdif}
    a^{(T)}_v-a^{(T)}_{v'}=(0,e_v-e_{v'})\,.
\end{equation}
Note that the right hand side no longer depends on $T$. Therefore, for any two tubes $T$ and $T'$ and $v$, $v'$ contained on both tubes, we have
\begin{equation}
a_v^{(T)}+a_{v'}^{(T')}=a^{(T)}_{v'}+a^{(T)}_v\,.
\end{equation}
A similar story holds for $a^{(T)}$ and $a^{(T)}_v$, resulting in a relation of the form
\begin{equation}   a_v^{(T)}+a^{(T')}=a^{(T)}+a^{(T)}_v\,.
\end{equation}
Both of these relations will give rise to toric operators. In particular, the above implies that for any tubes $T$, $T'$, and vertices $v$, $v'$ we have that
\begin{equation}\label{eq:cosmotoric}
    \begin{array}{rcr}
        v,v'\in T\cap T' &\implies &  \partial^{(T)}_{v} \partial^{(T')}_{v'}-\partial^{(T)}_{v'}\partial^{(T')}_{v} \simeq 0\, ,  \\
        v\in T\cap T' &\implies &  \partial^{(T)}_{v} \partial^{(T')}-\partial^{(T)}\partial^{(T')}_{v} \simeq 0\, ,
    \end{array}
\end{equation}
where we recall that $\simeq$ denotes that equality holds modulo the toric equivalence relations. Here, we have also introduced the notation 
\beq
\partial^{(T)} \equiv \frac{\partial}{\partial z^{(T)}}\ , \qquad \partial^{(T)}_v \equiv \frac{\partial}{\partial z^{(T)}_v}\ ,  
\eeq
for the partial derivatives with respect to $z$-variables. Note that a particular case of these relations arises when a tube is $T$ is completely contained in another tube $T'$. Then, it follows that there are toric operators such as the one above for every $v$ and $v'$ contained in $T$. These relations will be crucial in obtaining the reduction operators. 

Having found the toric operators, we can now turn our attention to the Euler operators. As we have seen before, the rows of $\A$ can be split into rows corresponding to the tubes and rows corresponding to the vertices. Since each row gives rise to an Euler operator, this implies that these can be split in a similar manner. In particular we obtain an operator $\E^{(T)}$ for each tube and an operator $\E_v$ for each vertex. This gives rise to $\mathcal{T}+N_{\rm v}$ operators which take the form
\begin{equation} \label{Euler_cosm}
    \E^{(T)}=\theta^{(T)}+\sum_{v\in T}\theta^{(T)}_v\,,\qquad  \E_v=\sum_{\{ T:v\in T \} }\theta^{(T)}_v  \, ,
\end{equation}
where $\theta^{(T)}\equiv z^{(T)}\partial^{(T)},\ \theta^{(T)}_v\equiv z^{(T)}_v\partial^{(T)}_{v}$. Let us stress that the sum in $\E_v$ is over all tubes that contain the vertex $v$. 

\paragraph{The GKZ system for the single exchange integral.} 

For completeness, let us return to the example of the single exchange integral and provide its GKZ system. From its tubing
\begin{equation}
\includegraphics[valign=c]{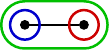}
\end{equation}
we find that the first vertex is enclosed by both the blue and the green tubes, while the second vertex is enclosed by the red and green tubes. Therefore, the GKZ system has toric relations of the form
\begin{equation}
\begin{array}{r}
    \partial^{(T_{\rm b})}_{1} \partial^{(T_{\rm g })}-\partial^{(T_{\rm b})}\partial^{(T_{\rm g })}_{1} \simeq 0\, ,\\
   \partial^{(T_{\rm r})}_{2} \partial^{(T_{\rm g })}-\partial^{(T_{\rm r})}\partial^{(T_{\rm g })}_{2} \simeq 0\,,
\end{array}
\end{equation}
and in fact, these are the only toric relations of this GKZ system. 
The Euler operators come in two parts. First we have the the operators from the tubes, which are given by
\begin{equation}
\begin{array}{rl}
    \E_{T_{\rm b}}&=\theta^{(T_{\rm b})}+\theta^{(T_{\rm b})}_1\, , \\
    \E_{T_{\rm r}}&=\theta^{(T_{\rm r})}+\theta^{(T_{\rm r})}_2\, , \\
    \E_{T_{\rm g }}&=\theta^{(T_{\rm g })}+\theta^{(T_{\rm g })}_1+\theta^{(T_{\rm g })}_2\, .
\end{array}
\end{equation}
Secondly, the Euler operators from the vertices are
\begin{equation}
    \E_1=\theta^{(T_{\rm b})}_1+\theta^{(T_{\rm g })}_1\, , \quad \E_2=\theta^{(T_{\rm r})}_2+\theta^{(T_{\rm g })}_2\,.
\end{equation}
Together with the toric operators above, these completely describe the GKZ system for the single exchange integral.

\subsection{Simplifying GKZ systems: reduction operators
\label{GKZ-red-gen}}

In the previous two sections we have introduced the general 
construction of GKZ systems and applied it to obtain the GKZ systems for cosmological correlators. As we will show below, these cosmic GKZ systems actually admit various simplifications that can be detected by analyzing their defining data. This is a consequence of their so-called reducability, which is rather well-understood for general GKZ systems \cite{saito_irreducible_2011,beukers_irreducibility_2011,schulze_resonance_2012}. When a GKZ system is reducible, there exist additional differential operators that annihilate \textit{some} but not all of the solutions. These additional operators were termed reduction operators in \cite{grimm_reductions_2024}, and we will outline their construction again here. We begin by providing the conditions on the GKZ data for a GKZ system to be reducible.

\paragraph{Reducibility of GKZ systems.} 

Recall that a GKZ system is completely defined by its matrix $\A$ and the parameter $\nu$. We will then consider index sets $F$ that are subsets of $\{1,\cdots,N\}$, where $N$ is the number of columns of $\A$. As a technical prerequisite for reducibility, we have to require that $F$ is a \textit{face} of $\A$, defined as follows. The subset $F$ is a face of $\A$ if there exists a linear functional $L_F:\bbZ^N\rightarrow \bbZ$ such that
\beq \label{eq:LF}
\begin{array}{rl}
  L_F(a_I) =0 & \text{ for } I\in F \, , \\
  L_F(a_I)>0 & \text{ for } I\not\in F \, .
\end{array}
\eeq
This property also has a geometric interpretation. Let us consider the column vectors $a_I$ of $\A$ as generating a cone. Similarly, we can consider the cone generated by the column vectors of $a_I$ with $I$ in $F$. Then, $F$ defines a face of $\A$ if the cone generated by $F$ is a face of the cone, in the geometric sense. If $F$ is a face of codimension one, we will call it a \textit{facet}. Geometrically, this will also correspond to a facet of the cone of $\A$.

Given the notion of a face, we now introduce resonance. We say that $F$ is a \textit{resonance face} for $\nu$ if
there are complex numbers 
$c_I$ and integers $n_I$ such that
\begin{equation} \label{resonance_cond}
    \nu=\sum_{I\in F}c_I\; a_I+\sum_{I=1}^N n_I \;a_I\ ,
\end{equation} 
where we recall that $a_I$ are the column vectors of $\A$. In other words, the vector $\nu$ lies in the span of the face $F$, up to shifts given by integer multiples of the columns of $A$. A minimal resonance face for $\nu$ is also known as a \textit{resonance center} for $\nu$. Since \eqref{resonance_cond} contains general $c_I$, any face that contains a resonance face is also a resonance face and a resonance center picks out the smallest combination.

Finally, one needs to exclude the 
possibility that $\A$ defines a pyramid over $F$. While the notion of a pyramid is defined geometrically, it is equivalent to the requirement that none of the toric operators contain $\partial_I$ for $I\not \in F$. This extra condition excludes the trivial factorization of solutions with an overall prefactor depending on $z_I$ for $I$ not in $F$.

Given these definitions, we can now use Theorem 3.1 of \cite{schulze_resonance_2012} (see also \cite{grimm_reductions_2024}). It states that if one has a resonance center $F$ for $\nu$ and $\A$ is not a pyramid over $F$, then the GKZ system with data $\nu,\A$ is reducible. As we will discuss next, this ensures that one is able to construct additional differential equations of the form 
\beq
   Q^{(F)}_u f(z;\nu) = 0\ ,
\eeq
satisfied by \textit{some} of the solutions to 
the GKZ system. The linear operators $Q^{(F)}_u$ are the reduction operators associated to a resonance center $F$ and a vector $u$ that labels different operators with the same $F$. 

\paragraph{Constructing reduction operators.} Let us now briefly explain how a reduction operator can be constructed in the above setting. Our starting point is a resonance face $F$.\footnote{Note that we do not directly consider a resonance center, which is a minimal resonance face. This implies that our construction might admit further reductions.} We begin with the observation that reducibility as stated above is modulo integer shifts of $\nu$ by the column vectors of $\A$. This is somewhat problematic, as the exact form of the reduction operator can therefore change depending the parameter $\nu$, providing us with the reduction operator at other values of $\nu$. If such shifts are necessary, then they can be dealt with by either applying the algorithm below for some different $\nu$ satisfying the assumptions, or shifting $\nu$ using partial derivatives or their inverses as in \cite{dwork_generalized_1990, beukers_irreducibility_2011,caloro_ahypergeometric_2023}.\footnote{Interestingly, the constructions of~\cite{caloro_ahypergeometric_2023} share some similarities with the algorithm for obtaining reduction operators below. Partly, this is because both are obtained from a similar construction in~\cite{dwork_generalized_1990,beukers_irreducibility_2011}. As the reduction operators were originally introduced in~\cite{grimm_reductions_2024} from a different perspective, it would be interesting to explore further how the two constructions relate.} However, observe that shifts by the columns contained in $F$ will not change the procedure below. Because of this, and the precise form of the cosmic GKZ systems we will consider, we will not need to consider such shifts in this paper.

We will focus on the case where we have a resonant face $F$, with some fixed $I$ not in $F$, and consider the case where $\nu$ is such that $\nu-a_I$ is in the span of $F$. Equivalently, this implies that $L_F(\nu-a_I)=0$, where $L_F$ is a linear functional defining $F$. To construct a reduction operator associated to $F$ and $I$, we will first define the operator
\begin{equation}\label{eq:EFdef}
\E_F  = \sum_J L_F (a_J) \theta_J \,.    
\end{equation}
Then, we will construct a vector $u$ in $\N^N$ such that 
\begin{equation}\label{eq:PEFsimQd}
     \pd_1^{u_1}\cdots \pd_N^{u_N} \E_F \simeq  Q^{(F)}_u \pd_I,
\end{equation}
and from this obtain the reduction operator $Q^{(F)}_u$.\footnote{It may happen that $\partial^u$ and $\E_F$ do not commute, in which case a small additional step is required. If any terms of the form $\partial_J^k z_J \partial_J$ appear when expanding $\partial^u \E_F$, simply replacing $\partial_J^k$ with $\prod_{j=1}^k (\theta_J-j)$ will guarantee that the expression is still proportional to $\partial_I$. For the systems we consider in this paper, we will not need this though.} Note that this reduction operator will be valid at parameter $\nu$.

To obtain the vector $u$ we can proceed in two ways. If we can immediately construct such a vector $u$ by inspecting the GKZ system, then it results in a reduction operator and we are done. This is the approach we will take in the following sections of this paper. However, we will also provide a somewhat technical condition, proven originally in \cite{dwork_generalized_1990} but adapted from \cite[Theorem 2.1]{beukers_irreducibility_2011}, that allows us to obtain such a vector algorithmically. This condition can be stated as follows. Recall that a facet of $\A$ is a face of co-dimension one. If, for every facet $F'$ of $\A$ and $J$ not in $F$ we have that
\begin{equation}\label{eq:reducopineq}
    L_{F'}(\A u+a_J)\geq L_{F'}(a_I)\, ,
\end{equation}
then $u$ satisfies equation~\eqref{eq:PEFsimQd}. Here, $L_{F'}$ is the linear functional defining $F'$ and we note that, since $F'$ is a facet, $L_{F'}$ is unique up to a constant pre-factor. Note that, since the entries of $u$ are integers and the linear functionals $L_{F'}$ can be written in terms of matrices, the above turns into an integer linear programming problem allowing us to obtain $u$ algorithmically.

\subsection{Reduction operators for cosmic GKZ systems} \label{cosmic_red_op}

In this section we determine the reduction operators for 
GKZ systems associated to cosmological correlators. This 
connects the general discussion of section~\ref{GKZ-red-gen} 
with the GKZ systems introduced at the end of section~\ref{GKZ_basics+cosm}.

We begin by showing that every tube corresponds to a resonant face. Recall that the first $\vert \mathcal{T}\vert$ rows of the matrix $\A$ were associated to the tubes in $\mathcal{T}$. From this, we can obtain a linear functional $L_{F_{T_j}}$ projecting a vector on its $j$-th coordinate. These linear functionals will  satisfy
\beq
\begin{array}{rl}
  L_{F_{T_j}}(a^{(T)}_m) =0 & \text{ for } T\neq T_j \, , \\
  L_{F_{T_j}}(a^{(T)}_m)>0 & \text{ for } T=T_j \, .
\end{array}
\eeq
Therefore, these define a face of $\A$ containing all columns of $\A$ except those arising from the face $T_j$. In particular, this face will correspond to the integral $I_{\mathcal{T}\setminus\{T_j\}}$.

From the above, we see that any tube $T$ will define a face of $\A$. It turns out that these faces are all resonant as well. To see this, we will consider two cases: the case where $T$ is the maximal tube, as well as the case where $T$ is not the maximal tube. Let us consider the latter first. Observe from equation~\eqref{aT-def} that
\begin{equation}
    a^{(T_{\rm max})}_v-a^{(T_{\rm max})}=(\bold{0},e_v)\,,
\end{equation}
where we recall that $\bold{0}$ is a $\vert \mathcal{T}\vert$-dimensional vector of zeroes and $e_v$ is the $N_v$-dimensional unit vector in the $v$-th direction. Inspecting the explicit form of $\nu$ provided in equation~\eqref{nu_cos}, this implies that it is possible to write
\begin{equation}
    \nu = \sum_{v} \alpha_v (a^{(T_{\rm max})}_v-a^{(T_{\rm max})}) + \sum_{T \in \mathcal{T}} a^{(T)}\,,
\end{equation} 
where we recall that the $\alpha_v$ are the twists of the different vertices which will be complex in general. Now note that if $T\neq T_{\rm max}$, the face $F_{T}$ will contain the columns associated to $T_{\rm max}$. Therefore, we see that any non-maximal face is resonant. A similar story holds for the maximal face itself. However, here we must choose any collection of tubes in $\mathcal{T}\setminus \{T_{\rm max}\}$ covering every vertex and proceed along the same lines. If such a covering is not possible, the maximal face will not be resonant. As we will see when constructing the higher-order operators below, in this case we will also obtain no reduction operator for the maximal face. Note that for a complete tubing this is never the case. 

To obtain the actual reduction operators we will proceed as laid out in section~\ref{GKZ-red-gen}. We begin by constructing the operator $\E_{F^{(T)}}$ for every tube $T$. However, since the linear functional here is simply a projection, we find that 
\begin{equation}\label{eq:tuberedbase}
    \E_{F^{(T)}}=\E^{(T)}=z^{(T)}\partial^{(T)}+\sum_{v\in T} z^{(T)}_v \partial^{(T)}_{v}\,,
\end{equation}
where $\E^{(T)}$ is the Euler operator of the GKZ system that is associated to $T$, as given in equation~\eqref{Euler_cosm}. Now, recall that a reduction operator is obtained by fixing both a face $F$ and an index $I$ not contained in $F$. For the faces we consider, any column associated to $T$ will be sufficient. Therefore, we will proceed using the column $a^{(T)} $. This implies that we must find $u$ such that
\begin{equation}\label{eq:duETQ}
     \pd_1^{u_1}\cdots \pd_N^{u_N} \E^{(T)} \simeq  Q^{(F_T)}_u \pd^{(T)} \, ,
\end{equation}
although in what follows, we will index $Q$ in different ways.

As we will see, the reduction operators we find fall in to two classes, the first-order operators and the higher-order operators. The first-order operators come with some problems though, as in general it will not be possible to write these solely in terms of the physical variables $z^{(T)}$ and their derivatives. However, we will show that a special combination of the first-order operators can be written in terms of the physical variables only, resulting in a first-order operator for each tube. The higher-order operators do not have this problem, and we will rewrite these directly in terms of the physical variables. Note that we will diverge somewhat from the discussion in section~\ref{GKZ-red-gen} and solve equation~\eqref{eq:duETQ} directly, without having to solve equation~\eqref{eq:reducopineq} iteratively for $u$.

\paragraph{First-order reduction operators from a contained tube.}

Let us consider tubes $T$ and $T'$ such that $T$ is fully contained in $T'$. Recall from equation~\eqref{eq:cosmotoric} that this implies that, for every $v$ in $T$ there is a toric relation the form
\begin{equation}\label{eq:firstordertoric}
    \partial^{(T)}_{v} \partial^{(T')} -\partial^{(T)} \partial^{(T')}_{v} \simeq 0\,.
\end{equation}
It then follows that
\begin{equation}\label{eq:pit'0ET}
\begin{array}{rl}
    \partial^{(T')}  \E^{(T)}&=z^{(T)} \partial^{(T)} \partial^{(T')} +\sum_{v\in T} z^{(T)}_v \partial^{(T)}_{v}\partial^{(T')} \\
    &\simeq z^{(T)} \partial^{(T)} \partial^{(T')} +\sum_{v\in T}z^{(T)}_v \partial^{(T')}_v\partial^{(T)} \,,
\end{array}
\end{equation}
where we have inserted equations~\eqref{eq:tuberedbase} and~\eqref{eq:firstordertoric}.
In this equation, $\partial^{(T)}$ can be factored out implying that we have obtained a relation of the form in equation~\eqref{eq:duETQ} and can read off the reduction operator. Writing this reduction operator as $Q^{(T)}_{T'}$, we find that
\begin{equation} \label{QTT'}  
    Q^{(T)}_{T'}=z^{(T)} \partial^{(T')}+\sum_{v\in T} z^{(T)}_v \partial^{(T')}_{v}\ . 
\end{equation}
Thus, we have found a first-order reduction operator whenever a tube $T$ is contained in another tube $T'$. Note that this implies that every non-maximal tube has at least one first-order reduction operator associated to it while the maximal tube has none. 

\paragraph{Physical restriction of first-order reduction operators.}

To use the reduction operators \eqref{QTT'}, we first have to deal with a fundamental challenge that arises when using GKZ systems. In the process of defining the GKZ system we had to introduce many additional parameters $z^{(T)}_v$ that are not present in the physical integral which is evaluated on the slice \eqref{physical_slice}. As mentioned in the general discussion of section~\ref{GKZ_basics+cosm}, the Euler operators impose restrictions on the variables, naturally leading to a choice of homogeneous variables. This can be used to eliminate partial derivatives with respect to some of the  $z^{(T)}_v$. However, the constraint \eqref{physical_slice} is more severe and it turns out to be impossible to write a general reduction operator $Q^{(T)}_{T'}$ only in terms of the physical variables. 
To circumvent this problem, we propose to introduce new operators 
\begin{equation}\label{eq:QTdef_t}
    Q^{(T)}\coloneqq \sum_{T'\supsetneq T} Q^{(T)}_{T'}\Big|_{\rm phys}\ .
\end{equation}
Here the sum is over all tubes $T'$ which contain $T$, excluding $T$ itself and $\vert_{\rm phys}$ means that we restrict to the slice \eqref{physical_slice} and act on solutions of the GKZ system. 
We show in appendix \ref{red_op_phys} that, using the Euler operators \eqref{Euler_cosm}, $Q^{(T)}$ can be written as
\begin{equation}\label{eq:QTresult}
     \boxed{\rule[-.5cm]{0cm}{1.3cm} \quad 
     Q^{(T)}=z^{(T)}\sum_{ T'\supsetneq T}\partial^{(T')}+\sum_{T'\subseteq T}(\theta^{(T')}+\nu^{(T')})-\sum_{v\in T}\alpha_v\ ,\quad }\,
\end{equation}
which only involves the physical derivatives $\partial^{(T)}$.

Note that in the construction above it was crucial that $T$ was not a maximal tube. We will now show that, by simply inserting the maximal tube $T_{\rm max}$ into equation~\eqref{eq:QTresult}, we obtain an operator $Q^{(T_{\rm max})}$ satisfying
\begin{equation}\label{eq:QTmaxaction}
    Q^{(T_{\rm max})}\simeq_{\E+\nu}0\,,
\end{equation}
where we indicated that it annihilates the integral $I_{\mathcal{T}}$ due to the Euler equations~\eqref{GKZ-diff_eq}. Technically, this operator is not a reduction operator of the GKZ system, but we will treat as such due to the property \eqref{eq:QTmaxaction}. To check this identity, we insert $T_{\rm max}$ into \eqref{eq:QTresult}
to find 
\begin{align} \label{QTmax}
   Q^{(T_{\rm max})} &= \sum_{T\in \mathcal{T}}\big(\theta^{(T)}+ \nu^{(T)}\big) -\sum_{v\in T_{\rm max}} \alpha_v \\
   &= \sum_{T\in \mathcal{T}} \big(\mathcal{E}^{(T)}+ \nu^{(T)}\big) -\sum_{v\in T_{\rm max}} \big(\mathcal{E}_v + \alpha_v\big)\ , \nonumber 
\end{align}
where we inserted the definitions 
of $\mathcal{E}^{(T)}$ and $\mathcal{E}_v$ given in \eqref{Euler_cosm} to obtain the second line. We now see that the expression on the second line is a sum of the Euler operators and therefore annihilates solutions to the GKZ system.

\paragraph{Higher-order reduction operators.} Having established how a reduction operator for a tube $T$ can be obtained by considering the tubes $T'$ that contain $T$, we now show that there are also reduction operators corresponding to the tubes $T'$ \textit{contained in} $T$. To be precise, we will consider a partition of $T$, i.e.~a collection of tubes $S_\alpha$ contained in $T$ such that every vertex in $T$ is in exactly one of the $S_\alpha$. The tube $T$ can then be recovered as the disjoint union

\begin{equation} \label{T-partition}
    T=\bigsqcup_{\alpha=1}^n S_\alpha\ .
\end{equation}
Furthermore, we can collect the $S_\alpha$ into a set $\pi$ as
\begin{equation}\label{eq:pidef}
    \pi=\{\,S_\alpha\, \vert \, 1\leq \alpha \leq n\,\}\,,
\end{equation}
which we will also refer to as the partition. Note that every partition $\pi$ is also a tubing, in fact it is a minimal tubing containing each vertex in $T$. We will show that, from every partition, we can obtain a new reduction operator. Furthermore, this reduction operator can be written in terms of only the physical derivatives, provided we restrict ourselves to the physical slice.

As for the first-order reduction operators, we will start by considering derivatives acting on the Euler operators \eqref{Euler_cosm}. We first note that, using similar arguments as before, there are toric operators of the form
\begin{equation}
    \partial^{(T)}_{v} \partial^{(S)} -\partial^{(T)} \partial^{(S)}_{v} \simeq 0
\end{equation}
for every $v$ in $S$ and $S$ in $\pi$. Therefore, we find that
\begin{equation}\label{eq:highereulerpart}
    \partial^{(S)} \sum_{v\in S} \theta^{(T)}_v\simeq \sum_{v\in S} z^{(T)}_v \partial^{(S)}_v \partial^{(T)} 
\end{equation}
for each $S$ in $\pi$. Now, we can simply use the decomposition of $T$ to write
\begin{equation}\label{eq:ETdecomp}
    \E^{(T)}=\theta^{(T)} +\sum_{S\in \pi}\sum_{v\in S} \theta^{(T)}_v\,.
\end{equation}
Combining equations~\eqref{eq:highereulerpart} and~\eqref{eq:ETdecomp} we obtain
\begin{equation}
    \prod_{S\in \pi} \partial^{(S)} \E^{(T)} \simeq \Bigg( z^{(T)} \prod_{S\in \pi}\partial^{(S)}  +\sum_{S\in \pi} \sum_{v\in S} z^{(T)}_v\partial^{(S)}_{v}\prod_{\substack{S'\in \pi\\\S'\neq S}} \partial^{(S')} \Bigg)\partial^{(T)} \,,
\end{equation}
from which we can immediately read of the reduction operator associated to $F^{(T)}$ and $a^{(T)}$.

However, we now again run into the issue that this operator involves derivatives with respect to the unphysical variables. To fix this, we will use that on the physical slice we have that $z^{(T)}_v=z^{(S)}_v=1$ for all $S$. Therefore, it is possible to rewrite
\begin{equation}\label{eq:highorderdef}
    \sum_{v\in S} z^{(T)}_v\partial^{(S)}_{v}=\sum_{v\in S} \theta_{S,v}+\cdots\simeq_{\E+\nu} -\theta_{S,0}-\nu^{(S)}+\cdots\, ,
\end{equation}
where the dots denote terms that go to zero in the physical limit, and we made use of the Euler operator $\E_{S}$. Therefore, we find a reduction operator associated to the partition $\pi$ which can be written as 
\begin{equation}\label{eq:highred}     
    Q^{(T)}_{\pi}=\Big(z^{(T)} -\sum_{S\in \pi} z^{(S)} \Big)\prod_{S'\in \pi} \partial^{(S')} -\sum_{S\in \pi}\nu^{(S)}\prod_{\substack{S'\in \pi,\\ S'\neq S}} \partial^{(S')} \ .
\end{equation}
Here we stress that this expression holds only on the physical slice \eqref{physical_slice}, while the existence of the operator is guaranteed for any $z_I$. Interestingly, we will always have $\nu^{(S)}=1$ for each $S\in \pi$. If this is the case, equation~\eqref{eq:highred} can be written as
\begin{equation} \label{higher-order-red_final}
 \boxed{\rule[-.5cm]{0cm}{1.3cm} \quad     Q^{(T)}_{\pi}=\Big(\prod_{S'\in \pi} \partial^{(S')} \Big)\Big(z^{(T)} -\sum_{S\in \pi} z^{(S)} \Big)\,,\quad }
\end{equation}
where the derivatives act on everything to their right. We will see that this form has interesting implications on the singularity structure of the integrals.

\paragraph{A simple example.}

To illustrate the discussions above, let us briefly consider a simple example. We will again consider the single-exchange integral with the tubing
\begin{equation}
\includegraphics[valign=c]{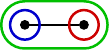}
\end{equation}
and obtain its reduction operators. We begin by obtaining the first-order reduction operator associated to the blue tube $T_{\rm b}$, noting that the reduction operator for $T_{\rm r}$ can be obtained in an almost identical manner. Inspecting equation~\eqref{eq:QTresult} we find that we must consider the tubes contained in $T_{\rm b}$, as well as those that contain it. Here, there are no tubes contained in $T_{\rm b}$. However, it is contained in the green tube $T_{\rm g}$. Thus we find that equation~\eqref{eq:QTresult} reduces to \begin{align} \label{red-op_inexample}
    Q^{(T_{\rm b})}&=z^{(T_{\rm b})}  \partial^{(T_{\rm g})} +z^{(T_{\rm b})} \partial^{(T_{\rm b })}  +\nu^{(T_{\rm b})}-\alpha_1\,\ ,\qquad Q^{(T_{\rm r})} = Q^{(T_{\rm b})}\big|_{T_{\rm b},\alpha_1\rightarrow T_{\rm r},\alpha_2}\ , \nonumber\\
    Q^{(T_{\rm g})}& = \sum_{T\in \{ T_{\rm g},T_{\rm r},T_{\rm b}\}} \big( z^{(T)} \partial^{(T)}  +\nu^{(T)}\big)-\alpha_1-\alpha_2\ . 
\end{align}
Note that it is also possible to write these operators in terms of the physical coordinates $X_v$ and $Y$ as
\begin{align}
     Q^{(T_{\rm b})}&=(X_1+Y)\frac{\partial}{\partial X_1}+\nu^{(T_{\rm b})}-\nu_1\ , \qquad Q^{(T_{\rm r})}=(X_2+Y)\frac{\partial}{\partial X_2}+\nu^{(T_{\rm r})}-\nu_2\ , \nonumber \\
     Q^{(T_{\rm g})}&= X_1\frac{\partial}{\partial X_1}+X_2\frac{\partial}{\partial X_2}+Y\frac{\partial}{\partial Y}+\nu^{(T_{\rm b})}+\nu^{(T_{\rm r})}+\nu^{(T_{\rm g})}-\alpha_1-\alpha_2
\end{align}
as was also found in~\cite{grimm_reductions_2024}. 

For the higher-order reduction operator, we note that $T_{\rm g }$ admits the decomposition $T_{\rm g }=T_{\rm b}\sqcup T_{\rm r}$. Following the notation of equation~\eqref{eq:pidef}, we will denote this partition by
\begin{equation}
\pi=\{T_{\rm b},T_{\rm r}\}=\{ \{1\},\{2\}\}\,.
\end{equation}
From such a decomposition we can obtain a higher-order reduction operator using equation~\eqref{higher-order-red_final} taking the form 
\begin{equation}\label{eq:QTgexplicit}
    Q^{(T_{\rm g })}_\pi= \partial^{(T_{\rm b})} \partial^{(T_{\rm r})}  (z^{(T_{\rm g })} -z^{(T_{\rm b})} -z^{(T_{\rm r})} )\ .
\end{equation}
It is straightforward to write this neatly in terms of $X_v$ and $Y$ as
\begin{equation}\label{eq:singexchangeQhigh}
    Q^{(T_{\rm g })}_\pi= \frac{1}{2}\left( \left(\frac{\partial}{\partial X_1}-\frac{\partial}{\partial X_2}\right)^2 - \frac{\partial^2}{\partial Y^2}\right) Y\,
\end{equation}
where the derivatives act on everything to their right. 

We are now ready to discuss the implications of acting with the reduction operators found in this section on the space of solutions to the cosmic GKZ system. 

\section{From reductions to relations, cuts, and contractions}\label{sec:physics}

In this section we discuss how 
the reduction operators derived in section \ref{cosmic_red_op} can be used 
to connect and simplify cosmological correlators. We will first show in section~\ref{ssec:tuberemoval} how the reduction operators remove tubes from a tubing. Subsequently, we will describe in sections~\ref{ssec:contractions} and~\ref{ssec:factors} that their action can be interpreted as either contracting or cutting an edge in the diagram. This leads to relations among integrals associated to different diagrams that are realized via differential operators. Our findings can also be understood diagrammatically via the removal of a tubes, which either 
results in a contraction or a factorization of integrals. 
The arising relations form the foundation for the algorithm to determine cosmological correlators that we develop in sections~\ref{sec:differentialchain} and \ref{sec:complexity}.

\subsection{Removing tubes using reduction operators}\label{ssec:tuberemoval}

We begin by describing the action of a reduction operator on the integral $I_\mathcal{T}$ and we will see that acting with a reduction operator removes a tube, up to twists in the integrand realized by partial derivatives. To derive this, we will first consider the reduction operators in representations that include derivatives with respect to the unphysical coordinates, as in this form the action of the reduction operator is the simplest. To obtain the action of the physical operators, we note that the unphysical derivatives have been removed using the Euler relations. In a GKZ system, two operators are equivalent modulo an Euler relation if they act equivalently on the integrand of the GKZ integral, modulo a total derivative in one of the integration variables. From this we find that the reduction operators in the physical coordinates must act in the same manner as the unphysical ones.

\paragraph{Tube-removal using $Q^{(T)}$.} We begin by considering the first-order reduction operators. It is useful to first consider the reduction operator $Q^{(T)}_{T'}$ as given in equation~\eqref{QTT'}. Acting on the integrand of~\eqref{eq:tubingintegral_t}, we easily verify the identity
\begin{equation}\label{eq:QTaction}
    Q^{(T)}_{T'}\frac{\prod_{i=1}^n x_v^\epsilon}{\prod_{T\in \mathcal{T}}p_T}=\frac{-p_T}{p_{T'}}\frac{\prod_{i=1}^n x_v^\epsilon}{\prod_{T\in \mathcal{T}}p_T}=\partial^{(T')} \frac{\prod_{i=1}^n x_v^\epsilon}{\prod_{T\in \mathcal{T}\setminus \{T\}}p_T}\,.
\end{equation}
Observe that, in effect, acting with $Q^{(T)}_{T'}$ has removed the tube $T$ from the diagram and replaced it with a derivative in $\partial^{(T')} $. This action generalizes for the operator $Q^{(T)}$ in physical variables up to total derivatives in the integration variables. Because these total derivatives vanish when performing the integrations, the integrals must satisfy \footnote{Note that this equation is consistent with the action~\eqref{eq:QTmaxaction} of the first-order reduction operator $Q^{(T_{\rm max})}$, even though it is, strictly speaking, not a reduction operator of the GKZ system.}
\begin{equation}\label{eq:QTonint}
    \boxed{\rule[-.5cm]{0cm}{1.3cm} \quad   Q^{(T)} I_\mathcal{T}=\bigg( \sum_{T'\supsetneq T} \partial^{(T')}  \bigg)I_{\mathcal{T}\setminus \{T\}}\,, \quad }
\end{equation}
where $Q^{(T)}$ is as in equation~\eqref{eq:QTdef_t}, and the sum is over all tubes $T'$ that strictly contain $T$. We will see later that, in the special case that $T$ is a minimal tube in a complete tubing $\mathcal{T}$, these equations imply contraction identities at the diagrammatical level.

\paragraph{Tube-removal using $Q^{(T)}_{\pi}$.}
Let us now show that there are similar relations for the higher-order reduction operators $ Q^{(T)}_{\pi}$. The procedure to obtain these is  similar to what we did before. However, now we find that the factorization depends on the partition $\pi$ of $T$, where this partition is defined as in equation~\eqref{T-partition}. To be precise, acting with $Q^{(T)}_{\pi}$ on the integral results in
\begin{equation}\label{eq:highordtuberemov}
     \boxed{\rule[-.5cm]{0cm}{1.3cm} \quad  Q^{(T)}_{\pi} I_\mathcal{T}=\left(\prod_{S\in \pi} \partial^{(S)}  \right)I_{\mathcal{T}\setminus \{T\}}\,. \quad}
\end{equation}
We will see that this identity implies an interesting factorization when taking $T$ to be the maximal tube in a tubing.

\subsection{Contractions using reduction operators}\label{ssec:contractions}

In this section, we  further investigate equation \eqref{eq:QTonint} and 
develop an associated diagrammatical interpretation resulting from the action of $Q^{(T)}$. More precisely, we investigate the properties of the integrals with removed tubes, such as $I_{\mathcal{T}\setminus \{T\}}$, and characterize the situations in which they can be represented by 
another integral that arises from a contracted diagram. 

\paragraph{Contractions and tubings.} 
To begin with, let us consider a sub-diagram within a tubing $\mathcal{S}$, with the following properties.  
We consider an edge connecting two vertices $v_1$ and $v_2$. The tubing $\mathcal{S}$
is assumed to contain a tube $T_{\rm g} =\{v_1,v_2\}$ containing both vertices, but both individual vertices are `bare' in the sense that  $\mathcal{S}$ does not contain 
the minimal tubes only encircling $v_1$ and $v_2$, respectively. 
Such a situation can arise, for example, by acting with reduction operators 
$Q^{(T)}$ on a complete tubing $\mathcal{T}$ in such a way that two vertices are bare, as we will discuss after \eqref{two-vertices_circled}. Diagrammatically, we thus consider the following partial tubing
\begin{equation} \label{two-vertices_non-circled}
\includegraphics[valign=c]{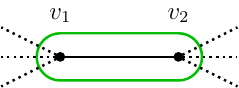}
\end{equation}
The dotted lines denote an arbitrary number of edges that connect to the rest of the diagram. Note that there can also be additional tubes that fully enclose this part of the diagram but these have not been drawn. We denote the integral associated to this tubing by $I_\mathcal{S}$.

We now want to show that $I_\mathcal{S}$ can be computed by evaluating the integral associated to the \textit{contracted diagram}, where the edge is shrunk to a point, and the two vertices $v_1,v_2$ coalesce. 
Diagrammatically, we want to establish an equality 
\begin{equation} \label{simple_two_vertex_conctraction}
\includegraphics[valign=c]{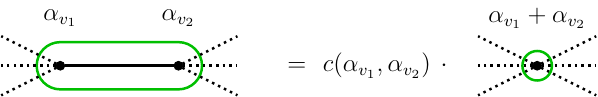}
\end{equation}
where we have displayed the weights associated to each vertex. In this expression $c(\alpha_{v_1},\alpha_{v_2})$ is a universal function of the initial weights of the vertices.
To show this, we first note that since all polynomials $p_T$, $T\in \mathcal{S}$, must enclose both $v_1$ and $v_2$, they can only depend on the combination $ x_{{v_1}}+x_{{v_2}}$. Therefore, changing coordinates to $x_+\equiv x_1+x_2$, $t \equiv x_2/(x_1+x_2)$, we obtain an integral of the form
\begin{equation}\label{int_red_1}
    \int_{\R_+} dx_{v_1}dx_{v_2}\, f(x_+)\, x_{v_1}^{\alpha_{v_1}-1} x_{v_2}^{\alpha_{v_2}-1}= \int_{\R_+} dx_{+} f(x_+) \, x_+^{\alpha_{v_1}+\alpha_{v_2}-1}\int_0^1 dt\, t^{\alpha_2-1}(1-t)^{\alpha_1-1}\,.
\end{equation}
Identifying the integral over $t$ as the Beta function $B(\alpha_1,\alpha_2)$, we can apply this logic to the integral associated to $\mathcal{S}$ to obtain

\begin{equation} \label{contr_integral}
        I_{\mathcal{S}} =B(\alpha_{v_1},\alpha_{v_2}) \cdot \int_{\R_+} d^{N_{\rm v}-2}x dx_+\frac{\prod_{v\neq 1,2} x_v^{\alpha_v-1}}{\prod_{T\in \mathcal{
        S}}p_T} \, x_+^{\alpha_{v_1}+\alpha_{v_2}-1} \,,  
\end{equation}
with the right-hand side being an integral for a diagram with only $N_{\rm v}-1$ vertices and $B$ denoting the Beta function. 

\paragraph{More general contractions.}
It turns out that the above discussion can be generalized further and equally applies to diagrams in which the vertices $v_1,v_2$ are not only connected by an edge, as we considered in \eqref{two-vertices_non-circled}. In fact, a contraction depicted in \eqref{simple_two_vertex_conctraction} generalizes to a tube $T$ that contains any two bare vertices $v_1,v_2$. Repeating the integration steps similar to \eqref{int_red_1} and \eqref{contr_integral} we thus infer the identity
\begin{equation} \label{involved_two_vertex_conctraction}
\includegraphics[valign=c]{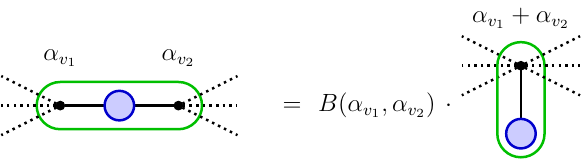}
\end{equation}
where the shaded blue circle denotes an arbitrary sub-tubing. Crucially, this observation applies regardless of the topology of the diagram. This implies that any tubing $\mathcal{T}$ of an $n$-point diagram can be reduced to a tubing of a $\vert \mathcal{T}\vert$-point diagram, relating the functions of higher-point diagrams to the ones of lower-point diagrams. These observations combined allow us to obtain the minimal representation necessary to calculate any $n$-point amplitude, which we will explain in section~\ref{sec:complexity}.

\paragraph{Relation to locality of the theory.} It is interesting to point out that using the reduction operators acting via \eqref{eq:QTonint} and the contractions \eqref{simple_two_vertex_conctraction} leads, at least in the simplest situation, to second-order differential equations that are reminiscent of locality constraints. To see this, we start from a complete tubing $\mathcal{T}$ which contains the two tubes $T_{\rm b}=\{v_1\}$ and $T_{\rm r}=\{v_2\}$ that encircle the individual vertices $v_1,v_2$
as
\begin{equation} \label{two-vertices_circled}
\begin{tikzpicture}
\begin{feynman}
        \vertex [dot](v1) {};

        \vertex[above=.7cm of v1] {$v_1$};
        \vertex [right=2 cm of v1,dot] (v2) {};  
        \vertex[above=.7cm of v2] {$v_2$};        
        \vertex[left=1cm of v1] (lc);
        \vertex[above=.5cm of lc] (lu);
        \vertex[below=.5cm of lc] (ld);

        \vertex[right=1cm of v2] (rc);
        \vertex[above=.5cm of rc] (ru);
        \vertex[below=.5cm of rc] (rd);    
        
        \diagram*  {
          (v1) --[very thick] (v2);
          (v1) --[connect,green!75!black,very thick] (v2);
          
          (v1)--[very thick,dotted] (lu);
          (v1)--[very thick,dotted] (lc);
          (v1)--[very thick,dotted] (ld);
          
          (v2)--[very thick,dotted] (ru);
          (v2)--[very thick,dotted] (rc);
          (v2)--[very thick,dotted] (rd);          
          };
        
        \vertex [above=0 cm of v1,shape=circle,draw=blue!80!black,fill=none,very thick,minimum size=.5cm] (b1) {};
        \vertex [above=0 cm of v2,shape=circle,draw=red!80!black,fill=none,very thick,minimum size=.5cm] (b2) {};
\end{feynman}
\end{tikzpicture}
\end{equation}
The diagram \eqref{two-vertices_non-circled} can be obtained from this tubing $\mathcal{T}$ by the action of the reduction operators $Q^{(T_{\rm b})}$ and  $Q^{(T_{\rm r})}$. In fact, applying \eqref{eq:QTonint} twice, 
we infer the relation 
\begin{equation}\label{eq:Qe1Qe2}
    Q^{(T_{\rm b})} Q^{(T_{\rm r})} I_\mathcal{T}=\bigg( \sum_{T\supsetneq \{v_1\}} \partial^{(T)}  \bigg)\bigg( \sum_{T'\supsetneq \{v_2\}} \partial^{(T')}  \bigg) I_{\mathcal{T}\setminus \{T_{\rm b},T_{\rm r}\}}\,.
\end{equation}
The integral $I_{\mathcal{T}\setminus \{T_{\rm b},T_{\rm r}\}}$ appearing on the right-hand side of this expression, is now associated to a tubing $\mathcal{S} = \mathcal{T}\setminus \{T_{\rm b},T_{\rm r}\}$, that contains a sub-diagram of the type \eqref{two-vertices_non-circled}. 
We now use \eqref{contr_integral} in \eqref{eq:Qe1Qe2}, and 
note that the derivatives on the right-hand side of \eqref{eq:Qe1Qe2} can be replaced by 
partial derivatives $\partial_{X_{v_1}}$ and $\partial_{X_{v_2}}$. Performing a partial integration and dropping boundary terms, we then find that these derivatives merely lead to a modification of the vertex weight $\alpha_{v_1}+\alpha_{v_2}$ to $\alpha_{v_1}+\alpha_{v_2}-2$. The resulting expression can be diagrammatically summarized as
\begin{equation}\label{eq:Qcontraction}
\includegraphics[valign=c]{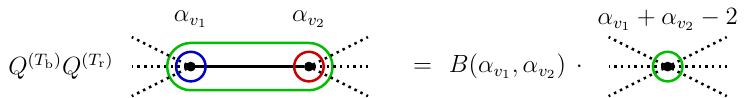}
\end{equation}
Since the reduction operators are first-order operators, we thus determined a second order differential equation relating the integral $I_{\mathcal{T}}$ to 
its contracted version.\footnote{Note that only in de Sitter space, i.e.~for $\epsilon=0$, the right hand side results in the correct weight for the vertex degree as determined from the 
same underlying model. This implies that in this case the integrals are directly related and further simplifications occur.} 

This differential equation is reminiscent of the differential equation obtained 
by using the properties of the propagator $G_e(Y_e,\eta_{v_1},\eta_{v_2})$ in a local quantum field theory. In fact, we can consider a Feynman diagram and replace the propagator $G_e(Y_e,\eta_{v_1},\eta_{v_2})$ by $\delta(\eta_{v_1}-\eta_{v_2})$. Due to the fact that the propagator satisfies the Green's function equation
\begin{equation}
     (\partial_{\eta_{v_1}}^2+Y_e^2)G_e=(\partial_{\eta_{v_2}}^2+Y_e^2)G_e=i\delta(\eta_{v_1}-\eta_{v_2})\ ,
\end{equation}
one can use integration-by-parts relations to equally derive a second-order differential equation relating different diagrams. From this, one finds an equality where on one side a second order differential operator acts on the original integral, while on the other side there is a contracted diagram with one less propagator, similar to~\eqref{eq:Qcontraction}.

\subsection{Cuts and factorizations}\label{ssec:factors}

In section \ref{ssec:contractions} we have considered diagrams that can be contracted due to the absence of minimal tubes encircling individual vertices. We next turn to the case where a maximal tube $T_{\rm max}$ is absent 
and study \textit{factorization identities} and the associated cuts in a diagram. In analogy to section~\ref{ssec:contractions}, we can remove a tube 
by using the reduction operators. In case of a maximal tube $T_{\rm max} \in \mathcal{T}$, however, we will use the higher-order reduction operator $Q^{(T_{\rm max})}_\pi$ and rely on the identity~\eqref{eq:highordtuberemov}. 

\paragraph{Factorization identities.}

While a general integral associated to 
a tubing does not factorize, it is not hard to identify tubings for which the integral splits.
To illustrate this, let us begin with a tubing $\mathcal{S}$ that can be decomposed 
as 
\begin{equation}
    \mathcal{S}=\mathcal{T}_1\sqcup \mathcal{T}_2\ ,
\end{equation}
with the crucial feature that $\mathcal{T}_1$, $\mathcal{T}_2$ are two disjoints tubings that are not inside a bigger tube. Diagrammatically, this can be represented as
\begin{equation}
\includegraphics[valign=c]{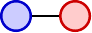}
\end{equation}
where the shaded blue and red circles denote tubings of their associated sub-diagrams.
Recall that for any tube $T$, the polynomial $p_T$ only depends on the integration variables that it encircles. This implies that, for any $T \in \mathcal{T}_1$, $p_T$ cannot depend on the integration variable of any vertex encircled by $\mathcal{T}_2$, and vice versa. This implies that the integral must factorize as 
\begin{equation} \label{max_fac}
    I_{\mathcal{S}}= \left(\int_{\R_+^k}d^kx \, \frac{\prod_{v=1}^k x_v^{\nuv-1}}{\prod_{T\in {\mathcal{T}_1}}p_T}\right)\left(\int_{\R_+^{n-k}}d^{\Nv-k}x \, \frac{\prod_{v=k+1}^{\Nv} x_v^{\nuv-1}}{\prod_{T\in {\mathcal{T}_2}}p_T}\right) = I_{\mathcal{T}_1} I_{\mathcal{T}_2}\,,
\end{equation}
where we have split the vertices such that the first $k$ are encircled by $\mathcal{T}_1$ while the others are encircled by $\mathcal{T}_2$. The resulting integrals 
 $I_{\mathcal{T}_1}$, $I_{\mathcal{T}_2}$ are associated to the blue and red subdiagrams and their respective tubings, leading to the diagrammatical representation
\begin{equation}
\includegraphics[valign=c]{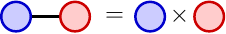}
\end{equation}
We can think of the edge connecting the two sub-diagrams as being cut. For a more general tubing, a similar factorization holds. Assuming a disjoint splitting of a tubing $\mathcal{S}$, we find 
\beq
   \mathcal{S} = \mathcal{T}_1\sqcup \ldots \sqcup \mathcal{T}_n:\qquad I_{\mathcal{S}}=I_{\mathcal{T}_1} \cdot \ldots \cdot I_{\mathcal{T}_n}\ . 
\eeq
In this case, it will result in multiple edges being cut at the same time.

\paragraph{Factorization formulas using reduction operators.}

Having established factorization identities for tubings consisting of disjoint sub-tubings, we next want to show that this situation can always be reached when applying a reduction operator. Let us start with a complete tubing $\mathcal{T}$. For such a tubing, there always is a maximal tube, which then contains two sub-tubings connected by a single edge. Diagrammatically, this can be represented as
\begin{equation}
\includegraphics[valign=c]{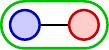}
\end{equation}
where the green tube is the maximal tube. 
From this, we find that it is possible to decompose $\mathcal{T}$ as 
\begin{equation}
    \mathcal{T}=\mathcal{T}_1\sqcup \mathcal{T}_2\sqcup \{T_{\rm max}\}\,,
\end{equation}
where $\mathcal{T}_1$, $\mathcal{T}_2$ are two disjoint tubings of the sub-diagram as above.
We can now use an appropriate reduction operator to remove $T_{\rm max}$. Since the maximal tube is not contained in another tube, it has no first-order reduction operator associated to it and we must consider the higher-order operators $Q^{(T_{\rm max})}_\pi$ associated to partitions of $T_{\rm max}$. There will be multiple of these, and for each such reduction operator it will realize the factorization described above. Here, we will use that $\mathcal{T}_1$ and $\mathcal{T}_2$ both have their own maximal tubes $T_1$ and $T_2$. and that every vertex is enclosed by either of the two. Thus, there is a natural partition
\begin{equation}
    T_{\rm max}=T_{1}\sqcup T_{2}\, , \quad \pi=\{T_1,T_2\}\ .
\end{equation}
Evaluating the higher-order reduction operator~\eqref{higher-order-red_final} with this partition, we find 
\begin{equation}\label{eq:QTmax}
    Q^{(T_{\rm max})}_\pi=\partial^{(T_1)} \partial^{(T_2)} \big(z^{(T_{\rm max})}  -z^{(T_1)} -z^{(T_2)}  \big)\ .
\end{equation}
Inserted in \eqref{eq:highordtuberemov}, this operator will then satisfy
\begin{equation} \label{apply_red_op}
    Q^{(T_{\rm max})}_\pi I_\mathcal{T}= \big(\partial^{(T_1)}  I_{\mathcal{T}_1}\big) \big(\partial^{(T_2)}  I_{\mathcal{T}_2}\big)\ .
\end{equation}
Note that there are many other reduction operator that might realize this factorization. For example, we can consider
\begin{equation}
    T_{\rm max}=\bigsqcup_{v=1}^{N_{\rm v}} \{v\}
\end{equation}
in order to obtain a higher-order reduction operator of degree $N_{\rm v}$ realizing the same factorization.

\paragraph{Singularity structure.}
Interestingly, the higher-order reduction operators have implications for the singularity structure of the cosmological correlators. Using \eqref{eq:QTmax} in~\eqref{apply_red_op}, we find
\begin{equation}  \partial^{(T_1)} \partial^{(T_2)} \big(z^{(T_{\rm max})} -z^{(T_1)} -z^{(T_2)} \big) I_\mathcal{T}=\partial^{(T_1)} \partial^{(T_2)} I_{\mathcal{T}_1}I_{\mathcal{T}_2}\, .
\end{equation}
This equation can be integrated directly, implying that $I_\mathcal{T}$ can be written as
\begin{equation} \label{singularity_split}
    I_\mathcal{T}=\frac{I_{\mathcal{T}_1} I_{\mathcal{T}_2}+f_{T_1}+f_{T_2}}{z^{(T_{\rm max})} -z^{(T_1)} -z^{(T_2)} } = - \frac{I_{\mathcal{T}_1} I_{\mathcal{T}_2}+f_{T_1}+f_{T_2}}{2 Y_e}\ ,
\end{equation}
where $f_{T_1}$ and $f_{T_2}$ are functions independent of $z^{(T_2)} $ and $z^{(T_1)} $ respectively.  Here we have rewritten the denominator in terms of the physical variables using \eqref{physical_slice}, where $Y_e$ is the momentum flowing along the edge connecting $\mathcal{T}_1$ and $\mathcal{T}_2$. Furthermore, since $I_{\mathcal{T}_1} I_{\mathcal{T}_2}$ is independent of $z^{(T_{\rm max})} $, the above equations imply that
\begin{equation}
    {\rm Res}_{Y_e\rightarrow 0}(I_{\mathcal{T}}) = I_{\mathcal{T}_1} I_{\mathcal{T}_2} +\ldots \,,
\end{equation}
where $\rm Res$ denotes the residue around $Y_e\rightarrow 0$ and the dots are some unknown terms due to $f_{T_1}$ and $f_{T_2}$. Factorizations as described above seem related to those obtained for amplitudes in both quantum field theory \cite{cachazo_sharpening_2008,arkani-hamed_locality_2018,travaglini_sagex_2022} as well as for cosmological correlators \cite{frellesvig_decomposition_2021,goodhew_cosmological_2021,melville_cosmological_2021,jazayeri_locality_2021,pietro_analyticity_2022,baumann_linking_2022,albayrak_perturbative_2024}. Note also that similar factorization formulae will hold for the other higher-order reduction operators.

\section{Differential chains from first-order operators}\label{sec:differentialchain}

One of the crucial observations in the results of the previous section is that acting with a reduction operator effectively removes a tube from a tubing. In this section, we will show that this allows us to use the first-order reduction operators to write derivatives of $I_\mathcal{T}$ as a sum of integrals $I_\mathcal{S}$ with $\cS \subseteq \cT$. This will enable us to determine a system of differential equations for the integral $I_\cT$ with a remarkable similarity to the kinematic flow algorithm of~\cite{arkani-hamed_kinematic_2023}. In section~\ref{ssec:chainconstruction} we present an algorithm to construct the general form of this differential chain. We then illustrate the involved steps in an explicit example in section~\ref{ssec:chainexamples}. Finally, we show in section~\ref{Pfaffian_system} that the general first-order differential system actually takes the form of a Pfaffian system. This observation yields a well-defined measure of complexity for cosmological correlators constructed from the integrals $I_{\mathcal{T}}$ using first-order reduction operators.

\subsection{Algorithmic construction of the differential chains}\label{ssec:chainconstruction}

\paragraph{A system of differential equations.}

We begin the construction of the differential equations by recalling the key insights from section~\ref{sec:physics}. Consider a tubing $\cT$ and a tube $T$, the corresponding integral $I_\cT$, and the first-order reduction operator $Q^{(T)}$. Now, depending on whether $T\in \cT$ or not, there are two possibilities. Either acting with $Q^{(T)}$ removes the tube as in equation~\eqref{eq:QTonint}, or the tube is already removed and we have
\begin{equation}
    \partial^{(T)} I_\cT=0\,,
\end{equation}
since the polynomial involving $z^{(T)} $ has been removed. Because there is a first-order reduction operator $Q^{(T)}$ for every non-maximal tube $T \in \cT$, this implies that there is the following system of equations
\begin{align} \label{eq:ITsystem}
        Q^{(T)}I_\mathcal{T}&=\sum_{T'\supsetneq T} \partial^{(T')}  I_{\mathcal{T}\setminus \{T\}} & \text{ if } T \in \mathcal{T}\,,\\
        \partial^{(T)} I_\mathcal{T}&=0 &\text{ if } T\not\in \mathcal{T} 
\end{align}
for every tubing $\cT$. Here, for the convenience of the reader, we recall that the first-order reduction operator takes the form  
\begin{equation}\label{eq:recallQ}
    Q^{(T)}=z^{(T)}\sum_{ T'\supsetneq T}\partial^{(T')}+\sum_{T'\subseteq T}(\theta^{(T')}+\nu^{(T')})-\sum_{v\in T}\alpha_v\ .
\end{equation}
This system of equations is the starting point for iteratively constructing a solution.

\paragraph{A differential chain.}
An essential property of the system of differential equations in \eqref{eq:ITsystem} is that the right-hand side only involves tubings containing strictly fewer tubes than $\cT$. This means that the equation can be iterated, leading to an expression in terms of increasingly smaller tubings. The only tube that can not be removed in this manner is the maximal tube $T_{\rm max}$, as from equation~\eqref{eq:ITsystem} it follows that
\begin{equation}\label{eq:QTmaxIcT}
    Q^{(T_{\rm max})}I_\cT=0\,.
\end{equation}
Therefore, the tube removal continues until only the maximal tube remains. 

Then, equation~\eqref{eq:QTmaxIcT} implies that associated integral must satisfy
\begin{equation}\label{eq:ITmaxdif}
    \partial^{(T_{\rm max})} I_{\{T_{\rm max}\}}=
    \frac{1}{z^{(T_{\rm max})}} \left(\sum_{v\in T_{\rm max}}\alpha_v -1\right) I_{\{T_{\rm max}\} }\,.
\end{equation}
Note that all of the other derivatives vanish, since $I_{\{T_{\rm max}\}} $ only depends on $z^{(T_{\rm max})}$. This allows us to solve for $I_{\{T_{\rm max}\}}$, which will consist of $z^{(T_{\rm max})}$ raised to a complex power.

From here, we can iteratively add tubes. In particular, let us first add a single tube tube $T$, and use the system~\eqref{eq:ITsystem} combined with equation~\eqref{eq:ITmaxdif} to write partial derivatives acting on $I_{\{T_{\rm max},T\}}$ in terms of the function itself and $I_{\{T_{\rm max}\}}$. Then, adding another tube $T'$, we can write partial derivatives acting on the new integrals in terms of itself and the functions $I_{\{T_{\rm max},T\}}$, $I_{\{T_{\rm max},T'\}}$ and $I_{\{T_{\rm max}\}}$. Continuing in this manner, we obtain a chain of first-order differential equations which starts with $I_{\{T_{\rm max}\}}$ and ends with the desired integral $I_{\cT}$. As a result, for any tubing $\cT$, the derivatives of $I_{\cT}$ can be expressed in the general form 
\begin{equation}\label{eq:ITpartialfin}
    \partial^{(T)}  I_\mathcal{T} = \sum_{\substack{\mathcal{S}\subseteq \mathcal{T},\\ T_{\rm max}\in \mathcal{S}}} r^{(T)}_\mathcal{S} (\mathcal{T}) \, I_{\mathcal{S}} \,
\end{equation}
where the $r^{(T)}_{\cS}(\cT)$ are rational functions of $\alpha_v$ and $z^{(T)}$ and the sum is over all sub-tubes of $\mathcal{T}$ which contain $T_{\rm max}$. 

In the remainder of this subsection, we will aim to make the structure of this differential chain as explicit as possible. We stress that the procedure is fully algorithmic and can be easily implemented computationally. Nevertheless, one needs to introduce some extra notation if one wants to write down closed-form expressions.

\paragraph{Notation for tube structure.}
 In the following, it is necessary to carefully keep track of the structure of tubes and tubings, and for this purpose we introduce the following notation. Given a tubing $\cT$ and two tubes $S,T\in \cT$, we write 
\begin{equation}
    S \prec_\cT T\ ,
\end{equation}
whenever $S\subsetneq T$ and there exists no $T'\in \cT$ such that $S\subsetneq T' \subsetneq T$. The interpretation is that if we sort the tubes in $\cT$ by inclusion, then $S$ is the precursor of $T$, and $T$ is the successor of $S$. Note that a tube may have any number of precursors, but the successor of a tube is unique.\footnote{The ordering $\prec_\cT$ gives the tubes in $\cT$ the structure of an ordered tree. This tree will be rooted if $\cT$ contains the maximal tube $T_{\rm max}$, which will always be the case for us. Additionally, complete tubings are in one-to-one correspondence with full binary trees. This perspective will be useful for the combinatorial analysis performed later.} We denote the successor of a tube $T$ in a tubing $\cT$ by $T^+_\cT$. Finally, we write
\begin{equation}
    S \sim_\cT T\ ,
\end{equation}
whenever $S$ and $T$ have the same successor in $\cT$. 

To illustrate this notation, consider the tubing $\cT$ given by
\begin{equation}
\includegraphics[valign=c]{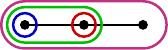}
\end{equation}
where we denote the red, blue, green and magenta tubes by $T_{\rm r}$, $T_{\rm b}$, $T_{\rm g}$ and $T_{\rm m}$ respectively.
Here we have $T_{\rm b} \sim_\cT T_{\rm r}$, since they have the same successor $T_{\rm g}$, which we can express as $(T_{\rm b})_\cT^+=(T_{\rm r})_\cT^+=T_{\rm g}$. Similarly we find  $T_{\rm b}\prec_\cT T_{\rm g} \prec_{\cT} T_{\rm m}$. On the contrary, we see that $T_{\rm m}$ is not a successor of $T_{\rm b}$ or $T_{\rm r}$, i.e.~$T_{\rm{b}} \nprec_{\cT} T_{\rm m}$.

\paragraph{Derivation of the differential chain.}
With this notation, we can derive an explicit form of the differential chain. We begin by observing that, for any $T\in \cT$, there is the following identity:
\begin{equation}
    \sum_{T'\subseteq T}(\theta^{(T')}+\nu^{(T')}) = \theta^{(T)}+\nu^{(T)} +\sum_{S \prec_\cT T}  \sum_{T'\subseteq S}(\theta^{(T')}+\nu^{(T')})\,.
\end{equation}
Similarly, for any $S$ with $S\prec_\cT T$ we have 
\begin{equation}\label{eq:partialSsum}
    \sum_{T'\supsetneq S} \partial^{(T')}=\partial^{(T)}+\sum_{T'\supsetneq T} \partial^{(T')}\,.
\end{equation}
Comparing this with the definition of the first-order reduction operator recalled in \eqref{eq:recallQ} above, it follows that 
\begin{equation}\label{eq:QT-QS}
    Q^{(T)}-\sum_{S \prec_\cT T }Q^{(S)}= \left(z^{(T)}-\sum_{S\prec_\cT T}z^{(S)}\right)\left(\sum_{T'\supseteq T}\partial^{(T')} \right)+\gamma^{(T)}_\mathcal{T}\,,
\end{equation}
where $\gamma^{(T)}_\cT$ is a constant defined by 
\begin{equation}\label{eq:gammadef}
    \gamma^{(T)}_\mathcal{T}= \nu^{(T)}-\sum_{\substack{v\in T,\\ v\not\in S\subsetneq T}} \alpha_v \,.
\end{equation}
Here the sum is over all vertices in $T$ that are not enclosed by any of the sub-tubes of $T$. The relation derived above can be rewritten to 
\begin{equation}\label{eq:partialTinQs1}
\boxed{\rule[-.6cm]{0cm}{1.5cm} \quad
    \sum_{T'\supseteq T}\partial^{(T')}= \frac{    Q^{(T)}-\sum_{S\prec_\cT T }Q^{(S)}-\gamma^{(T)}_\mathcal{T}}{z^{(T)}-\sum_{S\prec_\cT T}z^{(S)}}\,,\quad
    }
\end{equation}
which holds for any tube $T$.\footnote{One might worry that, since these expressions depend heavily on the tubing $\mathcal{T}$, these expressions only hold when acting on $I_\mathcal{T}$. In particular as, when acting on an arbitrary tubing $\mathcal{T}$, these expressions will involve terms of the type $Q^{(T)}I_\mathcal{T}$ with $T$ not in the tubing $\mathcal{T}$. However, one can use the fact that $\partial^{(T)}I_\mathcal{T}=0$ to fix these extractions and one obtains a result  compatible with the above. Note that, for us we will not need such expressions in any case.} This result can straightforwardly be translated to an expression for $\pd^{(T)}I_\cT$ in terms of the reduction operators by noting that
\begin{equation}
    \pd^{(T)} I_\cT = \left(\sum_{T' \supseteq T}\pd^{(T')} - \sum_{T'\supseteq T^+_\cT }\pd^{(T')}   \right) I_\cT
\end{equation}
and using \eqref{eq:partialTinQs1} for the two sums. For brevity we do not display the resulting equation here.

Now, observe that the operator on the left-hand side closely resembles the operator appearing on the right-hand side of equation \eqref{eq:ITsystem}; the difference is that the latter has one extra term. To connect these two equations, note that 
\begin{equation}
    \sum_{T'\supsetneq T}\partial^{(T')}=\sum_{T'\supseteq T^+_\cT}\partial^{(T')} \,.
\end{equation}

Combining this with equations \eqref{eq:ITsystem} and \eqref{eq:partialTinQs1}, we find that the action of a reduction operator $Q^{(T)}$ on $I_\cT$ can be written as  
\begin{equation}\label{eq:QTreduction1}
    Q^{(T)} I_\cT = \frac{  Q^{(T^+_\cT)}-\sum_{S \prec_{\cT \setminus\{T\}} T^+_\cT    }Q^{(S)}-\gamma^{(T^+_\cT) }_{\cT\setminus\{T\}}}{z^{(T^+_\cT)}-\sum_{S \prec_{\cT \setminus\{T\}} T^+_\cT  }z^{(S)}} I_{\cT\setminus \{T\}}\,.
\end{equation}
In this equation the iterative nature of the reduction operator is made manifest; using this equation the expression for $Q^{(T)} I_\cT$ can be recursively reduced until it is a linear combination of integrals $I_\cS$ with $\cS\subseteq \cT$ with rational coefficients. Finally, we note that, as the maximal tube $T_{\rm max}$ has no successor, the right-hand side of~\eqref{eq:QTreduction1} is not well-defined when evaluated for $T_{\rm max}$. Therefore, in this case we must separately replace $Q^{(T_{\rm max})}I_\cT$ by zero.

\paragraph{Compact form of differential chain.}
The equations derived above are explicit, but somewhat complicated. To increase its usability, we now rewrite \eqref{eq:QTreduction1} in a more compact form. We do this by first introducing the matrices 
\begin{align}\label{eq:Mdef}
    M^{T,S}_{\cT} &= \begin{cases}
    1&  \text{if } T \prec_\cT S\,, \\
    -1& \text{if } T\succ_\cT S \text{ or } T\sim_{\cT} S \\
    0  & \text{else}\,.
    \end{cases} 
\end{align}
for each $\cT$. Then, using this notation, we define the functions
\begin{equation}\label{eq:letterM}
    \ell^{(T)}_\cT = \bigg( \sum_{S\in \cT\setminus \{T\}}M^{T,S}_\cT z^{(S)}  \bigg)^{-1} \,.
\end{equation}
as well as the constants
\begin{equation}\label{eq:cdef}
    c^{(T)}_{\cT} = \sum_{S\in \cT\setminus T} M^{T,S}_\cT \sum_{v\in S} \alpha_v-1\,.
\end{equation}
Note that $c^{(T)}_{\cT}=-\gamma^{(T^+_\cT)}_\cT$, with $\gamma$ as in equation~\eqref{eq:gammadef}. With this new notation, equation \eqref{eq:QTreduction1} can be compactly written as
\begin{equation}\label{eq:QTiteration}
\boxed{\rule[-.6cm]{0cm}{1.4cm} \quad 
    Q^{(T)}I_\mathcal{T} = \ell_{\cT}^{(T)} \bigg(\sum_{S\in\cT\setminus \{T\}}M^{T,S}_{\cT} Q^{(S)} +c^{(T)}_{\cT} \bigg)I_{\cT\setminus\{T\}} \,.\quad
}
\end{equation}
Note that, as in equation \eqref{eq:QTreduction1}, this expression does not hold for $Q^{(T_{\rm max})}$, in which case we must impose $Q^{(T_{\rm max})}I_\cT =0$. Furthermore, note that $M^{T,S}_\cT$ is a purely combinatorial object, and can be found algorithmically using the index set representation of the tubes.

In summary, to obtain $Q^{(T)}I_\cT$ one must first apply equation~\eqref{eq:QTiteration}. Then, there will be terms of the form $Q^{(S)}I_{\cT\setminus \{T\}}$ for various $S$ and equation~\eqref{eq:QTiteration} can again be used on these terms to remove yet another tube from the tubing. This procedure can be recursively applied until only the maximal tube remains. As we know the remaining integral satisfies $Q^{(T_{\rm max})}I_{\{T_{\rm max}\}}$, this signals the end of the recursion. Inserting all of this into the original expression for $Q^{(T)}I_\cT$, one is left with an algebraic expression for the action of $Q^{(T)}$ in terms of the other integrals in the chain. Repeating this for each tube in $\cT$, one can then apply equation~\eqref{eq:partialTinQs1} relating the reduction operators with the partial derivative and hence determine the coefficients $r^{(T)}_\cS(\cT)$ in \eqref{eq:ITpartialfin}. As an alternative, we show in appendix~\ref{ap:matrixrep} how the iteration in equation~\eqref{eq:QTiteration} can also be rewritten and solved by interpreting it as a matrix equation on a suitable vector space, resulting in a direct expression of $Q^{(T)} I_\cT$ in terms of the other integrals.

\subsection{An example: the single-exchange diagram  } \label{ssec:chainexamples}

In order to illustrate the construction above, we will again return to the example of the single-exchange integral and construct its differential chain explicitly. Doing this, we will see the iterative nature of this differential chain, motivating the nomenclature of recursive reductions. As the purpose of this section is to illustrate the results above, we will treat this simple example with the general technology, even though directly solving the system~\eqref{eq:ITsystem} would be more efficient in this case.

\paragraph{Functions in the chain.}

To construct the chain, recall that the single-exchange integral arises from the tubing
\begin{equation}\label{eq:singexchagnetubingS5S2}
\includegraphics[valign=c]{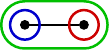}
\end{equation}
and that we have labeled the blue, red and green tubes as $T_{\rm b}$, $T_{\rm r}$ and $T_{\rm g}$ respectively. Furthermore, this GKZ system has three first-order reduction operators, each associated to a tube. As discussed in section~\ref{ssec:chainconstruction}, we can construct the differential chain by studying the action of these reduction operators. In particular, we know that acting with the reduction operators $Q^{(T_{\rm r})}$ and $Q^{(T_{\rm b})}$ will remove the red or blue tube from the tubing, while $Q^{(T_{\rm g})}$ will annihilate the functions in the chain, since $T_{\rm g}$ is the maximal tube. Thus we find that there are four tubings we must consider, organized as 
\begin{equation}\label{eq:singexchdifchain}
\includegraphics[valign=c]{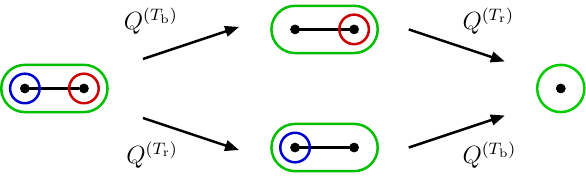}
\end{equation}
where the arrows indicate that the reduction operator acts on the integral associated to the left tubing can be written in terms of derivatives of the right tubing. Note that in the right-most diagram here, we have already used the discussion from section~\ref{ssec:contractions} to contract the edge. Interestingly, the structure of these differential chains is quite similar to the kinematic flow algorithm of \cite{arkani-hamed_differential_2023}.

From the above, we find that we must consider four functions in our differential chain, given by
\beq 
I_\text{\includegraphics[valign=c]{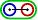}}\,,\quad  
I_\text{\includegraphics[valign=c]{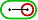}}\,,\quad 
I_\text{\includegraphics[valign=c]{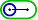}}\,,\quad 
I_\text{\includegraphics[valign=c,scale=.8]{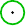}}\ ,
\eeq 
where, in order not to clutter the notation, we have drawn the tubings explicitly in the subscripts. 

\paragraph{General approach.}

The next step is to construct the differential chain. We are now ready to obtain the action of the reduction operators on the functions found above. We will begin with the function $I_\text{\includegraphics[valign=c]{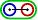}}$ 
and consider the action of $Q^{(T_{\rm b})}$. Recall from section~\ref{ssec:chainconstruction} that, in general, the action of a reduction operator is given in terms of the symbol $M^{T,S}_\cT$, which is determined by the successor structure of $\cT$. Therefore, the first step will be to determine this successor structure for the tubing of interest. Afterwards, we obtain an iterative equation relating the action of $Q^{(T)}$ on $I_\cT$ with reduction operators acting on $I_{\cT \setminus\{T\}}$. Repeating the above procedure we are eventually left with a linear combination of integrals in the differential chain with rational pre-factors. Then, one can use equation \eqref{eq:partialTinQs1} to solve for the partial derivatives in terms of the reduction operators.

\paragraph{The first reduction.}

In general, the successor structure of a diagram can be computed algorithmically using the fact that we can represent tubes as index sets and tubings as sets of tubes. However, given a diagrammatical representation of a tubing, it can also be observed immediately. For us, considering the tubing~\eqref{eq:singexchagnetubingS5S2} we find that the only successor relations are
\begin{equation}
    T_{\rm b} \prec_\text{\includegraphics[valign=c]{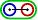}} T_{\rm g}\, , \quad   T_{\rm r} \prec_\text{\includegraphics[valign=c]{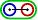}} T_{\rm g}\, , \quad   T_{\rm b} \sim_\text{\includegraphics[valign=c]{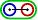}} T_{\rm r}\, .
\end{equation}
From this and the definition of $M^{T,S}_\cT$ in~\eqref{eq:Mdef}, we can immediately read off the non-zero elements of $M^{T,S}_\text{\includegraphics[valign=c]{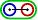}}$, which are given by
\begin{equation}
\begin{array}{rrrrr}
  &  M^{T_{\rm b},T_{\rm b}}_\text{\includegraphics[valign=c]{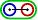}}&=M^{T_{\rm r},T_{\rm r}}_\text{\includegraphics[valign=c]{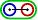}}&= M^{T_{\rm g},T_{\rm g}}_\text{\includegraphics[valign=c]{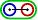}}&=-1\, ,\\[5pt]
       M^{T_{\rm b},T_{\rm r}}_\text{\includegraphics[valign=c]{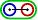}} &=M^{T_{\rm r},T_{\rm b}}_\text{\includegraphics[valign=c]{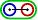}}&= M^{T_{\rm g},T_{\rm b}}_\text{\includegraphics[valign=c]{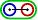}}&= M^{T_{\rm g},T_{\rm r}}_\text{\includegraphics[valign=c]{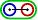}}  &=-1\, ,\\[5pt]
 & &  M^{T_{\rm b},T_{\rm g}}_\text{\includegraphics[valign=c]{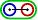}}&= M^{T_{\rm r},T_{\rm g}}_\text{\includegraphics[valign=c]{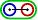}}&=  1 \,.
\end{array}
\end{equation}
The letters $\ell^{(T)}_\text{\includegraphics[valign=c]{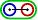}}$, as well as the constants $c^{(T)}_\text{\includegraphics[valign=c]{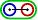}}$ can be readily obtained from these matrices using equations~\eqref{eq:letterM} and~\eqref{eq:cdef}. Note that, as acting with the reduction associated to the maximal tube will always result in zero, we will not need to obtain the letters $ \ell^{(T_{\rm g})}_\cT$ or constants $ c^{(T_{\rm g})}_\cT$ for any of the tubings of the single-exchange diagram. Thus, we find that the remaining letters are given by
\begin{equation}\label{eq:singexchangeletters1}
    \ell^{(T_{\rm b})}_\text{\includegraphics[valign=c]{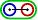}}=\frac{1}{z^{(T_{\rm g})}-z^{(T_{\rm r})}}\,, \quad  \ell^{(T_{\rm r})}_\text{\includegraphics[valign=c]{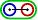}}=\frac{1}{z^{(T_{\rm g})}-z^{(T_{\rm b})}}\,,
\end{equation}
while the constants are given by
\begin{equation}\label{eq:singexchangeconsts}
      c^{(T_{\rm b})}_\text{\includegraphics[valign=c]{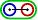}}=\alpha_1-1\,, \quad  c^{(T_{\rm r})}_\text{\includegraphics[valign=c]{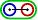}}=\alpha_2-1\,,
\end{equation}
where we recall that $\alpha_1$ and $\alpha_2$ are the twists of the vertices encircled by the blue tube and red tube respectively.

Using the equations above, obtaining the action of the reduction operators on $I_\text{\includegraphics[valign=c]{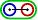}}$ comes from a straightforward application of equation~\eqref{eq:QTiteration}, which yields
\begin{equation}\label{eq:QTbactsingexchange2}
\begin{split}
    Q^{(T_{\rm b})}I_\text{\includegraphics[valign=c]{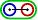}}=&\, \frac{Q^{(T_{\rm g})}-Q^{(T_{\rm r})}+\alpha_1-1}{z^{(T_{\rm g})}-z^{(T_{\rm r})}}I_\text{\includegraphics[valign=c]{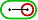}}\,, \\
            Q^{(T_{\rm r})}I_\text{\includegraphics[valign=c]{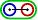}}=& \,\frac{Q^{(T_{\rm g})}-Q^{(T_{\rm b})}+\alpha_2-1}{z^{(T_{\rm g})}-z^{(T_{\rm b})}}I_\text{\includegraphics[valign=c]{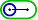}}\,, \\
            Q^{(T_{\rm g})}I_\text{\includegraphics[valign=c]{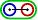}}=& \,0\,,
\end{split}
\end{equation}
where, for the last equality, we have used the fact that acting with the reduction operator of the maximal tube always results in zero.

From equation \eqref{eq:QTbactsingexchange}, the recursive nature of the reduction operators immediately becomes clear. We see that, in order to obtain the action of $Q^{(T_{\rm b})}$ on $I_\text{\includegraphics[valign=c]{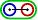}}$, we must now proceed by obtaining the action of the reduction operators on the sub-tubings of~\eqref{eq:singexchagnetubingS5S2}. For general diagrams, this procedure will continue until all tubes but the maximal one are removed.

\paragraph{The second reduction.}
Thus, the next task at hand is to obtain the action of the reduction operators on the sub-tubings of~\eqref{eq:singexchagnetubingS5S2}. Here, we will consider $I_\text{\includegraphics[valign=c]{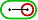}}$ and note that the actions on $I_\text{\includegraphics[valign=c]{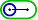}}$ can be obtained by permutations. We must consider two reduction operators now, $Q^{(T_{\rm g})}$ and $Q^{(T_{\rm r})}$. The action of $Q^{(T_{\rm g})}$ must still be zero as $T_{\rm g}$ is the maximal tube. The only successor relation of this diagram is 
\begin{equation}
 T_{\rm r} \prec_\text{\includegraphics[valign=c]{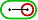}} T_{\rm g}
\end{equation}
resulting in
\begin{equation}
       M^{T_{\rm g},T_{\rm g}}_\text{\includegraphics[valign=c]{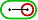}}= M^{T_{\rm r},T_{\rm r}}_\text{\includegraphics[valign=c]{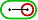}}= M^{T_{\rm g},T_{\rm r}}_\text{\includegraphics[valign=c]{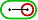}}=-1\, , \quad M^{T_{\rm r},T_{\rm g}}_\text{\includegraphics[valign=c]{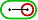}}=1
\end{equation}
for the symbol $M$. Then, proceeding along the same lines as above we obtain
\begin{equation}\label{eq:singexchangelettersandcoefs2}
      \ell^{(T_{\rm r})}_\text{\includegraphics[valign=c]{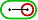}}=\frac{1}{z^{(T_{\rm g})}}\,, \quad       c^{(T_{\rm r})}_\text{\includegraphics[valign=c]{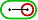}}=\alpha_1+\alpha_2-1\,,
\end{equation}
where again, we note that it is not necessary to obtain the corresponding expressions for $T_{\rm g}$ as it is the maximal tube. Inserting the above into equation~\eqref{eq:QTiteration} we obtain
\begin{equation}\label{eq:QTrInored}
\begin{split}
    Q^{(T_{\rm r})} I_\text{\includegraphics[valign=c]{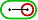}}&=\frac{Q^{(T_{\rm g})}+\alpha_1+\alpha_2-1}{z^{(T_{\rm g})}}I_\text{\includegraphics[valign=c]{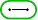}}\,, \\
Q^{(T_{\rm g})}  I_\text{\includegraphics[valign=c]{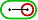}}&= 0\,.
\end{split}
\end{equation}
The corresponding equations can be obtained for $I_\text{\includegraphics[valign=c]{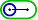}}$ can be obtained by permuting $T_{\rm r}$ with $T_{\rm b}$. We know from section~\ref{ssec:contractions} that $I_\text{\includegraphics[valign=c]{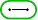}}=B(\alpha_1,\alpha_2)I_\text{\includegraphics[valign=c,scale=.8]{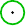}}$, with $B$ the beta-function. This allows us to rewrite the first equation in~\eqref{eq:QTrInored} in terms of $I_\text{\includegraphics[valign=c,scale=.8]{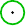}}$. Furthermore, we know that this integral must satisfy
\begin{equation}
    Q^{(T_{\rm g})}  I_\text{\includegraphics[valign=c,scale=.8]{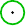}}= 0
\end{equation}
and we find that the iteration terminates here.

Finally, we can simply insert~\eqref{eq:QTrInored} in equation~\eqref{eq:QTbactsingexchange}, combined with the corresponding equations for $T_{\rm b}$, and obtain
\begin{equation}\label{eq:QTbactsingexchange}
\begin{split}
    Q^{(T_{\rm b})}I_\text{\includegraphics[valign=c]{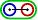}}=&\, \frac{\alpha_1-1}{z^{(T_{\rm g})}-z^{(T_{\rm r})}}I_\text{\includegraphics[valign=c]{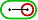}}- \frac{(\alpha_1+\alpha_2-1)B(\alpha_1,\alpha_2)}{z^{(T_{\rm g})}(z^{(T_{\rm g})}-z^{(T_{\rm r})})}I_\text{\includegraphics[valign=c,scale=.8]{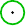}}\,,    \\
            Q^{(T_{\rm r})}I_\text{\includegraphics[valign=c]{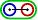}}=& \frac{\alpha_2-1}{z^{(T_{\rm g})}-z^{(T_{\rm b})}}I_\text{\includegraphics[valign=c]{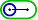}}- \frac{(\alpha_1+\alpha_2-1)B(\alpha_1,\alpha_2)}{z^{(T_{\rm g})}(z^{(T_{\rm g})}-z^{(T_{\rm b})})}I_\text{\includegraphics[valign=c,scale=.8]{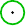}}\,,     \\
            Q^{(T_{\rm g})}I_\text{\includegraphics[valign=c]{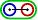}}=& \,0\,,
\end{split}
\end{equation}
Using these expression, we can now obtain the action of the partial derivatives on $I_\text{\includegraphics[valign=c]{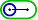}}$.

\paragraph{Partial derivatives.}

Now that we have found the action of all the reduction operators, the next step is to apply equation \eqref{eq:partialTinQs1} to rewrite the partial derivatives in terms of the reduction operators. Considering this equation for all $T$, one can straightforwardly solve for the partial derivatives.  In the following, we will focus on $\partial^{(T_{\rm b})}$, although the process will be similar for the other derivatives in the chain. 
To obtain the partial derivatives for the single exchange integral, let us insert $T=T_{\rm b}$ in equation~\eqref{eq:partialTinQs1} and act with it on $I_\text{\includegraphics[valign=c]{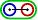}}$. 
In this case, we find
\begin{equation}
\begin{split}
  (\partial^{(T_{\rm b})}+\partial^{(T_{\rm g})})I_\text{\includegraphics[valign=c]{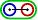}}=&
        \frac{\alpha_1-1}{z^{(T_{\rm b})}}\gamma^{(T)}_\text{\includegraphics[valign=c]{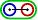}}I_\text{\includegraphics[valign=c]{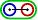}}     +\frac{\alpha_1-1}{z^{(T_{\rm b})}(z^{(T_{\rm g})}-z^{(T_{\rm r})})}I_\text{\includegraphics[valign=c]{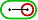}}\\
        &- \frac{(\alpha_1+\alpha_2-1)B(\alpha_1,\alpha_2)}{z^{(T_{\rm b})}z^{(T_{\rm g})}(z^{(T_{\rm g})}-z^{(T_{\rm r})})}I_\text{\includegraphics[valign=c,scale=.8]{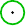}}\,.        
\end{split}
\end{equation}
Again, the corresponding equation for $T_{\rm r}$ can be found in an identical manner.

The final equation we need is obtained by inserting $T_{\rm g}$ in equation~\eqref{eq:partialTinQs1}. The combined system of equations is easily solved for the partial derivatives, giving
\begin{equation}\label{eq:singexchange3}
    \partial^{(T_{\rm b})}   I_\text{\includegraphics[valign=c]{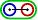}} 
        = 
        r_\text{\includegraphics[valign=c]{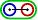}}I_\text{\includegraphics[valign=c]{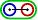}}        +         r_\text{\includegraphics[valign=c]{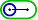}}I_\text{\includegraphics[valign=c]{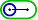}}         +         r_\text{\includegraphics[valign=c]{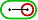}}
        I_\text{\includegraphics[valign=c]{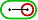}}         +         r_\text{\includegraphics[valign=c,scale=.8]{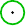}}I_\text{\includegraphics[valign=c,scale=.8]{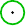}}   
\end{equation}
where the coefficients are given by
\begin{equation}
\begin{array}{ll}
     r_\text{\includegraphics[valign=c]{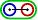}}&= \frac{\alpha_1 -1}{z^{(T_{\rm b})} }+\frac{1}{z^{(T_{\rm g})} -z^{(T_{\rm b})} -z^{(T_{\rm r})} }  \, ,\\
   r_\text{\includegraphics[valign=c]{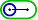}}  &=  \frac{\alpha_2-1}{(z^{(T_{\rm g})} -z^{(T_{\rm b})} )(z^{(T_{\rm g})} -z^{(T_{\rm b})} -z^{(T_{\rm r})} )}\, ,\\
   r_\text{\includegraphics[valign=c]{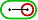}} &= \frac{\alpha_1-1}{z^{(T_{\rm b})} (z^{(T_{\rm g})} -z^{(T_{\rm b})} -z^{(T_{\rm r})} )}\, , \\
r_\text{\includegraphics[valign=c,scale=.8]{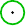}} &= \frac{(\alpha_1+\alpha_2-1)B(\alpha_1,\alpha_2)}{z^{(T_{\rm b})} (z^{(T_{\rm g})} -z^{(T_{\rm b})} )(z^{(T_{\rm g})} -z^{(T_{\rm b})} -z^{(T_{\rm r})} )}\,,
\end{array}
\end{equation}
Using the same methods, similar expressions can be found for $\partial^{(T_{\rm r})}$ and $\partial^{(T_{\rm g})}$, as well as how the derivatives act on other functions in the chain.

\subsection{Interpretation as Pfaffian chains} \label{Pfaffian_system}
The chain of differential equations established in this section actually has a particular form, known as a Pfaffian chain. In this subsection we first give a general discussion of Pfaffian chains, and then apply them in the context of cosmological correlators.

\paragraph{Definition of Pfaffian chains.}
A Pfaffian function is a function which is defined by a triangular system of algebraic differential equations. More precisely, given a domain $U\subseteq \bbR^n$, a Pfaffian chain is a finite sequence of functions $\zeta_1,\ldots,\zeta_r:U\to \bbR$ which satisfies 
\begin{equation} \label{Pfaffian_chain-ODEs}
    \frac{\pd\zeta_i}{\pd x_j} =  P_{ij}(x_1,\ldots,x_n,\zeta_1,\ldots,\zeta_i)  \qquad \text{for all }i,j,
\end{equation}
where each $P_{ij}$ is a polynomial of $n+i$ variables. The triangularity condition, i.e.~the assumption that the derivatives of $\zeta_i$ depends only on $\zeta_1,\ldots,\zeta_i$ and not on $\zeta_{i+1},\ldots,\zeta_{r}$, is essential to ensure that the functions in the chain are sufficiently well-behaved. Given such a chain, a Pfaffian function is a function of the form
\begin{equation}
    f(x_1,\ldots,x_n) = P(x_1,\ldots,x_n,\zeta_1,\ldots,\zeta_r)
\end{equation}
where $P$ is a polynomial in $n+r$ variables.

As an example, consider the function $\zeta(x_1,\ldots,x_n)=x_1^{m_1}\cdots x_n^{m_n}$, which satisfies
\begin{equation}
        \frac{\pd \zeta}{\pd x_j} =  m_j \zeta_j \zeta \,,
\end{equation}
for each $j$, where $\zeta_j$ are the functions $\zeta_j(x_1,\ldots,x_n)=1/x_j$ which satisfy
\begin{equation}\label{eq:expfaff}
        \frac{\pd \zeta_j}{\pd x_k} = -\delta_{jk}\zeta_j^2 .
\end{equation}
In this way, the functions $(\zeta_1,\ldots,\zeta_n,\zeta)$ form a Pfaffian chain.

The relevance of Pfaffian functions is that they have finiteness features which can be precisely quantified using their Pfaffian chain description. In particular, the data of the chain can be used to define a notion of complexity for Pfaffian functions. It consists of four numbers, namely the number of variables, $n$; the length of the chain, $r$ (also called the \textit{order}); the degree of the chain, $\alpha$, defined by $\alpha=\max_{i,j}(\deg(P_{ij}))$; and the degree of the Pfaffian function, $\beta$, defined by $\beta=\deg(P)$. These together define the \textit{Pfaffian complexity} of $f$, schematically denoted by
\begin{equation} \label{PfaffC}
    \cC(f) = (n,r,\alpha,\beta).
\end{equation}

The Pfaffian complexity $\mathcal{C}(f)$ can be viewed as giving a measure of how much information is needed to define the function $f$. An essential feature is that it depends on the description of the function. Since a given function may have several different descriptions, its complexity is not uniquely defined. In particular, this feature can be used to compare the complexity of different descriptions, which will provide us with a way of quantifying the effect of the reduction operators on cosmological correlators.

\paragraph{Interpretation of Pfaffian complexity.} Before discussing the applications to cosmological correlators, it is worthwhile to discuss the interpretation of Pfaffian complexity in slightly more detail. The four numbers comprising the Pfaffian complexity are not arbitrary, but in fact part of a larger mathematical program aiming to assign a meaningful notion of complexity to large classes of functions, called sharp o-minimality \cite{binyamini_sharply_2022,binyamini_tameness_2023}. The meaning comes from generalizing the computational and topological properties of algebraic functions. In the algebraic case, i.e.~when the functions of interest are polynomials, these properties can be captured in terms of the maximum degree of the polynomials, $\cD$, and number of variables $\cF$. Crucially, the computational complexity of algorithms performed on these algebraic functions then admit bounds which are polynomial in $\cD$ and exponential in $\cF$ \cite{binyamini_sharply_2022}. 

The aim of sharp o-minimality is to assign a suitable pair $(\F,\D)$ to more general functions, while keeping similar bounds on computational complexity. For the Pfaffian functions, the right generalization turns out to be given by $\F=n+r$, and $\D = \deg P + \sum_{i,j}\deg P_{ij}$ \cite{binyamini_sharply_2022}. These ideas have previously been applied to various physical settings \cite{grimm_complexity_2024,grimm_structure_2024,grimm_complexity_2024a}, where these concepts are explained in more detail. For our purposes, it suffices that the Pfaffian complexity is a measure of the complexity of a function which can be given a computational meaning. 

\paragraph{Pfaffian chain for cosmological correlators.}
The differential chain structure found earlier in this section bears a striking resemblance to the Pfaffian chains reviewed above, and indeed we will show below that it is possible to write this representation of cosmological correlators in the form of a Pfaffian chain. In an earlier work \cite{grimm_structure_2024} it was already shown that this is possible, relying on the kinematic flow algorithm \cite{arkani-hamed_differential_2023}. However, with having a Pfaffian chain in terms of first-order reduction operators will allow us to explicitly study the reduction of complexity implemented by the higher-order reduction operators in section \ref{sec:complexity}. 

Let us now discuss how to set up a Pfaffian chain for the cosmological correlators, based on equation \eqref{eq:QTiteration}. To begin, we observe that we need to include the functions $\ell^{(T)}_\cT$ in the Pfaffian chain. We refer to these functions as letters. In analogy to equation \eqref{eq:expfaff}, the required Pfaffian differential equations take the simple form
\begin{equation}\label{eq:PCletters}
\pd^{(S)} \ell^{(T)}_\cT  = -M^{S,T}_\cT\left(\ell_\cT^{(T)}\right)^2  \,. 
\end{equation}
The Pfaffian complexity will then depend on the number $n_{\rm L}$ of letters which we need to specify. Recalling the definition of $\ell^{(T)}_\cT$ in terms of the matrices $M^{T,S}_\cT$, we see that the number of letters $N_{\rm L}$ is determined by the number of pairs $(T,\cT)$ with distinct precursors and successors. In general this is a complicated counting problem which depends on the topology of the Feynman graph and the chosen complete tubing. It is bounded by the number of pairs $(T,\cT)$, which grows exponentially in the number of vertices. 

The recursive nature of the differential chain found earlier in this section, expressed in the form of equation \eqref{eq:ITpartialfin} guarantees that it is a Pfaffian chain. In this chain we need a differential equation for every function $I_\cS$ with $T_{\rm max}\in\cS\subseteq\cT$, and there are $2^{|\cT|-1}-1=2^{N_{\rm v}-1}-1$ such functions; this determines the order $r$ of the Pfaffian chain. The degree $\alpha$ depends on the number of iterations of the recursion equation \eqref{eq:QTiteration} that are required. In turn, the number of recursions depends on the depth of the tubing $\cT$, i.e.~the length of the longest ascending chain of tubes in $\cT$. For a complete tubing, the depth is always equal to the number of vertices $N_{\rm v}$. Finally, the degree $\beta$ is equal to $1$, since the function of interest $I_\cT$ is already part of the chain. From these observations we deduce that the Pfaffian complexity of $I_\cT$ is bounded by
\begin{equation}
    \cC(I_\cT) =  (N_{\rm v}, N_{\rm L}+2^{N_{\rm v}-1}-1 , N_{\rm v} ,1) \,.
\end{equation}

The precise growth depends on the topology of the graph and the chosen tubing $\cT$ through the number $N_{\rm L}$. In the next section we will see how this complexity can be reduced by implementing the higher-order reduction operators considered in section \ref{sec:GKZ} and \ref{sec:physics}.

\section{Algebraic relations and the recursive reduction algorithm}
\label{sec:complexity}

In the previous section we have shown that to parameterize any tree-level cosmological correlator one can construct a basis of functions that is closed under partial derivatives by merely using the first-order reduction operators. The next natural step is to consider the role of the higher-order reduction operators. In this section we will study these operators in more detail and argue that they imply algebraic relations between various basis functions.

We begin in section~\ref{ssec:higherrelations} by showing explicitly how the higher-order reduction operators  lead to algebraic relations. Afterwards, we will showcase some examples of such relations in section~\ref{ssec:higherrelsexample}. Then, we will explain in section~\ref{ssec:minimalbasis} how these relations help to obtain a more minimal set of basis functions. Furthermore, we will illustrate the reduction in complexity by showing that the full double-exchange correlator can be expressed in terms of only four such functions. Finally, we will provide the exact counting of these minimal representation functions in section~\ref{ssec:minrepcounting}.

\subsection{Algebraic relations from higher-order operators}\label{ssec:higherrelations}

In this section we will explain how the higher-order reduction operators lead to algebraic relations between different integrals. Concretely, this follows from two observations. Firstly, as we have seen already in section~\ref{sec:physics}, acting with higher-order reduction operators removes tubes, similar to first-order reduction operators. Secondly, we have shown in section~\ref{sec:differentialchain} that acting on an integral $I_\mathcal{T}$ with a differential operator must result in a linear combination of integrals associated to sub-tubings of $\mathcal{T}$ with rational coefficients. Using this, the derivatives of a higher-order reduction operator acting on an integral $I_\mathcal{T}$ can be rewritten in terms of these integrals resulting in a purely algebraic relation between the various basis functions. The exact terms that can appear in these relations will vary, depending on whether the higher-order reduction operator comes from a maximal tube in which case the factorization relations of section~\ref{ssec:factors} become important. Therefore, we will treat the two cases separately, beginning with the non-maximal case.

\paragraph{Algebraic relations from non-maximal tubes.}

Recall that, if a tube $T$ admits a partition $\pi$, it is possible to obtain a higher-order reduction operator $Q^{(T)}_\pi$ using equation~\eqref{eq:highred}. Furthermore, when acting on an integral $I_\mathcal{T}$ with $T$ contained in this tubing, we have seen in section~\ref{ssec:tuberemoval} that the reduction operator will act as
\begin{equation}
    Q^{(T)}_\pi I_\mathcal{T} = \prod_{S\in \pi}\partial^{(S)}  I_{\mathcal{T}\setminus \{T\}}\,.
\end{equation}
Now, using equation~\eqref{eq:ITpartialfin} to iteratively rewrite the derivatives acting on an integral in terms of sub-tubings, it is possible to turn this differential relation in to an algebraic one. Furthermore, this equation can be solved for $I_\mathcal{T}$ resulting in an algebraic relation between $I_\mathcal{T}$ and integrals $I_\mathcal{S}$ for sub-tubings $\mathcal{S}$ of $\mathcal{T}$.

If the tube $T$ is not a maximal tube, this algebraic relation will only involve sub-tubings $\mathcal{S}$ that contain the maximal tube, and therefore are already included in the differential chain constructed in section~\ref{ssec:chainconstruction}. Therefore, we find that not all the functions in the differential chain are algebraically independent and we do not actually need to solve the differential equation for all of these functions. Instead, we can solve the differential equations only for a subset of these functions and obtain the others using the algebraic relations.

In conclusion, we find that if a tubing $\mathcal{T}$ contains \textit{any} non-maximal tube $T$ that admits a partition, there is a relation of the form
\begin{equation}\label{eq:ITalgebraic}     \boxed{\rule[-.5cm]{0cm}{1.3cm} \quad 
    I_\mathcal{T}=\sum_{\substack{\mathcal{S}\subsetneq \mathcal{T},\\ T_{\rm max} \in \mathcal{S}}} \tilde{r}_\mathcal{S}(\cT) I_{\mathcal{S}} \quad}
\end{equation}
where the $\tilde{r}_\mathcal{S}$ are rational functions of $z$ and $\alpha$ and the sum is over all strict sub-tubings of $\mathcal{T}$ that contain $T_{\rm max}$. In other words, the sum is over all sub-tubings of $\mathcal{T}$ that are contained in the differential chain. Note that the functions $\tilde{r}$ can be obtained explicitly using the procedure above. However, we leave a general explicit expression for these coefficients for future work.

\paragraph{Factorization relations.}

We now turn our attention to the case where the maximal tube admits a partition, which will lead to similar algebraic relations to the ones found above. However, as explained in section~\ref{ssec:factors}, removing the maximal tube results in a factorization formula. Thus, acting with the corresponding reduction operators results in
\begin{equation}
    Q^{(T_{\rm max})}_\pi I_\mathcal{T} = \prod_{S\in \pi}\partial^{(S)}  \prod_{\alpha=1}^{k} I_{\mathcal{T}_\alpha}\,,
\end{equation}
where we have labeled the tubings of the different factors as $\mathcal{T}_\alpha$, and denoted the number of such factors by $k$. 

Similar to the approach above, the derivatives on both sides of this equation can be rewritten in terms of functions that belong to a differential chain. However, there is one key difference: the integrals $I_{\cT_\alpha}$ do not include the maximal tube $T_{\rm max}$. As a result, these integrals and their derivatives are not part of the original chain. Instead, they form their own separate differential chains.
This slightly changes the algebraic relations, as now these new functions and their derivatives have to be incorporated as well. Thus we see that the higher-order reduction operators will not allow us to immediately decrease the number of functions we need to solve for. However, one should keep in mind that the diagrams associated to each tubing $I_{\cT_\alpha}$ are much simpler than the original diagram. Therefore, the resulting algebraic relation will still result in an algebraic relation that simplifies $I_\mathcal{T}$. Furthermore, as we will see in section~\ref{ssec:minimalbasis}, these new functions can be written in terms of the same set of minimal representation functions as the ones already part of the chain.

To conclude, given a tubing $\mathcal{T}$ we find that whenever the maximal tube $T_{\rm max}$ admits a partition $\pi$, the integral $I_\mathcal{T}$ will satisfy an algebraic relation of the form
\begin{equation}\label{eq:ITmaxalgebraic}     \boxed{\rule[-.5cm]{0cm}{1.3cm} \quad 
    I_\mathcal{T}=\sum_{\substack{\mathcal{S}\subsetneq \mathcal{T},\\ T_{\rm max} \in \mathcal{S}}} \tilde{r}_\mathcal{S} I_{\mathcal{S}} +\prod_{\alpha=1}^k \bigg(\sum_{\substack{\mathcal{S}\subsetneq \mathcal{T}_\alpha,\\ T_{\rm max,\alpha} \in \mathcal{S}}} \tilde{r}_\mathcal{S} I_{\mathcal{S}}\bigg) \quad}
\end{equation}
where again, $\mathcal{T}_\alpha$ are the different factors appearing after removing $T_{\rm max}$, the first sum is over all sub-tubings of $\mathcal{T}$ that contain $T_{\rm max}$ while the second sum is over all sub-tubings of the factor $\mathcal{T}_\alpha$ containing its respective maximal tube.

\subsection{Some algebraic relations for the single- and double-exchange integrals} \label{ssec:higherrelsexample}

To make the above more explicit, we will now showcase how these algebraic relations can be obtained in two examples, in particular one for a maximal tube and one for a non-maximal tube. We will first derive a factorization relation for the single-exchange integral. Afterwards, we will derive an algebraic relation for functions in the differential chain of the double-exchange integral. We have chosen somewhat simple examples here in order to keep the formulas from becoming too involved. However, the same procedures generalize to any tree-level cosmological correlator.

\paragraph{Factorization relation for the single-exchange integral.}

We begin by considering the single-exchange integral again, since here the formulas will be the most simple. As we have seen in equation~\eqref{eq:singexchangeQhigh}, the maximal tube in 
\begin{equation}
\includegraphics[valign=c]{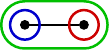}
\end{equation}
admits a partition by the blue and red tubes. Therefore, there is a reduction operator \begin{equation}
    Q^{(T_{\rm g})}_\pi=\partial^{(T_{\rm r})} \partial^{(T_{\rm b})} \big(z^{(T_{\rm g})}  -z^{(T_{\rm r})} -z^{(T_{\rm b})}  \big)\,,
\end{equation}
where have denoted the partition by $\pi$, recall that $T_{\rm r}$, $T_{\rm b}$ and $T_{\rm g}$ are the red, blue and green tubes respectively and note that the derivatives act on everything to their right. As discussed in section~\ref{ssec:factors}, this reduction operator will lead to a differential relation of the form
\begin{equation}\label{eq:singexchangehighfact}
    \partial^{(T_{\rm r})} \partial^{(T_{\rm b})} \big(z^{(T_{\rm g})}  -z^{(T_{\rm r})} -z^{(T_{\rm b})}  \big)
    I_\text{\includegraphics[valign=c]{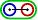}}=\big(\partial^{(T_{\rm b)}}   I_\text{\includegraphics[valign=c,scale=.8]{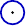}} \big) \big(\partial^{(T_{\rm r})}I_\text{\includegraphics[valign=c,scale=.8]{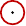}}\big)\,.\end{equation}
in accordance with equation~\eqref{apply_red_op}.

\paragraph{Algebraic relation for the single-exchange integral.}

Now, we will use the differential chain constructed in section~\ref{ssec:chainexamples} to rewrite this differential relation into an algebraic one. We will begin by rewriting the right-hand side. 

In section~\ref{ssec:chainconstruction} we have seen that, if a tubing only consists of a single tube, the only non-zero differential equation it satisfies can be obtained from equation~\eqref{eq:ITmaxdif}. This implies that
\begin{equation}
    \begin{array}{rl}
    \partial^{(T_{\rm b)}}   I_\text{\includegraphics[valign=c,scale=.8]{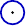}}&  \displaystyle  =\frac{\alpha_1-1}{z^{(T_{\rm b})}} I_\text{\includegraphics[valign=c,scale=.8]{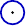}}\, , \\
     \partial^{(T_{\rm r})}I_\text{\includegraphics[valign=c,scale=.8]{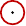}}&  \displaystyle   =\frac{\alpha_2-1}{z^{(T_{\rm r})}} I_\text{\includegraphics[valign=c,scale=.8]{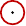}} \, ,
    \end{array}
\end{equation}
where we recall that the vertices are ordered such that the blue tube encircles the first vertex while the red tube encircles the second. Inserting these identities into equation~\eqref{eq:singexchangehighfact} results in
\begin{equation}
     \partial^{(T_{\rm r})} \partial^{(T_{\rm b})} \big(z^{(T_{\rm g})}  -z^{(T_{\rm r})} -z^{(T_{\rm b})}  \big)
    I_\text{\includegraphics[valign=c]{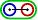}}=\frac{(\alpha_1-1)(\alpha_2-1)}{z^{(T_{\rm b})}z^{(T_{\rm r})}}I_\text{\includegraphics[valign=c,scale=.8]{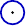}}I_\text{\includegraphics[valign=c,scale=.8]{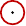}}\ ,
\end{equation}
and we see already that one side of the equation is now purely algebraic.

Using a similar reasoning, we apply the strategy of section~\ref{ssec:chainexamples} to rewrite the derivatives acting on $I_\text{\includegraphics[valign=c]{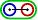}}$ 
in terms of functions in the differential chain. This process, while somewhat tedious, is straightforward and results in
\begin{equation}
\begin{split}
     \partial^{(T_{\rm r})} \partial^{(T_{\rm b})} \big(z^{(T_{\rm g})}  -z^{(T_{\rm r})} -z^{(T_{\rm b})}  \big)
    I_\text{\includegraphics[valign=c]{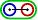}}=&\frac{(\alpha_1-1)(\alpha_2-1)\big(z^{(T_{\rm g})}-z^{(T_{\rm b})}-z^{(T_{\rm r})}\big) I_\text{\includegraphics[valign=c]{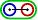}}}{z^{(T_{\rm b})} z^{(T_{\rm r})}}\\
        & + \frac{(\alpha_1-1)(\alpha_2-1)\left(I_\text{\includegraphics[valign=c]{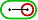}}+ I_\text{\includegraphics[valign=c]{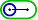}}\right)}{z^{(T_{\rm b})} z^{(T_{\rm r})}}\,.
\end{split}
\end{equation}
Inserting this equation into~\eqref{eq:singexchangehighfact} and solving for $ I_\text{\includegraphics[valign=c]{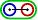}}$, we obtain
\begin{equation}\label{simplified-full-I}
     I_\text{\includegraphics[valign=c]{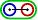}}=\frac{ I_\text{\includegraphics[valign=c,scale=.8]{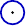}}   I_\text{\includegraphics[valign=c,scale=.8]{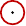}}
-  I_\text{\includegraphics[valign=c]{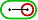}}-I_\text{\includegraphics[valign=c]{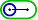}}}{z^{(T_{\rm g})}-z^{(T_{\rm b})}-z^{(T_{\rm r})}}
  \,,
\end{equation}
which is the algebraic relation within the single-exchange chain due to the higher-order reduction operator $Q^{(T_{\rm g})}_\pi$.

Interestingly, equation~\eqref{simplified-full-I} implies a concrete simplification for 
$I_\text{\includegraphics[valign=c]{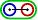}}$
at the functional level. The differential chain~\eqref{eq:singexchdifchain} for the single exchange integral results in a coupled system of second-order differential equations satisfied by 
$I_\text{\includegraphics[valign=c]{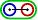}}$. 
In general, one would expect that the solution would be some two-variable generalized hypergeometric function such as an Appell function. However, from the explicit form of the single exchange integral obtained in~\cite{arkani-hamed_differential_2023,grimm_reductions_2024} it follows that it can be written as a sum of single-variable hypergeometric functions, as well as polynomials raised to complex powers. The functions 
$I_\text{\includegraphics[valign=c]{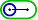}} $ and $I_\text{\includegraphics[valign=c]{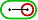}} $ 
take the form of such single-variable hypergeometric functions while 
$I_\text{\includegraphics[valign=c,scale=.8]{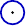}}   I_\text{\includegraphics[valign=c,scale=.8]{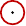}}$ 
can be written in terms of polynomials raised to complex powers. Therefore, equation~\eqref{eq:singexchangehighfact} encodes exactly this simplification. We will see in section~\ref{ssec:minimalbasis} that, for general diagrams, many such functional simplifications will happen. In section~\ref{ssec:recursivereduction}, we will explain how to obtain the minimal set of such functions necessary.

\paragraph{A relation for the double-exchange integral.}

As a second example, let us briefly examine the type of algebraic relations that appear when considering reduction operators that are not associated to a maximal tube. In this case, it is necessary to introduce an example that is slightly more involved than the single exchange integral, namely the double-exchange integral. In particular, we will consider the double-exchange diagram with the tubing
\begin{equation}\label{diag:doubexchforred}
\includegraphics[valign=c]{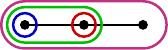}
\end{equation}
and, since the green tube admits a partition, obtain an algebraic relation for $I_\text{\includegraphics[valign=c]{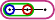}}$.

In order to obtain this relation, let us first provide the functions in the differential chain needed to construct $I_\text{\includegraphics[valign=c]{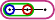}}$. These can be obtained simply by considering the sub-tubings of~\eqref{diag:doubexchforred}, resulting in the differential chain
\begin{equation}
\includegraphics[valign=c]{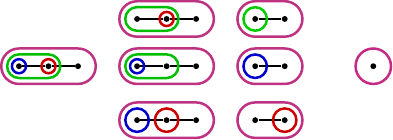}
\end{equation}
where we have not drawn the arrows relating different diagrams in order to avoid clutter and again have contracted any edges using the arguments of section~\ref{ssec:contractions}. We emphasize that the integrals being part of this differential chain implies that a partial derivative acting on any of the integrals
\begin{equation}
\begin{array}{llll}
     I_\text{\includegraphics[valign=c]{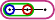}}\, ,& I_\text{\includegraphics[valign=c]{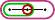}} \,, &
  I_\text{\includegraphics[valign=c]{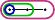}}\,,&
 I_\text{\includegraphics[valign=c]{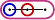}}\, ,\\
I_\text{\includegraphics[valign=c]{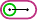}}\,,&
I_\text{\includegraphics[valign=c]{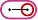}}\,,&
I_\text{\includegraphics[valign=c]{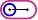}}\,,&
I_\text{\includegraphics[valign=c,scale=.8]{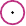}}\, ,     
\end{array}
\end{equation}
can be expressed as a linear combination of the others with rational coefficients.

Since, in the left-most diagram, the green tube admits a partition using the blue and red tubes, there will be a differential relation
\begin{equation}\label{eq:doubleexchangereductionop}
    Q^{(T_{\rm g})}_\pi I_\text{\includegraphics[valign=c]{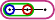}}= \partial^{(T_{\rm r})}  \partial^{(T_b)}  I_\text{\includegraphics[valign=c]{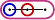}} \,,
\end{equation}
where $Q^{(T_{\rm g})}_\pi$  takes the form\footnote{Note that the expression for $Q_\pi^{(T_{\rm g})}$ is the same here as for the single-exchange integral, for which $T_{\rm g}$ admits the same partition.}
\begin{equation}
   Q^{(T_{\rm g})}_\pi=  \partial^{(T_{\rm r})} \partial^{(T_{\rm b})} \big(z^{(T_{\rm g})}  -z^{(T_{\rm r})} -z^{(T_{\rm b})}  \big)\,.
\end{equation}
Now, again using the fact that derivatives acting on any of these functions can be expressed as other functions in the differential chain, this will lead to an algebraic relation. The process of constructing the differential chain, as well as solving for the partial derivatives is computationally more involved in this case. However, there is no fundamental difficulty and one can proceed along the same lines as above. 

Interestingly, the resulting algebraic relation is remarkably similar to the equation~\eqref{simplified-full-I}, being of the form
\begin{equation}
    I_\text{\includegraphics[valign=c]{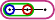}}= \frac{ I_\text{\includegraphics[valign=c]{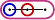}}-I_\text{\includegraphics[valign=c]{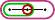}} - 
  I_\text{\includegraphics[valign=c]{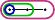}}} {z^{(T_{\rm g})}-z^{(T_{\rm b})}-z^{(T_{\rm r})}}\,.
\end{equation}
Note that in this case, all the functions on the right-hand side of this equation are already part of the differential chain. Therefore, this algebraic relation actually reduces the number of functions that one has to determine for the double-exchange integral.

\subsection{Minimal representation functions}\label{ssec:minimalbasis}

Having discussed the algebraic relations, we observe that 
they often lead us to consider integrals of the factorized diagrams.  
Staying within the differential chain, this would not bring a simplification since the number of functions one has to determine has not decreased. 
This leads us to consider a change of perspective: instead of simply counting how many functions appear in a certain differential chain, we consider the types of functions that can appear. In effect, this implies that we must consider functions equivalent when they merely differ by permuting or shifting inputs. Computationally, one only needs to obtain each such function once, since the permutations and shifts are simple operations. If we implement all such simplifications, we find a minimal set of functions necessary to describe tree-level cosmological correlators, which we call the minimal representation functions.

\paragraph{Permutations.}

During the construction of the differential chains in the examples above, we have already seen many functions appear multiple times with differently permuted inputs. For example, for the single-exchange integral the functions $I_\text{\includegraphics[valign=c]{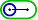}}$ and $I_\text{\includegraphics[valign=c]{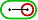}}$ where 
permuting the inputs $z^{(T_{\rm b})}$ and $z^{(T_{\rm r})}$ as well as the twists $\alpha_1$, and $\alpha_2$ results in an equivalence
\begin{equation}
    I_\text{\includegraphics[valign=c]{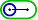}} \vert_{z^{(T_{\rm b})} ,\alpha_1 \rightarrow z^{(T_{\rm r})} ,\alpha_2} = I_\text{\includegraphics[valign=c]{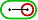}}
\end{equation}
as one can immediately see from the diagrams themselves. However, this procedure similarly works for more complicated diagrams. For example, one can obtain an equivalence of the tubings
\begin{equation}
\includegraphics[valign=c]{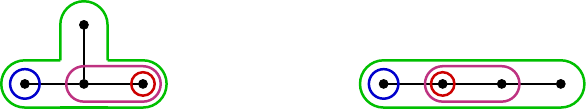}
\end{equation}
now involving four edges, simply by permuting the inputs $z^{(T)} $ and $\alpha_v$. This is a consequence of the fact that, as described in section~\ref{sec:tubings}, the GKZ system is agnostic of any topological properties of the diagram. Instead, the only information that enters the GKZ system is combinatorial, consisting strictly of the vertices contained in each tube.

\paragraph{Permutations for factorizations.}

Let us note that symmetries are also prevalent in the factorization relations for the single-exchange integral. In this case three integrals $I_\text{\includegraphics[valign=c,scale=.8]{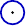}}$, $I_\text{\includegraphics[valign=c,scale=.8]{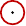}}$, and $I_\text{\includegraphics[valign=c,scale=.8]{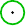}}$  appear, which merely differ by permuting the variables and the twists, with possibly some additional shifts. In fact, this behavior is rather general. Let us consider an integral $I_\cT$ admitting a factorization of the form
\begin{equation}
    Q^{(T_{\rm max})}_\pi I_\cT = \prod_{S\in \pi} \partial^{(S)} \prod_{\alpha=1}^k I_{\cT_\alpha}\,
\end{equation}
where there are $k$ factors $\cT_{\alpha}$. Then there are two options, either a non-maximal tube in $\cT$ also admits a partition, implying that there is an algebraic relation relating $I_\cT$ in terms of functions part of the differential chain, or the factors $I_{\cT_\alpha}$ are permutations of functions already part of the differential chain. This implies that any integral of a tubing with a factorization relation can be fully written in terms of permutations of its sub-tubings.

To see this, we will assume that $I_{\cT}$ does not contain a non-maximal tube that admits a partition. Then, let us choose any of the factors $\cT_{\alpha}$. We will show that there is a tubing $\cS_\alpha$ such that $I_{\cT_\alpha}$ is a permutation of $I_{\cS_\alpha}$ and $\cS_\alpha$ is a sub-tubing of $\cT$. We begin by removing all non-maximal tubes from $\cT$ that are not contained in $\cT_\alpha$, note that removing non-maximal tubes will result in a function that is in the differential chain. Furthermore, we will consider the maximal tube of $\cT_{\alpha}$ and also remove it, we will denote the resulting tubing by $\cS_\alpha$. Note that the maximal tube of $\cT_{\alpha}$ is not the maximal tube of $\cT$, therefore $I_{\cS_\alpha}$ will be contained in the differential chain of $\cT$. Since $I_{\cT_\alpha}$ does not admit a partition, its maximal tube must contain at least one bare edge. Therefore, we can use the contraction identities from section~\ref{ssec:contractions} to contract all edges in $\cS_\alpha$ that are not fully contained in $\cT_\alpha$. The resulting tubing will be a permutation of $\cT_\alpha$ in which only the maximal tubes are permuted.

Let us illustrate the above with an example. We will consider the double-exchange integral with the tubing
\begin{equation}\label{eq:minimalfactor}
\includegraphics[valign=c]{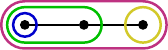}
\end{equation}
The magenta tube admits a partition by the green and yellow tubes, and the resulting algebraic relation will involve the factors
\begin{equation}\label{eq:minimalfactors}
\includegraphics[valign=c]{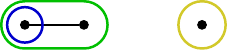}
\end{equation}
which naively should be added to the differential chain separately. However, removing the yellow and green tube from~\eqref{eq:minimalfactor} results in
\begin{equation}
\includegraphics[valign=c]{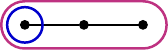}
\end{equation}
which can be contracted to
\begin{equation}
\includegraphics[valign=c]{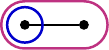}
\end{equation}
that is equivalent to the left factor in~\eqref{eq:minimalfactors} by a permutation of their maximal tubes. Similarly, removing all tubes except for the maximal tube in~\eqref{eq:minimalfactor} would result in the right factor of~\eqref{eq:minimalfactors}.

\paragraph{Minimal representation functions.}

The permutation symmetry above, as well as the algebraic relations found throughout this section, lead us to a natural question: what is the set of function that remains after all redundancy has been removed? The resulting functions, which we dub the minimal representation functions, will have as their defining property that these are the minimal set of functions that must be solved using their differential equations, as there can be no further algebraic or permutative identities. In other words, these functions are the building blocks that all other functions in the differential chain can be constructed from, using the algebraic and permutative relations. 

Interestingly, the minimal representation functions are shared for all tree-level cosmological correlators, independent of any particular tubing or topology in a diagram. As described in section~\ref{sec:tubings}, this is rooted in the fact that the GKZ system is agnostic to this information. To signify that we only care about the functions themselves and are agnostic to the particular tubing or diagram that they arise from, we will denote the minimal representation functions by removing the color from their tubings, as in $   I_\text{\includegraphics[valign=c]{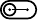}}$ and $I_\text{\includegraphics[valign=c,scale=.8]{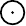}}$.
Note that, in order  to solve the differential equations satisfied by a minimal representation function, it may be necessary to color in these tubes again.

The minimal representation functions also give an intuitive handle on the complexity of the functions that can appear. For example, for the single-exchange integral one shows, see e.g.~\cite{Grimm:2024tbg}, that it consists only of polynomials to complex powers and $_2F_1$ hypergeometric functions. We can motivate the expected complexity by the minimal chain these functions can be contained in. For example, the function $   I_\text{\includegraphics[valign=c]{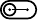}}$ can be minimally contained in the chain
\begin{equation}
  \includegraphics[valign=c]{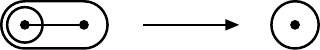}
\end{equation}
Here, the length of the chain will describe the order of the full differential equations, while the different arrows at each layer are related to the number of variables each function depends on. Note that each function also implicitly satisfies the differential equation due to the maximal tube. However, this differential equation will only fix an overall scaling of the variables, the arrows here then denote how many remaining variables the function depends on. From this, we find that the chain above corresponds to a second order differential equation in one variable, giving rise to a $_2F_1$ hypergeometric function.

\paragraph{Tubings for minimal representation functions.}

Even though the minimal representation functions no longer correspond to any particular tubing, they can still be represented by a tubing. Now, it turns out that these tubings must have a few specific properties. In particular, note that, if any tube contains no bare vertices, the tubing admits a partition and can therefore be removed. Conversely, if a tube contains multiple bare vertices, these can be contracted using the methods of section~\ref{ssec:contractions}. This implies that, for a minimal representation function, each tube in the corresponding tubing must contain \textit{exactly} one bare vertex. This property will help us greatly in section~\ref{ssec:minrepcounting}, where we will provide counting formulas for the minimal representation functions. 

Furthermore, we find that these conditions imply that a diagram must contain exactly the same number of vertices as the number of tubes. From this, we find that the minimal representation functions are naturally ordered by the number of vertices. Moreover, since acting with reduction operators removes tubes, we find that a derivative acting on a minimal representation function with $n$ vertices can be expressed in terms of the function itself, as well as minimal representation functions with $n-1$ vertices. Therefore, we find that the differential chain of a minimal representation function with $n$ vertices consists of itself, alongside minimal representation functions that have strictly fewer vertices.

\subsection{The recursive reduction algorithm}\label{ssec:recursivereduction}

In this section, we will summarize the results obtained throughout this paper in to a single algorithm, the recursive reduction algorithm. We outline the key steps required for performing the reductions, referring to earlier sections for explicit formulas. We then illustrate the reduction process schematically for the double-exchange integral, demonstrating that it can be expressed in terms of just four minimal representation functions.

\paragraph{The recursive reduction algorithm.}

The recursive reduction algorithm is based on the idea that it is beneficial 
to decompose the cosmological correlators into the simplest set of building block functions. While this introduces combinatorial complexity, solving the differential equations for the building block functions will be significantly simpler. 

The algorithm proceeds in the following steps: 

\noindent
\textbf{Step 1:} Considering a cosmological correlator with a fixed number of external momenta, one first has to write down all tree-level diagrams that contribute. Each diagram is initially studied separately. Focusing on a diagram one needs to find all complete tubings $\cT$. The goal is then to construct the minimal representation functions for the sum $\sum_{\cT} I_{\cT}$. 

\noindent
\textbf{Step 2:} Next, one considers a specific tubing $\cT$. To obtain $I_{\cT}$ one constructs the differential chain in which $I_{\cT}$ resides. 
This requires finding all the sub-tubings of $\cT$ and then using the reduction operators as in section~\ref{ssec:chainconstruction}. Note that in this step, we are not required to obtain the different functions in the chain explicitly, only the differential equations they satisfy, which follow by recursively applying equations~\eqref{eq:partialTinQs1} and~\eqref{eq:QTiteration}.

\noindent
\textbf{Step 3:} Once the differential relations have been obtained, one uses the algebraic relations of section~\ref{ssec:higherrelations} to eliminate the integrals associated to any tubing admitting a partition. In addition, one also contracts any edges using the methods of section~\ref{ssec:contractions}, and identifies the remaining functions up to permutations in the variables, as described in section~\ref{ssec:minimalbasis}. The resulting set of functions will be the minimal representation functions.

\noindent
\textbf{Step 4:} It remains to find the minimal representation functions by solving the differential equations that they satisfy. 
This is computationally the most difficult step, as solving such coupled systems of differential equations is a hard problem. 
Note that, as described in section~\ref{ssec:minimalbasis}, acting with a reduction operator on a minimal representation function removes edges. These first-order relations suggest that the minimal representation functions could admit an iterated integral representation.

\noindent
\textbf{Step 5:}
In the final step, we invert all of the algebraic and permutation relations used in step 3, in order to express the integral $I_{\cT}$ in terms of the minimal representation functions. This step will consist of keeping track of a large number of identities between different functions, and will therefore be computationally tedious. However, no fundamental difficulties remain in this step.

To illustrate the algorithm above, we will now partly apply it to the double-exchange integral. Note that we will be somewhat schematic, as keeping track of and displaying such large numbers of identities is quite tedious and not very insightful. Instead, we will mostly focus on step 2 and step 3 to obtain the minimal representation functions. This illustrates the large number of symmetries and identities that can be found in this example.

\paragraph{Minimal representation for the double-exchange integral.}

To obtain the minimal representation of the double-exchange integral, we must first construct its differential chain. We will begin with the tubing
\begin{equation}\label{eq:doubleexchangecomptub}
\includegraphics[valign=c]{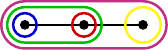}
\end{equation}
and note that the other tubing for the double-exchange integral can be obtained from this one by symmetry. Then, to construct the differential chain, we must find all sub-tubings of this tubing. There are 16 such tubings, given by
\begin{equation}
\includegraphics[valign=c]{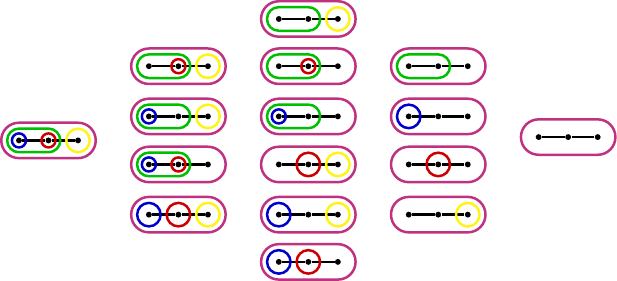}
\end{equation}
where again we have not drawn the arrows relating different diagrams in order to avoid clutter.

Now, we must eliminate all functions which can be algebraically removed using the higher-order reduction operators. This means that any tubing which contains a tube that can be partitioned must be removed. Note, if the reduction operator results in a factorization relation, one should in principle keep track of both of the factors. However, as we have seen in section~\ref{ssec:minimalbasis}, these will lead to the same minimal representation functions. Thus, we will ignore them here. Removing all of these redundant tubings, we are left with the following set:
\begin{equation}
\includegraphics[valign=c]{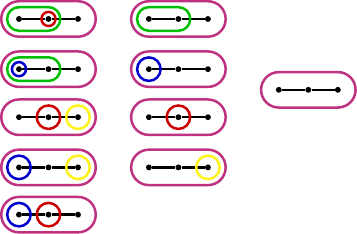}
\end{equation}
Note that the remaining tubes are localized in the right-most columns. This is general behavior as, when a tubing contains more tubes than vertices, it must contain a tube admitting a higher-order reduction operator.

Then, before we identify the different functions up to symmetry, we must contract all possible edges using the techniques of section~\ref{ssec:contractions}. From this, we find that each tubing in the $n$-th column of our differential chain can be contracted to include only diagrams with $n$ edges. In particular, we find
\begin{equation}
\includegraphics[valign=c]{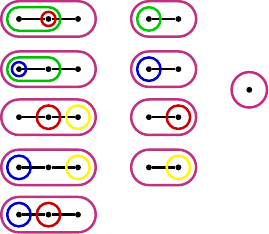}
\end{equation}
Note that this does not decrease the number of tubings, but will greatly increase the number of functions that can be identified up to permutations. Performing this identification is the final step of the algorithm, after which we are left with the four tubings
\begin{equation}
\includegraphics[valign=c]{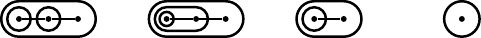}
\end{equation}
where we have again removed the colors of the tubings to represent that, in the actual correlator these will appear with differently permuted variables.

We would like to emphasize how drastic the decrease in necessary functions is after these reductions. Initially, we found that the double-exchange integral can be obtained using a differential chain containing sixteen different tubings, and thus required solving differential equations for sixteen different functions. Furthermore, analyzing the differential chain one would expect a solution of these equations to be some four-variable generalized hypergeometric function. However, applying all the possible simplifications and reductions, we are left with only four different minimal representation functions, using which it must be possible to express the original correlator. Additionally, two of those functions can already be obtained when solving for the single-exchange integral, while the other two are new two-variable generalized hypergeometric functions. These functions are substantially less complex than a generically expected solution of the original system, a direct consequence of the great number of relations present.

\subsection{Counting complexity}\label{ssec:minrepcounting}
In this section we have shown that the implementation of the higher-order reduction operators allows one to effectively remove a significant number of functions from the computation of a cosmological correlator. In the remainder of the section we will make this reduction precise by counting the number of minimal representation functions $N_{\rm m}(N_{\rm v})$. Afterwards, we discuss the minimal representation functions from the perspective of Pfaffian chains, and discuss some challenges related to the Pfaffian complexity.

\paragraph{Counting minimal representation functions.}
Recall that, when performing the recursive reduction algorithm, the remaining tubings are characterized by having exactly one bare vertex, no tubes which admit a partition, and all tubings related by permutations removed. The following table displays the graphs needed for $N_{\rm v}=1,2,3,4$.

\begingroup 
\def\arraystretch{1.5}
\begin{table}[h]
    \centering
    \begin{tabular}{c|l}
    $N_{\rm v}$ & Minimal representation functions\\
    \hline
        $1$ & $
\begin{tikzpicture}[baseline=-0.5ex]
    \begin{feynman}
        \vertex [circle,fill,inner sep=1pt](v51) {};
        \diagram*  {
};
    	\vertex [above=0 cm of v51,shape=circle,draw=black,fill=none,very thick,minimum size=.6 cm] (bv51) {};
    \end{feynman}
\end{tikzpicture}$ \\
         $2$ & $
\begin{tikzpicture}[baseline=-0.5ex]
    \begin{feynman}
        \vertex [circle,fill,inner sep=1pt](v411) {};
        \vertex [right=.5cm of v411,circle,fill,inner sep=1pt] (v412) {};            
        \diagram*  {
	     (v411) --[very thick] (v412);
          (v411) --[connect=3mm,black,very thick] (v412);        
};
        \vertex [above=0 cm of v411,shape=circle,draw=black,fill=none,very thick,scale=1] (bv411) {};
    \end{feynman}
\end{tikzpicture}$ \\
         $3$ & $
\begin{tikzpicture}[baseline=-0.5ex]
    \begin{feynman}
        \vertex [circle,fill,inner sep=1pt](v211) {};
        \vertex [right=.5cm of v211,circle,fill,inner sep=1pt] (v212) {};    \vertex [right=.5cm of v212,circle,fill,inner sep=1pt] (v213) {}; 
        \vertex[right=1.5 cm of v213] (s2); 
        
        \vertex [above=0cm of s2,circle,fill,inner sep=1pt](v311) {};
        \vertex [right=.5cm of v311,circle,fill,inner sep=1pt] (v312) {};    \vertex [right=.5cm of v312,circle,fill,inner sep=1pt] (v313) {};         
        \diagram*  {
          (v211) --[very thick] (v212) --[very thick] (v213);
          (v211) --[connect=3mm,black,very thick] (v213);     
        
          (v311) --[very thick] (v312) --[very thick] (v313);
          (v311) --[connect=2mm,black,very thick] (v312);
          (v311) --[connect=3mm,black,very thick] (v313);           
};
        \vertex [above=0 cm of v211,shape=circle,draw=black,fill=none,very thick,scale=1] (bv211) {};
        
        \vertex [above=0 cm of v212,shape=circle,draw=black,fill=none,very thick,scale=1] (bv212) {};
        
        \vertex [above=0 cm of v311,shape=circle,draw=black,fill=none,very thick,scale=.6] (bv311) {};
    \end{feynman}
\end{tikzpicture}$\\
         $4$ & $
\begin{tikzpicture}[baseline=-0.5ex]
    \begin{feynman}
        \vertex [circle,fill,inner sep=1pt](v11) {};
        \vertex [right=.5cm of v11,circle,fill,inner sep=1pt](v12) {};
        \vertex [right=.5cm of v12,circle,fill,inner sep=1pt](v13) {};
        \vertex [right=.5cm of v13,circle,fill,inner sep=1pt](v14) {};
        
        \vertex [right=1cm of v14, circle,fill,inner sep=1pt](v21) {};
        \vertex [right=.5cm of v21,circle,fill,inner sep=1pt](v22) {};
        \vertex [right=.5cm of v22,circle,fill,inner sep=1pt](v23) {};
        \vertex [right=.5cm of v23,circle,fill,inner sep=1pt](v24) {};  

        \vertex [right=1cm of v24, circle,fill,inner sep=1pt](v31) {};
        \vertex [right=.5cm of v31,circle,fill,inner sep=1pt](v32) {};
        \vertex [right=.5cm of v32,circle,fill,inner sep=1pt](v33) {};
        \vertex [right=.5cm of v33,circle,fill,inner sep=1pt](v34) {};     

        \vertex [right=1cm of v34, circle,fill,inner sep=1pt](v41) {};
        \vertex [right=.5cm of v41,circle,fill,inner sep=1pt](v42) {};
        \vertex [right=.5cm of v42,circle,fill,inner sep=1pt](v43) {};
        \vertex [right=.5cm of v43,circle,fill,inner sep=1pt](v44) {};     
        
        \diagram*  {
          (v11) --[very thick] (v12) --[very thick] (v13)--[very thick] (v14);
          (v11) --[connect=3mm,black,very thick] (v14);    

          (v21) --[very thick] (v22) --[very thick] (v23)--[very thick] (v24);
          (v21) --[connect=3mm,black,very thick] (v24);  
          (v22) --[connect= 2mm,black,very thick] (v23);

          (v31) --[very thick] (v32) --[very thick] (v33)--[very thick] (v34);
          (v31) --[connect=3mm,black,very thick] (v34);
          (v31) --[connect=2mm,black,very thick] (v33);  

          (v41) --[very thick] (v42) --[very thick] (v43)--[very thick] (v44);
          (v41) --[connect=3mm,black,very thick] (v44);
          (v41) --[connect=2.3mm,black,very thick] (v43);
          (v41) --[connect=1.6mm,black,very thick] (v42);
          
};
        \vertex [above=0 cm of v11,shape=circle,draw=black,fill=none,very thick,scale=1] (bv11) {};
       \vertex [above=0 cm of v12,shape=circle,draw=black,fill=none,very thick,scale=1] (bv12) {};
        \vertex [above=0 cm of v13,shape=circle,draw=black,fill=none,very thick,scale=1] (bv13) {};

        \vertex [above=0 cm of v21,shape=circle,draw=black,fill=none,very thick,scale=1] (bv21) {};
       \vertex [above=0 cm of v22,shape=circle,draw=black,fill=none,very thick,scale=.6] (bv22) {};

        \vertex [above=0 cm of v31,shape=circle,draw=black,fill=none,very thick,scale=.6        ] (bv31) {};
       \vertex [above=0 cm of v32,shape=circle,draw=black,fill=none,very thick,scale=.6] (bv32) {};

        \vertex [above=0 cm of v41,shape=circle,draw=black,fill=none,very thick,scale=.45        ] (bv41) {};       
    \end{feynman}
\end{tikzpicture}$
    \end{tabular}
    \caption{The minimal representation functions for all diagrams up to four vertices.}
 \label{tab:minrep}
\end{table}
\endgroup

To obtain an expression for $N_{\rm m}$, we derive a recursion relation as follows. We start with a chain of $N_{\rm v}$ vertices, and encircle all vertices by the maximal tube. 
Since this maximal tube must have exactly one bare vertex, which by permutation symmetry can be taken to be the right-most vertex, the remaining $N_{\rm v}-1$ vertices must be encircled by adding more tubes. In order to count in how many ways this can be done, we note that the counting receives contributions from all possible ways of partitioning the $(N_{\rm v}-1)$-chain into smaller chains. For these smaller chains, the same counting problem holds. This observation implies the following recursion relation:
\begin{equation}
    N_{\rm m}(N_{\rm v}) = \sum_{\pi\in P(N_{\rm v}-1)} \prod_{j\in \pi} N_{\rm m}(j) \,.
\end{equation}
Here $P(N_{\rm v}-1)$ denotes the set of integer partitions of $N_{\rm v}-1$. Note that this formula denotes the number of minimal representation functions with \textit{exactly} $N_{\rm v}$ vertices, and therefore this counting does not include the functions with fewer edges. To incorporate these one would simply take the sum of $N_{\rm m}(n)$ from $n=1$ to $N_{\rm v}$.

To clarify the meaning of this formula, consider for example $n=5$. Then sum then runs over all integer partitions of $4$, which are given by 
\begin{equation}
    \{4\},\,\{3,1\},\,\{2,2\},\,\{2,1,1\},\,\{1,1,1,1\} \,.
\end{equation}
The number of minimal representation functions of the 5-chain is then given by
\begin{align*}
    N_{\rm m}(5)  =& N_{\rm m}(4) + N_{\rm m}(3)N_{\rm m}(1) + N_{\rm m}(2)N_{\rm m}(2) \\
    & + N_{\rm m}(2)N_{\rm m}(1)N_{\rm m}(1) + N_{\rm m}(1)N_{\rm m}(1)N_{\rm m}(1)N_{\rm m}(1)  \\
    =& \,9.
\end{align*}
The sequence $N_{\rm m}(N_{\rm v})$ coincides with the one documented in \cite{OEIS}, and no closed-form expression is known.

For comparison, let us consider the number of functions $N_{\rm f}(N_{\rm v})$ needed to express a cosmological correlator $\psi$ in terms of the differential chain from section~\ref{sec:differentialchain}, i.e.~without the implementation of the higher-order reduction operators. Recall from equation~\eqref{eq:psiG} that, for a given graph, $\psi$ is given by a sum of the form
\begin{equation}
    \psi = \sum_{\cT\, \text{complete} } I_\cT\,,
\end{equation}
where this sum is over all complete tubings of the graph.
For each term $I_\cT$, we have to solve the differential chain from the previous section. However, many of the functions in the various chains will overlap, since a tubing $\cS$ can be a sub-tubing of several distinct complete tubings $\cT$. Effectively, this means that we need to solve for  $I_\cS$ for \textit{every} tubing $\cS$ containing the maximal tube. In other words, the number $N_{\rm f}(N_{\rm v})$ is given by the number of such tubings. 

This counting depends on the topology of the graph, so for concreteness let us consider a chain of $N_{\rm v}$ vertices. In this case, the counting problem is equivalent to the number of ways in which a list of $N_{\rm v}$ items can be grouped into nested sublists, which is discussed in \cite{OEIS2}. The first few values of this sequence are compared to the number of minimal representation functions in table \ref{tab:growths}.

\begin{table}[h]
    \centering
    \begin{tabular}{c|c|c|c|c|c|c|c|c|c}
        $N_{\rm v}$ & 1 & 2 & 3 & 4 & 5 & 6 & 7 & 8 & 9   \\ \hline
        $N_{\rm m} $ & 1 & 1 & 2 & 4 & 9 & 20 & 49 & 117 & 297   \\ \hline
        $N_{\rm f}$ & 1 & 4 & 24 & 176 & 1440 & 12,608 & 115,584 & 1,095,424 & 10,646,016    \\ \hline
        $N_{\rm k}$ & 1 & 4 & 16 &  64 &256 & 1024 & 4096 & 16,384 & 65,536    \\
    \end{tabular}
    \caption{Comparison of the number of new functions needed to compute the contribution to a cosmological correlator from a graph of $N_{\rm v}$ vertices, in the minimal representation ($N_{\rm m}$), the differential chain ($N_{\rm f}$), and the kinematic flow ($N_{\rm k}$).}
    \label{tab:growths}
\end{table}

For further comparison, we also include the number of functions needed for a chain of $N_{\rm v}$ vertices in the kinematic flow algorithm of \cite{arkani-hamed_differential_2023,arkani-hamed_kinematic_2023}, which is given by $N_{\rm k}(N_{\rm v})=4^{N_{\rm v}-1}$. The table shows that, compared to the differential chain and the kinematic flow representations, the recursive reduction algorithm achieves a significant reduction in complexity. 

\paragraph{Pfaffian perspective on complexity reduction.}
To close this section, let us comment on how the reduction in complexity is quantified in the framework of Pfaffian chains. In principle, this is done by implementing the recursive reduction algorithm on the Pfaffian chain, thereby constructing a new Pfaffian chain with a lower Pfaffian complexity. Whenever there is an algebraic relation among functions in the chain, as in equation \eqref{eq:ITalgebraic} and \eqref{eq:ITmaxalgebraic}, a function can be eliminated from the chain, reducing the order $r$ by one. Since there are many such relations, the order $r$ will reduce significantly. 

However, there are two aspects of the reduction which are not captured by the Pfaffian framework. Firstly, the Pfaffian chain structure demands that we separately define all the letters $\ell^{(T)}_\cT$ by the differential equation \eqref{eq:PCletters}. The algebraic relations in the recursive reduction algorithm do not lead to a clear reduction in the number of letters needed for the minimal representation functions, so the order $r$ of the Pfaffian chain will still have a contribution $n_{\rm L}$ which grows exponentially in the number of vertices.

Secondly, part of the reductions in the recursive reduction algorithm require permutations among the variables in the integrals. This type of symmetry is however not detected by Pfaffian complexity, since it assumes a fixed ordering on the variables. For example, consider a Pfaffian chain containing a function $f(z_1,z_2)$. The function $g(z_1,z_2)=f(z_2,z_1)$, obtained by swapping the two variables, cannot be obtained as a Pfaffian function without adding it to the Pfaffian chain separately and increasing the complexity. This is closely related to the challenge in establishing a connection between complexity and symmetry, as pointed out in \cite{grimm_complexity_2024a}. We believe that this calls for a complexity framework in which symmetries of this form are more naturally detected, and we leave this as a future direction of research.

\section{Conclusion}\label{sec:conclusion}

In this paper, we have presented a novel strategy to study the space of perturbative cosmological correlators using reduction operators associated to GKZ systems. 
More precisely, we exploited the fact that the GKZ system derived for tree-level cosmological correlators of the model reviewed in section~\ref{sec:general} 
is highly reducible, and hence leads to a large set of additional differential operators, the reduction operators, that act on the solution space in a controlled way. We introduced an algorithm to construct these operators for any reducible GKZ system. For the cosmic GKZ system these operators are $Q^{(T)}$, $Q^{(T)}_{\pi}$ given in \eqref{eq:QTresult} and \eqref{higher-order-red_final}, which were constructed to only depend on the physical variables. We then showed that the first-order operators $Q^{(T)}$ lead to a closed system of first-order differential equations for the integrals of the tubings associated to a diagram. Afterwards, we have used the existence of the additional reduction operators $Q^{(T)}_{\pi}$ to obtain algebraic relations between various functions in this system. 

In our analysis it was essential to realize that reduction operators act by removing tubes from a tubing as shown with the key equations \eqref{eq:QTonint} and \eqref{eq:highordtuberemov}. This property allowed us to connect integrals connected to different tubings and led to a simple implementation of contractions and cuts of the diagram. This highlights the advantage of the operators $Q^{(T)}$, $Q^{(T)}_{\pi}$, since they interact nicely with the diagrammatic representation of the cosmological correlators using tubes and tubings. 
Taken all these insights together, we were led to propose a recursive reduction algorithm, which yields a set of minimal representation functions associated to each tree-level correlator.
Interestingly, this minimal representation turns out to be surprisingly small. For example, the double-exchange integral can be expressed in terms of just four of such functions as seen in table~\ref{tab:minrep}. This is in contrast to the sixteen basis functions necessary for the kinematic flow algorithm of \cite{arkani-hamed_differential_2023}.

We now highlight the key factors that led to the remarkably low number of basis functions required to parameterize a tree-level correlator for a fixed number of external momenta. The general approach begins with identifying all tree diagrams and their corresponding tubings. This leads to a vast combinatorial complexity, which is increasing with the number of interaction vertices in the action \eqref{model_action}. For each tubing, a GKZ system needs to be constructed, and the strategy outlined in this work applies to each case. However, the resulting space of integrals exhibits numerous non-trivial relations, which we systematically uncovered.
First, we demonstrated the existence of diagrammatical relations, where certain tubings of one diagram correspond to tubings of another diagram with an edge either contracted or cut. Second, we found that many remaining basis functions are related through permutations of the input variables $z^{(T)}$ and shifts of the vertex parameters $\alpha_v$, revealing that far fewer functions are genuinely independent. Such symmetries are natural in combinatorial problems, and we quantified their role in reducing the number of necessary functions. While our approach may introduce some additional combinatorial challenges, these are computationally straightforward. More importantly, the substantial reduction in the complexity of solving the differential equations for the remaining functions far outweighs these challenges.

Another goal of this work was to further develop the idea of establishing a quantifiable measure of complexity of any algorithm that is used to compute the cosmological correlators. As done for the kinematic flow algorithm in \cite{arkani-hamed_kinematic_2023}, we were again able to show that also the integrals associated to tubings fit into a Pfaffian chain. The Pfaffian framework then provides a measure of complexity, which can be used to give upper bounds on the number of zeros or poles of the function, but also on the computational complexity of the algorithm. It turned out that these bounds are rather weak and that the inclusion of the algebraic relations and the simplifications due to permutations and shifts into this construction is very challenging. There are two issues that we believe hinder us to present better estimates of the full complexity. Firstly, we know that the Pfaffian complexity is very sensitive to adding new functions, since the bounds also have to hold even for the worst-case solutions to a given Pfaffian system. However, a crucial part of the Pfaffian chain are the letters \eqref{eq:PCletters}. These are actually rather simple functions, but we did not succeed to find a simple representation to incorporate them. Secondly, we are not aware of a refined Pfaffian framework that incorporates symmetries and provides stronger bounds. We believe that both issues should be addressed in the future. Eventually, we hope to fully compare the complexity of algorithms. The algorithm giving the best bounds on the number of poles, which matches our physical exceptions, would then be have the most minimal representation of the information.

Even before addressing the challenging tasks surrounding complexity, there are many interesting future directions to explore. One interesting future direction is to further explore the space of tree-level correlators. There are interesting aspects about the interplay of first-order and higher-order operators that might simplify our discussion further. In particular, we expect that using the higher-order operators earlier in the reduction could be beneficial when focusing on the singularity structure of the correlator discussed in section~\ref{ssec:factors}. The next natural step is then to investigate  loop-level cosmological correlators, which can also be represented using tubings \cite{baumann_kinematic_2024}. Also at loop-level the main task will be to understand the space of reduction operators and which relations they impose on the amplitudes. We expect that much of our strategy carries over to these cases and it would be desirable to go through the construction in a follow-up investigation. Eventually, one can aim at finding a full-fledged recursion for the complete amplitude. Another natural extension is the application of reduction operators to cosmological correlators with massive propagators, or more generally, the examination of other phenomenological models replacing the conformally coupled scalar action \eqref{model_action}.

Finally, we emphasize that the approach developed in this work can also be applied in the study of perturbative quantum field theory amplitudes in flat space. As we will show in \cite{futurewRuth}, the results from Section \ref{GKZ-red-gen} can be used to identify reduction operators, which subsequently can be employed to construct a minimal set of functions to parameterize Feynman integrals. As seen for tree-level cosmological correlators, one challenge is finding reduction operators directly restricted to physical variables, while another is efficiently reducing the basis functions. However, given the extensive study of Feynman integrals, we believe these constructions are feasible for general diagrams. In the future work \cite{futurewRuth}, we will also explore how this formalism provides a systematic approach to uncovering algebraic relations between Feynman integrals.

\subsubsection*{Acknowledgements}

We are grateful to Daniel Baumann, Ruth Britto, Andrei Gabrielov, Arthur Lipstein, Paul McFadden, Dmitry Novikov,  Guilherme Pimentel, David Prieto, and Anna-Laura Sattelberger for insightful discussions and their useful comments. 
This research is supported, in part, by the Dutch Research Council (NWO) via a Vici grant.

\appendix

\section{First-order reduction operators in physical variables} \label{red_op_phys}

In this appendix we will show that the operator $Q^{(T)}$ from equation~\eqref{eq:QTdef_t} can be rewritten using only the physical coordinates. To show this, we first note that, when restricting to the physical slice $z^{(T)}_v\rightarrow1$, we have
\begin{equation}
    z^{(T)}_v\partial^{(T')}_{v}=\theta^{(T')}_{v}+\ldots
\end{equation}
for tubes $T$, $T'$ and all vertices $v$ in $T\cap T' $. Here, the $\ldots$ denote that this equality holds up to terms which go to zero under the restriction and we recall that $\theta^{(T)}_v=z^{(T)}_v\partial^{(T)}_v$. Then, considering the definition~\eqref{eq:QTdef_t} of $Q^{(T)}$ and inserting equation~\eqref{QTT'} for the various reduction operators, we find
\begin{equation}\label{eq:physQTderiv}
 \sum_{T'\supsetneq T}Q_{T,T'}=z^{(T)}  \sum_{T'\supsetneq T}\partial^{(T')} +\sum_{v\in T} \sum_{T'\supsetneq T} \theta^{(T')}_{v}+\ldots\,.
\end{equation}
Note that the first term on the right-hand side is already in terms of the physical variables only. Thus, we will now use the Euler operators of the GKZ system to rewrite the second term in this expression.

Recall that the Euler operators of the GKZ system associated to cosmological correlators are given by \eqref{Euler_cosm}. From these expressions, we construct the following useful combination of Euler operators
\begin{equation}\label{eq:physredeuler}
    \sum_{T'\subseteq T}\E_{T'}-\sum_{v\in T}\E_i=\sum_{T'\subseteq T} \theta^{(T')} +\sum_{T'\subseteq T}\sum_{v\in T'}\theta^{(T')}_{v}-\sum_{v\in T}\sum_{\{T':v\in T'\}}\theta^{(T')}_{v}\,,
\end{equation}
where the sum $\sum_{\{T':v\in T'\}}$ is over all tubes $T'$ containing $v$. Notice that, by the non-crossing condition, we have that $v\in T$ and $v\in T'$ if and only if either (1) $v\in T'$ and $T'\subseteq T$, or (2) $v\in T$ and $T'\supsetneq T$. Therefore, we find that
\begin{equation}
\sum_{v\in T}\sum_{\{T':v\in T'\}}\theta^{(T')}_{v}=\sum_{T'\subseteq T}\sum_{v\in T'}\theta^{(T')}_{v}+\sum_{T'\supsetneq T}\sum_{v\in T}\theta^{(T')}_{v}\,.
\end{equation}
Inserting this into equation~\eqref{eq:physredeuler}, we can solve for $\sum_{T'\supsetneq T}\sum_{v\in T}\theta^{(T')}_{v}$ and obtain
\begin{equation}
     \sum_{T'\supsetneq T}\sum_{v\in T}\theta^{(T')}_{v}=\sum_{v\in T}\E_i- \sum_{T'\subseteq T}\E_{T'}+\sum_{T'\subseteq T} \theta^{(T')} 
\end{equation}
When acting on solutions of the GKZ system, an Euler operator $\E_J$ may be replaced with $\nu_J$. Therefore, we have the equality
\begin{equation}
    \sum_{v\in T}\sum_{T'\supsetneq T}\theta^{(T')}_{v}\simeq_{\E+\nu}\sum_{T'\subseteq T}(\theta^{(T')} +\nu^{(T')})-\sum_{v\in T}\nu_i\,,
\end{equation}
where we recall that $\simeq_{\E+\nu}$ means that this equality holds only when acting on solutions of the GKZ system at the parameter $\nu$.

Finally, we insert this equation into equation~\eqref{eq:physQTderiv} and obtain
\begin{equation}
     Q^{(T)}\simeq_{\E+\nu}\sum_{T'\supsetneq T}Q_{T,T'}=z^{(T)}  \sum_{T'\supsetneq T}\partial^{(T')} +\sum_{T'\subseteq T}(\theta^{(T')} +\nu^{(T')})-\sum_{v\in T}\nu_i+\ldots\,.
\end{equation}
Recall that the $\ldots$ terms go to zero in the physical limit. Thus we find that, when acting on GKZ systems and in the physical limit, the combination of reduction operators will act as 
\begin{equation}
    \sum_{T'\supsetneq T}Q_{T,T'}\vert_{\rm phys}=z^{(T)}  \sum_{T'\supsetneq T}\partial^{(T')} +\sum_{T'\subseteq T}(\theta^{(T')} +\nu^{(T')})-\sum_{v\in T}\nu_i\,,
\end{equation}
which we identify as being $Q^{(T)}$ as stated in~\eqref{eq:QTresult}. 

\section{Matrix form of the first-order system}\label{ap:matrixrep}

This appendix is devoted to solve equation~\eqref{eq:QTiteration} as a matrix equation, thereby solving the iterative equation for $Q^{(T)}I_\cT$ in terms of linear combinations of integrals $I_\cS$ with rational coefficients. Thus, we must first fix a tubing $\cT$, such that we can obtain $Q^{(T)}I_\cT$ for each $T\in \cT$. Then, we must identify a suitable vector space, which enables us to keep track of both a tubing as well as a tube contained in this tubing. This leads us to construct a basis of vectors $e_{(S,\cS)}$, where $S$ is a tube contained in $\cS$, and $\cS$ is a subset of $\cT$. Furthermore, it will be useful to take linear combinations of such pairs, which we will denote as $c \cdot e_{(S,\cS)}$ where $c$ is some matrix of coefficients. Note that, in the end, we will take $c$ to be rational functions in the $z$. Using this notation, we obtain the vector space of all such formal combinations as
\begin{equation}
    \mathbf{V}\coloneqq \big\{ \, \sum_{(S,\cS)} c_{(S,\cS)} e_{(S,\cS)}\; : \;S\in \cS ,\,\cS \subseteq \cT\, \big\}\,.
\end{equation}
This vector space will be the key in solving for $Q^{(T)}I_\cT$.

To relate this vector space to the actual integrals we are trying to solve for, we must first define a mapping between the two. Thus, we begin by defining the integral mapping $\mathrm{int}$, which sends each basis element $\mathbf{V}$ to an integral and extending linearly. In other words, we have
\begin{equation}
  \mathrm{int}(e_{(S,\cS)})=I_{\cS}
\end{equation}
for the basis elements.  Note that this function does not take into account the tube $S$ in $e_{(S,\cS)}$.

Now, we will define a variety of operators on $\mathbf{V}$ such that we can rewrite~\eqref{eq:QTiteration} in terms of matrices acting on $\mathbf{V}$. We begin by defining the operator $\mathbf{Q}$ implicitly, using equation~\eqref{eq:QTiteration}. In particular, we will use that, as we know that equation~\eqref{eq:QTiteration} can be solved iteratively, each combination $Q^{(S)}I_{\cS}$ must be a linear combinations of integrals $I_{\cS'}$. Therefore, there must exist an operator $\mathbf{Q}$ such that
\begin{equation}
   \mathrm{int}(\mathbf{Q} \cdot e_{(S,\cS)})=Q^{(S)} I_\mathcal{S}
\end{equation}
for each combination $(S,\cS)$.
Note that, since the operator $\mathrm{int}$ does not take into account the tube $S$, there is some ambiguity in this definition of $\mathbf{Q}$. However, since eventually we will always apply $\mathrm{int}$ to the equations which we obtain, this ambiguity will not be important for us. Therefore, we are free to fix $\mathbf{Q}$ such that its image lies in the set
\begin{equation}
    \emptyset \times \mathcal{PT}=\big\{ \,(\emptyset,\cS)\;\vert\; \cS\subseteq \cT\,\}\,
\end{equation}
where $\mathcal{PT}$ is the power set of $\cT$.

It turns out that equation~\eqref{eq:QTiteration} implies that $\mathbf{Q}$ must satisfy an analogous matrix equation. In particular, we will define three operators $\mathbf{A}$, $\mathbf{L}$ and $\mathbf{G}$ on the vector space $\mathbf{V}$, and solve for $\mathbf{Q}$ in terms of these operators. We begin with a matrix $\mathbf{A}$, which is defined as 
\begin{equation}
    \mathbf{A}_{(S,\cS),(S',\cS')}=\left\{ \begin{array}{rll}
    1 & \text{if } S=S' &\text{and } \cS=\cS'\, ,\\
    -1 & \text{if } S\prec S' &\text{and } \cS=\cS'\, , \\
    0 & \text{otherwise}\, ,&
\end{array} \right.
\end{equation}
where we recall from section~\ref{ssec:chainconstruction} that $S\prec S'$ implies that $S$ is a maximal tube contained strictly in $S'$.  Additionally, we will define the matrix $\mathbf{L}$ as
\begin{equation}
    \mathbf{L}_{(S,\cS),({S' ,\cS')}}=\left\{ \begin{array}{ll}
    \ell^{(S')}_{\cS'} & \text{if } S\succ S' \text{ and } \cS=\mathcal{S}'\setminus\{T'\}\\
    0 & \text{otherwise}
\end{array} \right.\, ,
\end{equation}
where $\ell^{(S')}_{\cS'}$ are the letters from \eqref{eq:letterM}.\footnote{One can also define $\mathbf{L}$ in terms of $1/p^{(S')}_{\cS'}$, with $p^{(S')}_{\cS'}$ the denominator in equation~\eqref{eq:QT-QS}.} Observe that, while the actions of $\mathbf{L}$ and $\mathbf{A}$ on the whole of $\mathbf{V}$ is rather involved, its action can be obtained element-wise quite easily. Finally we will define the operator $\mathbf{G}$ acting as
\begin{equation}
   \mathbf{G}_{(S,\cS),{(S',\cS')}}=\left\{ \begin{array}{ll}
    \gamma^{(S')}_{\cS'} & \text{if } S=\emptyset \text{ and } \cS=\cS'\\
    0 & \text{otherwise}
\end{array} \right.\, ,
\end{equation}
with $\gamma^{(S')}_\mathcal{S'}$ as in equation~\eqref{eq:gammadef}.

With all of this notation, equation~\eqref{eq:QTiteration} can be written as an equation for $\mathbf{Q}$, which is given by
\begin{equation}
    \mathbf{Q}=\left(\mathbf{Q}\mathbf{A}-\mathbf{G}\right)\mathbf{L}\,.
\end{equation}
This can simply be solved for $\mathbf{Q}$, from which we find that
\begin{equation}
    \mathbf{Q}=-\mathbf{G} \mathbf{L}\left(1-\mathbf{A}\mathbf{L}\right)^{-1}\,.
\end{equation}
Note that, as $\mathbf{A}\mathbf{L}$ is nil-potent, this can also be written as 
\begin{equation}
    \mathbf{Q}=-\mathbf{G} \mathbf{L} \sum_{i=0}^k \left(\mathbf{A}\mathbf{L}\right)^i\,.
\end{equation}
where $k$ is the nil-potent degree of $\mathbf{A}\mathbf{L}$. 

Then, equation~\eqref{eq:QT-QS} can be written as
\begin{equation}
   Q^{(T)}I_\mathcal{T} = -\sum_{i=0}^k\sum_{(S,\cS)}\left( \mathbf{G} \mathbf{L} (\mathbf{A}\mathbf{L})^i\right)_{(S,\cS),(T,\cT)}I_{\mathcal{T'}}\,.
\end{equation}
Note that the image of $\mathbf{G}$ is contained in $\emptyset \times \mathcal{PT}$, with $\mathcal{PT}$ the set of all possible tubings, confirming that the image of $\mathbf{Q}$ is as well. Furthermore, recall that in the construction of $\mathbf{V}$, we only considered tubings $\cS$ that are subsets of $\cT$. Therefore, the equation above can be written as
\begin{equation}
           Q^{(T)}I_\mathcal{T} = -\sum_{i=0}^k\sum_{\cS\subseteq \cT}\left( \mathbf{G} \mathbf{L} (\mathbf{A}\mathbf{L})^i\right)_{(\emptyset,\cS),(T,\cT)}I_{\mathcal{S}}\,,
\end{equation}
and we have found that, using the matrices $\mathbf{G}$, $\mathbf{L}$ and $\mathbf{A}$, one can solve the iterative equation for $Q^{(T)}I_\cT$ in terms of the integrals $I_\cS$.

\bibliographystyle{utphys}

\providecommand{\href}[2]{#2}\begingroup\raggedright\endgroup

\end{document}